\documentclass[conference]{IEEEtran}

\usepackage[noadjust]{cite}
\usepackage{multirow} 
\usepackage{algpseudocode}
\usepackage{algorithm}
\usepackage{rotating}
\usepackage{kantlipsum} 
\usepackage{commath}
\allowdisplaybreaks
\usepackage{mathtools}  
\usepackage{verbatim}
\usepackage{xr-hyper} 
\usepackage{enumitem}

\usepackage{epstopdf}
\usepackage{wrapfig}
\usepackage{latexsym}
\usepackage{amssymb}
\usepackage{amsthm}
\usepackage{amsfonts}
\usepackage{amsmath} 
\usepackage{graphicx}
\usepackage{latexsym}
\usepackage{booktabs}
\usepackage{support-caption} 
\usepackage{subcaption} 
\usepackage{xspace}

\usepackage[normalem]{ulem} 

\usepackage[T1]{fontenc}

\usepackage{hyperref}

\usepackage[maxfloats=256]{morefloats}
\ifCLASSINFOpdf
\else
\fi
\usepackage{array}
\newcommand{\dq}[1]{``#1''}

\usepackage[dvipsnames]{xcolor}

\newcommand{\commentBy}[3]{\textcolor{#1}{\textbf{#2:} #3}}

\newif\ifcommentson
\commentsontrue

\newcommand{\ste}[1]{\ifcommentson \commentBy{blue}{SS}{#1} \fi}

\newcommand{\AM}[1]{\ifcommentson \commentBy{orange}{AM}{#1} \fi}
\newcommand{\pl}[1]{\ifcommentson \commentBy{green}{PL}{#1} \fi}

\newif\ifextended
\newif\ifshortver

\extendedtrue

\newcommand{\extended}[1]{\ifextended \ifshortver \textcolor{purple}{#1} \else \textcolor{black}{#1} \fi  \fi}
\newcommand{\shortver}[1]{\ifshortver \ifextended \textcolor{blue}{#1} \else \textcolor{black}{#1} \fi \fi}

\newif\ifrevision

\begin{document}

\bstctlcite{IEEEexample:BSTcontrol}


\title{PASTRAMI: Performance Assessment of SofTware Routers Addressing Measurement Inaccuracy}

\author{\IEEEauthorblockN{
Paolo Lungaroni\IEEEauthorrefmark{1},
Andrea Mayer\IEEEauthorrefmark{1}\IEEEauthorrefmark{2},
Stefano Salsano\IEEEauthorrefmark{1}\IEEEauthorrefmark{2},
Pierpaolo Loreti\IEEEauthorrefmark{1}\IEEEauthorrefmark{2},
Lorenzo Bracciale\IEEEauthorrefmark{1}\IEEEauthorrefmark{2}
}
\IEEEauthorblockA{
\IEEEauthorrefmark{1}University of Rome Tor Vergata,
\IEEEauthorrefmark{2}CNIT
}
\vspace{2ex}
\vspace{-4ex}
\\ 
\extended{\textbf{Extended version of submitted paper - v02 - February 2026}}
}

\markboth{Journal of \LaTeX\ Class Files,~Vol.~14, No.~8, August~2015}%
{Shell \MakeLowercase{\textit{et al.}}: Bare Demo of IEEEtran.cls for IEEE Journals}
%



\maketitle

\begin{abstract}
Virtualized environments offer a flexible and scalable platform for evaluating network performance, but they can introduce significant variability that complicates accurate measurement. This paper presents \textbf{PASTRAMI}, a methodology designed to assess the accuracy of performance measurements of software routers. In particular we address the accuracy of performance metrics such as the Partial Drop Rate at 0.5\% (PDR@0.5\%). While PDR@0.5\% is a key metric to assess packet processing capabilities of a software router, its reliable evaluation depends on consistent router performance with minimal measurement variability. Our research reveals that different Linux versions exhibit distinct behaviors, with some demonstrating non-negligible packet loss even at low loads and high variability in loss measurements, rendering them unsuitable for accurate performance assessments. This paper proposes a systematic approach to differentiate between stable and unstable environments, offering practical guidance on selecting suitable configurations for robust networking performance evaluations in virtualized environments.
\end{abstract}

\begin{IEEEkeywords}
Virtualized environments, software routers, Partial Drop Rate (PDR), network performance evaluation, performance metrics, network measurement variability.
\end{IEEEkeywords}

%
\IEEEpeerreviewmaketitle

\section{Introduction}
\label{sec:intro}

Virtualization is a technology that allows multiple virtual instances of computing resources—such as servers, storage devices, networks, or operating systems—to be created and managed on a single physical hardware system. This abstraction enables efficient utilization of resources, isolation, and flexibility in deploying and managing applications \cite{anand2021need}. 
A looser approach is containerization. Containerization is a technology that allows for the packaging of an application and its dependencies into a single, standardized unit called a container. A container can run consistently across different computing environments, such as development, testing, and production. Both virtualization techniques can be combined with each other by expanding the capabilities and resources available on a single hardware server. Another advantage brought by virtualization is the decoupling between software and hardware resources, allowing flexibility and resilience and scalability of services within the infrastructure in a dynamic way \cite{bhardwaj2021virtualization}.


In this specific context, networks have undergone a significant evolution, progressing from a physical to a virtual environment. Virtual networking allows for the dynamic interconnection of virtual resources and services through virtual/software switches and routers. Software routers facilitate the interconnection \cite{li2021joint} between a virtualized network and a physical one, enabling integration with other services or networks, e.g., Internet.
Virtual networking is a critical component in modern datacenter and Operator infrastructures. By accurately measuring networking performance, operators are able to optimize service delivery for high-demand resources and meet the stringent requirements imposed by large data flows \cite{li2018virtual}. 


Network performance is often assessed with expensive hardware and software. These devices are unsuitable for dynamic deployment in data centers due to their physical constraints. Virtualized environments are more efficient and flexible for assessing network performance. They enable organizations to test different configurations, simulate traffic scenarios, and enhance network performance and stability. While virtualization offers significant advantages, it is essential to address the challenges and constraints associated with this approach to ensure accurate assessments \cite{sonmez2018edgecloudsim, nardini2020simu5g}.

This paper presents PASTRAMI, a methodology designed to evaluate the accuracy of performance measurements in virtual environments and software routers. In particular, we focus on the accuracy the Partial Drop Rate at 0.5\% (PDR@0.5\%) performance metric. Although PDR@0.5\% is a key metric for evaluating the packet processing capabilities of a software router, its reliable evaluation depends on consistent router performance with minimal variability in measurements. Our research has shown that different versions of the Linux kernel exhibit distinct behaviors, with some experiencing non-negligible packet losses even at low loads and considerable variability in loss measurements, making them unsuitable for accurate performance evaluations. PASTRAMI provides a systematic approach to distinguish between stable and unstable environments, offering practical guidance on selecting the most suitable configurations for robust network performance evaluations in virtualized environments.

\section{Background and Related Work}
\label{sec:background}

The Partial Drop Rate (PDR) at 0.5\% (PDR@0.5\%) metric, provides a way to assess the performance of a system under test (SUT) by identifying the highest input rate that allows for a \textit{Delivery Ratio} (DR) of at least 99.5\%. DR is defined as the ratio between the output and input packet rates of the SUT. This metric, along with the No-Drop Rate (NDR), offers a useful tool to compare performance across different devices or configurations. While NDR represents the maximum rate without any packet loss, PDR@X\% incorporates an acceptable packet loss threshold, making PDR@0.5\% a more realistic measure for systems that may experience small packet losses due for example to limitations in CPU or NIC interrupt handling. The concept of PDR@0.5\% is described in a technical report \cite{csit-report} produced by the CSIT (Continuous System Integration and Testing) group of the FD.io project \cite{fdio}. The use of PDR@0.5\%metric is discussed in various studies, assessing the performance of software switches~\cite{osinski2022novel}, segment routing~\cite{abdelsalam2020srperf, team2020segment, MAYER2021107705} and Network Function Virtualization \cite {cotroneo2017nfv}.


By reducing the entire load-performance relationship to a single scalar value, PDR@0.5\% simplifies performance evaluation. This makes it easier to quantitatively compare various systems, configurations, or software implementations, without needing to analyze the full throughput curve. The typical example along this line is to use PDR@0.5\% to evaluate the performance of different variants or versions of packet processing functions in software routers, or the performance cost of introducing additional functionality in the packet processing operations \cite{konstantynowicz2020multiple}.


The PDR@0.5\% metric offers a convenient summary of performance, but may need to be limited to stable environments. Reducing complex performance to a single value can obscure nuances in the system's response to varying loads. The metric doesn't account for performance at lower or higher loads. In environments where performance may fluctuate, PDR@0.5\% is problematic. This metric overlooks fluctuations in throughput or other issues that may emerge along the load-loss curve. PDR@0.5\% should be applied with caution and supplemented with additional metrics.

Another limitation of using PDR@0.5\% exclusively is that it assumes the system operates stably up to that point, ignoring any variability in performance. In practice, however, many systems exhibit performance fluctuations under certain conditions, especially in virtualized environments, where factors like CPU contention or network stack inefficiencies can lead to inconsistent results. These issues are not captured by PDR@0.5\%, which may result in overly optimistic conclusions about the robustness of a system.


Note also that, by definition, PDR@0.5\% does not reflect system behavior under high loads, equal to or exceeding the system saturation throughput.

\section{Reference configurations}
\label{sec:reference_confs}

\begin{figure*}[ht!]
    \centering
    \includegraphics[width=1.0\linewidth]{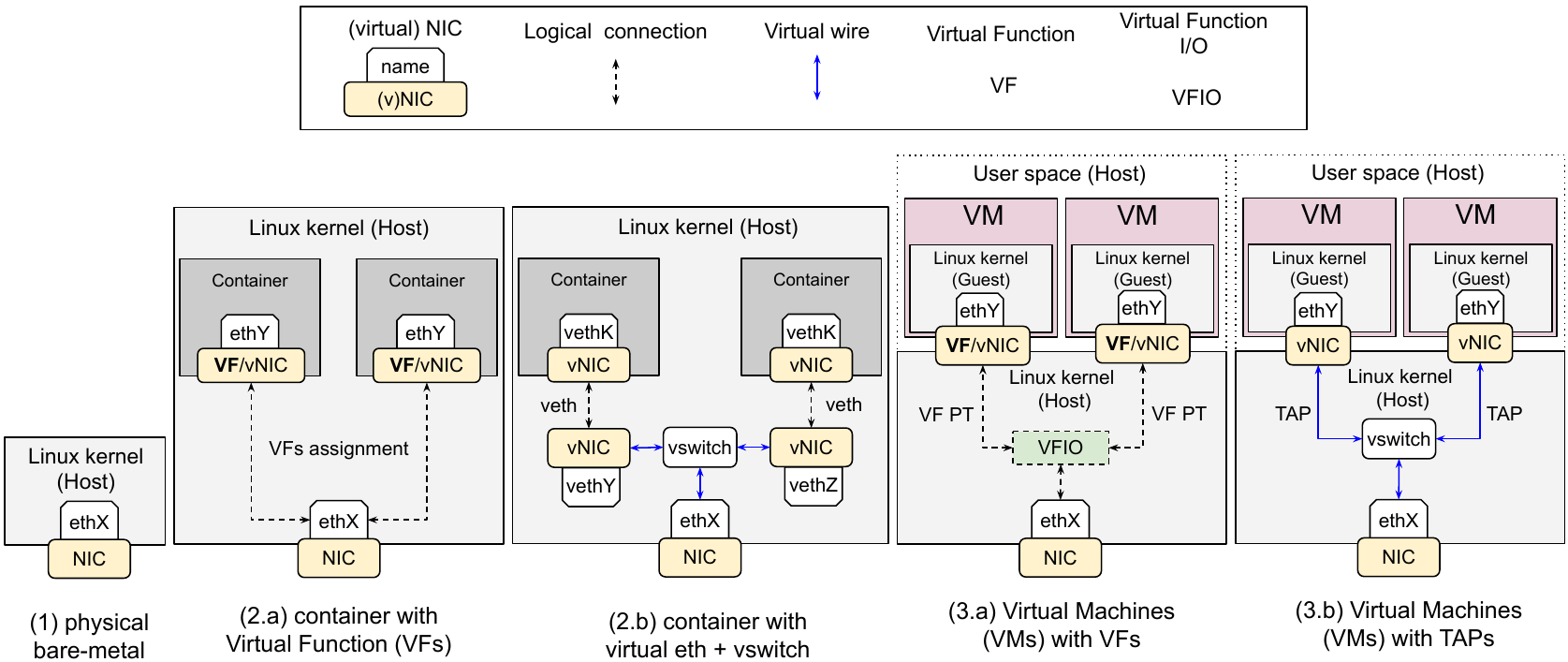}
    \caption{Reference configurations}
    \label{fig:reference-config-topo}
\end{figure*}

\begin{figure*}[ht!]
    \centering
    \includegraphics[width=1.0\linewidth]{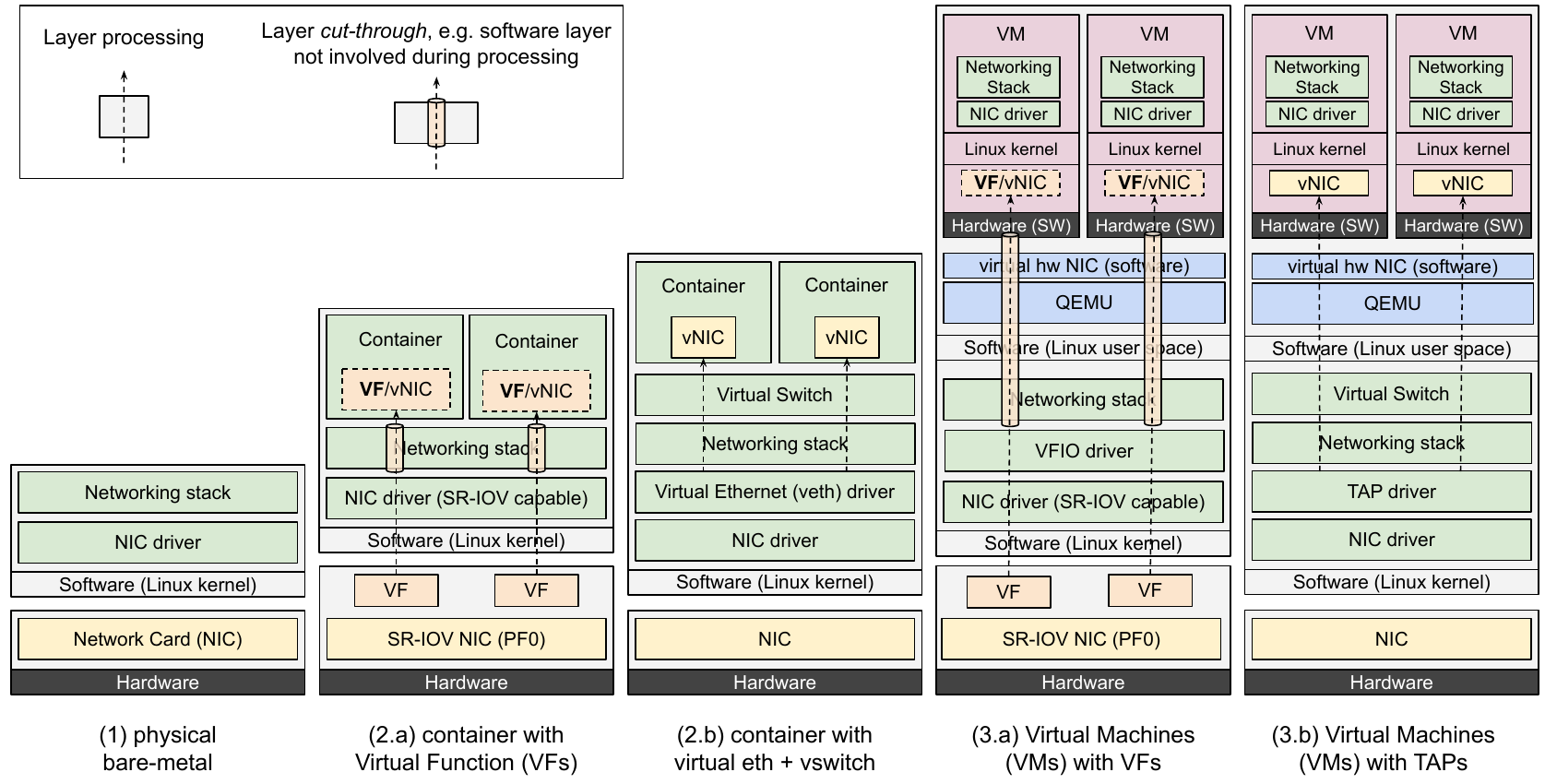}
    \caption{Layer impacting performance in different example configurations}
    \label{fig:layer-impacting-perf}
\end{figure*}

Virtualized environments are often used for evaluating network capabilities, but they pose significant challenges: dynamic resource allocation and virtualization overhead. Resource allocation in virtualized environments is flexible, with variations in CPU, memory, and bandwidth that can impact performance. Additionally, the overhead introduced by the virtualization layer can interfere with network measurements, making it difficult to isolate and assess individual components accurately. Fig.~\ref{fig:reference-config-topo} illustrates the different types of virtualization environments that should be considered when measuring network performance.

Considering Fig.~\ref{fig:reference-config-topo}-(1), the virtualization environment is very basic, as it uses only the networking capabilities provided by the Linux kernel that runs natively on the hardware. In this case, the Linux environment is used without any kind of network virtualization.

Referring to Fig.~\ref{fig:reference-config-topo}-(2.a), the virtualization environment is based on Linux Containers, e.g., LXC \cite{incuslxc}, Docker \cite{merkel2014docker}.
This technology allows applications and services to run in an environment isolated from the Host operating system, while still sharing resources such as memory, CPU, etc. In Fig.~\ref{fig:reference-config-topo}-(2.a), however, Containers use dedicated virtual NICs to communicate with each other and the outside world. Such virtual NICs are called Virtual Functions (VFs). By using VFs, a physical network adapter (supporting SR-IOV \cite{sriov}) can be partitioned into multiple virtual adapters. Each virtual adapter can operate independently and it appears to the Host as a physical network card, with its own PCI address.
Virtualization in this configuration adds minimal overhead as shown in Fig.~\ref{fig:layer-impacting-perf}-(2.a). Packet processing is primarily carried out by the network card driver, which supports SR-IOV and the Linux networking capabilities within the Container.

In consideration of the topology depicted in Fig.~\ref{fig:reference-config-topo}-(2.b), the virtualization environment relies on Virtual Ethernet Pairs (veths) \cite{veth} to facilitate communication between the Host and Containers. 
In Linux, a veth is defined as a pair of virtual Ethernet devices (endpoints) that are connected to each other. Packets are sent to one endpoint and received by the other. A veth pair functions like an Ethernet cable, enabling communication between network namespaces, e.g. foundation block of the Linux Containers.

Virtual Ethernet pairs are entirely software-based, so there is no need for costly hardware-enabled Virtual Functions to connect the Host and Containers. Fig.~\ref{fig:layer-impacting-perf}-(2.b) indicates that additional packet processing is needed to exchange packets between the Host and Containers. The Virtual Ethernet driver manages these interfaces and handles packets sending/receiving operations. A software-based virtual switch is needed to connect Containers to the Host network, increasing the overall processing overhead. 


Fig.~\ref{fig:reference-config-topo}-(3.a) illustrates an alternative virtual environment topology using Virtual Machines (VMs) instead of Linux Containers. A VM is typically implemented on Linux using QEMU/KVM, \cite{qemu, kvm} a robust virtualization solution that enables users to run multiple isolated operating systems (Guests) on a single machine. QEMU/KVM provides virtual hardware to Guests, including CPUs, memory management, and I/O virtualization. VMs running on Linux execute as processes and cannot directly interact with the Host hardware. VM access to Host hardware, including the NICs, is enabled by device pass-through. Fig.~\ref{fig:reference-config-topo}.(3.a) shows how Virtual Function I/O (VFIO) \cite{vfio} kernel driver allows a VM access hardware directly. In this scenario, the Host is equipped with a SR-IOV-capable NIC and the VFIO driver is used to make VFs available to the Guests. Fig.~\ref{fig:layer-impacting-perf}-(3.a) reveals that when VFs are passed through to VMs, the overall overhead is reduced to bare minimum as the VMs can directly access the NIC hardware resources of the Host.

When VFs or SR-IOV NICs are unavailable on the Host, VMs can use virtual NICs, as shown in Fig.~\ref{fig:reference-config-topo}-(3.b). QEMU can emulate multiple network cards, which Guests see as hardware NICs (e.g. PCI cards). Guest NICs can be connected to the Host network through a variety of mechanisms, frequently involving the use of virtual Host Ethernet devices (TAPs) \cite{tuntap}, as reported in Fig.~\ref{fig:layer-impacting-perf}-(3.b). The TAP device bridges the VM and Host network, enabling VM packet transmission and reception.
Fully software-emulated NIC solutions may have performance issues. QEMU provides optimization techniques (whenever possible), including VirtIO \cite{virtio} and paravirtualization, to reduce packet processing overhead and facilitate communication between the Guests and the Host. 

\section{Methodological aspects}
\label{sec:methodological_aspects}

\subsection{Metrics for load-loss and load-delivery analysis}

We define the offered load $O$ [p/s] as the rate of packets that are sent to an incoming interface of the system under test (SUT). For simplicity, we now assume that all packets have the same size at the IP layer $S$ [Bytes]. We also assume that the offered load is constant over time in our observation interval of duration $T$: $O(t)=O$ for each $t \in T$.

Let us now count the number $N$ of packets that are correctly forwarded by the SUT in the interval $T$. The average throughput of the SUT over $T$ is $R(O)=\frac{N}{T}$ [p/s], where $R(O) \leq O$.

We can count the number of Lost Packets $L$ over the interval $T$ as $L = (O \cdot T) - N$ (all packets that have NOT arrived are lost!). We define the Packet Loss Ratio (PLR) of the SUT for an offered load $O$ over an observation interval $T$ as:
\[
PLR(O) = \frac{L}{O \cdot T}
\]
The PLR can also be expressed as follows:
\[
PLR(O) = \frac{O \cdot T - N}{O \cdot T} = 1 - \frac{N}{O \cdot T} = 1 - \frac{R(O)}{O}
\]
We define the Delivery Ratio $DR$ as $DR(O) = \frac{R(O)}{O}$, and it can be expressed as a function of the Packet Loss Ratio:
\begin{equation}
DR(O) = 1 - PLR(O)
\label{eq:dr_plr}
\end{equation}

In the above discussion, we simply assumed that we can count the number $N$ of packets forwarded with an experiment of duration $T$. In the real world, when we perform a number of $K$ repetitions of an experiment (each with duration $T$) with the same offered load $O$, we will have in general $K$ different values of $N$ and therefore of $L$: $L(O, i)$ for $i \in [0,K-1]$. For this reason, we resort to a statistical characterization of our metrics $PLR(O)$, $T(O)$, and $DR(O)$. 

We define $PLR(O)$ as:
\[
PLR(O) = \frac{1}{K} \sum_{i=0}^{K-1} \frac{L(O, i)}{O \cdot T}
\]

The standard deviation of the measurements, denoted as $\sigma_{\text{PLR}}(O)$, based on a population sample is given by:
\begin{equation}
\sigma_{\text{PLR}}(O) = \sqrt{\frac{1}{K-1} \sum_{i=0}^{K-1} \left( \frac{L(O, i)}{O \cdot T} - PLR(O) \right)^2}
\label{eq:sigma_plr}
\end{equation}

We are interested in evaluating the reliability of our estimated metric $PLR(O)$ (the same reasoning can be applied to $T(O)$, and $DR(O)$). In particular, we consider the confidence interval at 95\%, defined as:

\[
P(\text{LCL}_{95} < PLR(O) < \text{UCL}_{95}) = 0.95
\]

where \(\text{LCL}_{95}\) and \(\text{UCL}_{95}\) are the lower and upper confidence limits for a 95\% confidence interval. They respectively represent the lower and upper bounds of the 95\% confidence interval, defining the range within which the true $PLR(O)$ is expected to lie with 95\% confidence.

When the standard deviation of the population \(\sigma\) is unknown and estimated using the sample standard deviation $\sigma_{\text{PLR}}(O)$ from \(K\) independent observations, the lower and upper confidence limit for the 95\% confidence interval of the true parameter $PLR(O)$ are given by:

\[
\text{LCL}_{95} = PLR(O) - t_{K-1, \, 0.025} \cdot \frac{\sigma_{\text{PLR}}(O)}{\sqrt{K}}
\]

\[
\text{UCL}_{95} = PLR(O) + t_{K-1, \, 0.025} \cdot \frac{\sigma_{\text{PLR}}(O)}{\sqrt{K}}
\]

where:
\begin{itemize}
    \item \(t_{K-1, \, 0.025}\) is the critical value from the Student's t-distribution with \(K-1\) degrees of freedom (97.5th percentile, reflecting the 95\% confidence level),
    \item $\sigma_{\text{PLR}}(O)$ is the standard deviation, calculated as in Eq.~\ref{eq:sigma_plr},
    \item \(K\) is the number of experiments.
\end{itemize}

The use of the \(t\)-distribution accounts for the additional uncertainty due to estimating the variance from a sample rather than knowing it exactly\footnote{In the evaluation sections, we used the table of the t-distribution available at \url{https://en.wikipedia.org/wiki/Student's_t-distribution}}.

To evaluate the uncertainty of our estimate \( PLR(O) \), we introduce a normalized metric \( \Delta_{\text{UCL}_{95}} \) that considers the upper bound of the 95\% confidence interval, defined as:

\[
\Delta_{\text{UCL}_{95}} = \frac{\text{UCL}_{95} - PLR(O)}{PLR(O)}
\]

The metric \( \Delta_{\text{UCL}_{95}} \) provides a dimensionless measure of the relative uncertainty in the estimate of \( PLR(O) \). A smaller value of \( \Delta_{\text{UCL}_{95}} \) indicates that the upper confidence limit is close to \( PLR(O) \), suggesting higher precision in the estimate. In contrast, a larger value indicates that there is a greater uncertainty in the estimation relative to its magnitude.

\subsection{Ideal vs. real load-loss curves}

\begin{figure}[ht!]
    \centering
    \includegraphics[width=0.8\columnwidth]{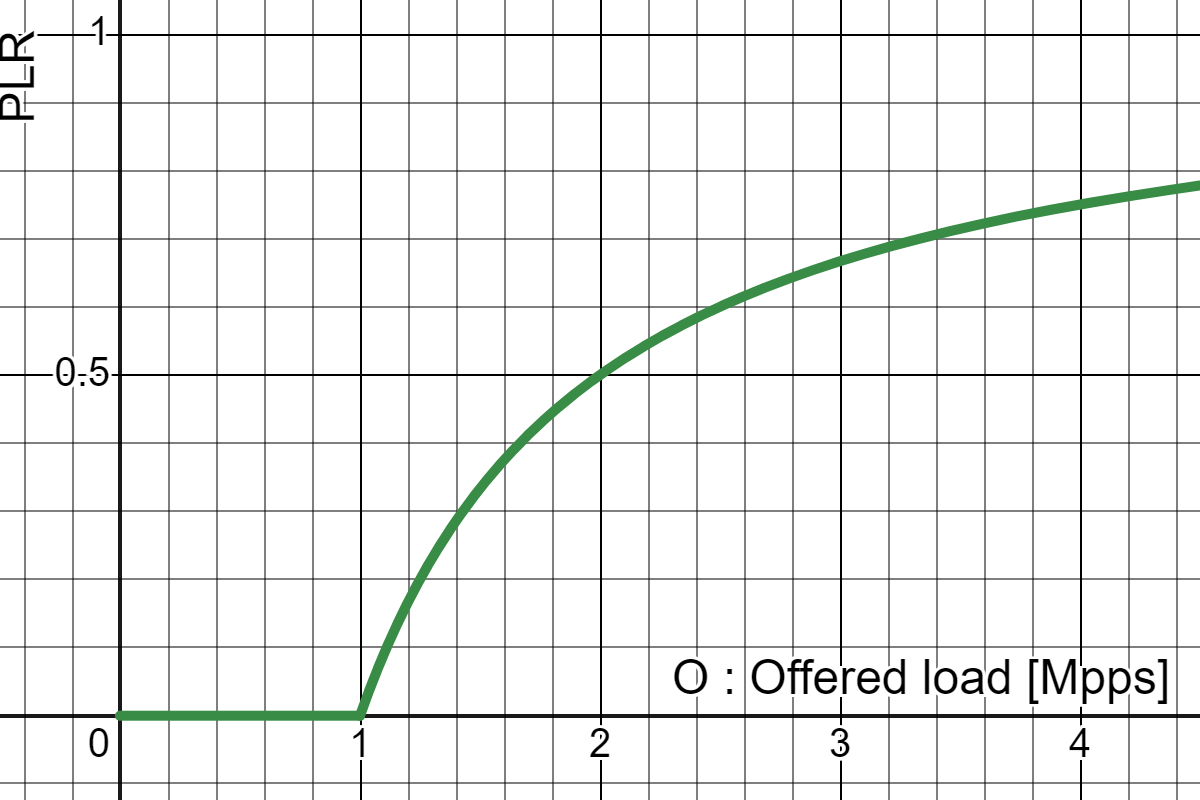}
    \caption{Ideal load-loss curve (saturation offered load is 1 Mpps, the saturation throughput equals the saturation offered load.}
    \label{fig:load_loss_curve}
\end{figure}

Let us define the concepts of \textit{saturation throughput} and  \textit{ideal load-loss curve}. The saturation throughput, denoted as $T^\text{sat}$, is the maximum throughput that a system can sustain before packet losses increase significantly. It is reached when the offered load $O$ equals the saturation offered load $O^\text{sat}$. For loads beyond this point, the throughput remains constant:

\[
T(O) = T^\text{sat},\quad O \geq O^\text{sat}
\]

where $T^\text{sat} = O^\text{sat} \cdot (1 - PLR(O^\text{sat}))$. In an ideal system, the load-loss curve starts with a packet loss ratio (PLR) of 0, which persists for all offered loads up to the saturation point \( O^\text{sat} \). For loads greater than $O^\text{sat}$, the PLR increases according to Eq.~\ref{eq:plr_sat} (see Fig.~\ref{fig:load_loss_curve}).

\begin{equation}
PLR(O) = \frac{O - T^\text{sat}}{O} = 1 - \frac{T^\text{sat}}{O}
\label{eq:plr_sat}
\end{equation}

In a realistic system, the packet loss starts at 0, then it initially remains on the order of $10^{-8}$ to $10^{-6}$, and then smoothly increases up to a loss ratio of $10^{-2}$ (i.e., 1\% loss) This behavior arises due to factors such as the interrupt-driven packet processing, limited buffering capacity, or other inefficiencies in the packet forwarding mechanisms. These small losses occur even at relatively low loads, before the system reaches \( O^\text{sat} \).

When the offered load exceeds \( O^\text{sat} \), the throughput in a realistic system may begin to decline, rather than staying constant as in the ideal case. This can be described by the following function:

\begin{equation}
T(O) = f(O - O^\text{sat}) \cdot T^\text{sat},\quad O > O^\text{sat}
\label{eq:realistic_T}
\end{equation}

where $f(0) = 1$, $f(x) > 0$, and $f(x)$ is a non-increasing function. 

For example, we can identify a threshold $O^\text{trash}$ for the offered load, such that the system goes into trashing and decreases its throughput with increasing offered load. Just as an example, this can be expressed by defining a stepwise linear $f(x)$ to be used in Eq.~\ref{eq:realistic_T} as follows:

\begin{equation}
f(x) =
\begin{cases} 
1 & O^\text{sat} < x < O^\text{trash}, \\
\max\left(0, 1 - \frac{x - O^\text{trash}}{M}\right) & x \geq O^\text{trash}
\end{cases}
\label{eq:realistic_f(x)}
\end{equation}

where $M > 0 \quad [Mpps]$ is a positive constant with dimension mega packets/s. This adjustment reflects that, in real-world conditions, factors such as congestion, buffer overflow, or resource limitations can cause the throughput to decrease when the load exceeds the saturation threshold.

\begin{figure}[ht!]
    \centering
    \includegraphics[width=0.9\columnwidth]{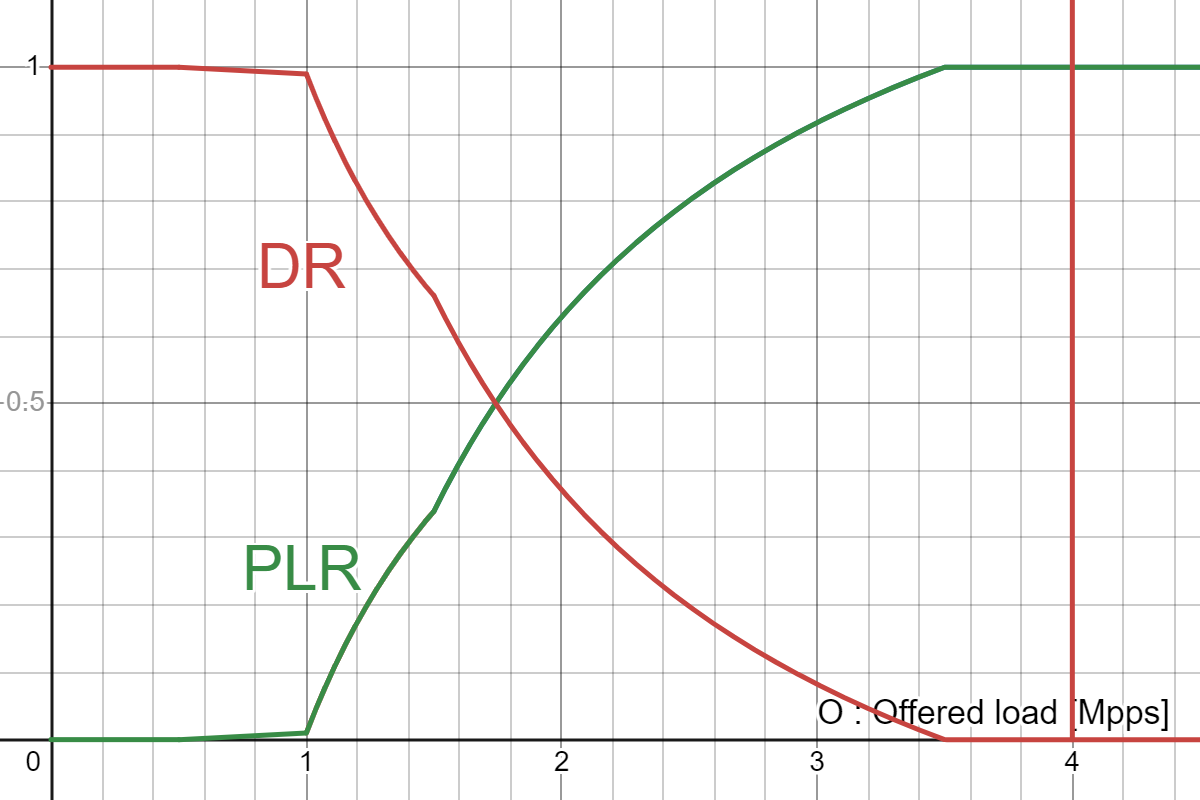}
    \caption{A more realistic synthetic load-loss curve (in green) along with the load-delivery curve (in red). The saturation offered load is 1 Mpps, the saturation throughput equals 0.99 of the saturation offered load (PLR at saturation is 1\%), the trashing threshold $O^\text{trash}$ is 1.5 Mpps and $M$ is 1 Mpps.}
    \label{fig:real_loss_curve}
\end{figure}

To sum up, in a realistic system the load-loss curve is characterized by non-zero PLR values before \( O^\text{sat} \), by a non-zero \( PLR(O^\text{sat}) \) and its throughput may even decline when the offered load exceeds the saturation point. In Fig.~\ref{fig:real_loss_curve} the green line represents a realistic synthetic loss curve $PLR(O)$ generated using Eq.~\ref{eq:realistic_T} and Eq.~\ref{eq:realistic_f(x)}.

\begin{figure}[ht!]
    \centering
    \includegraphics[width=\columnwidth]{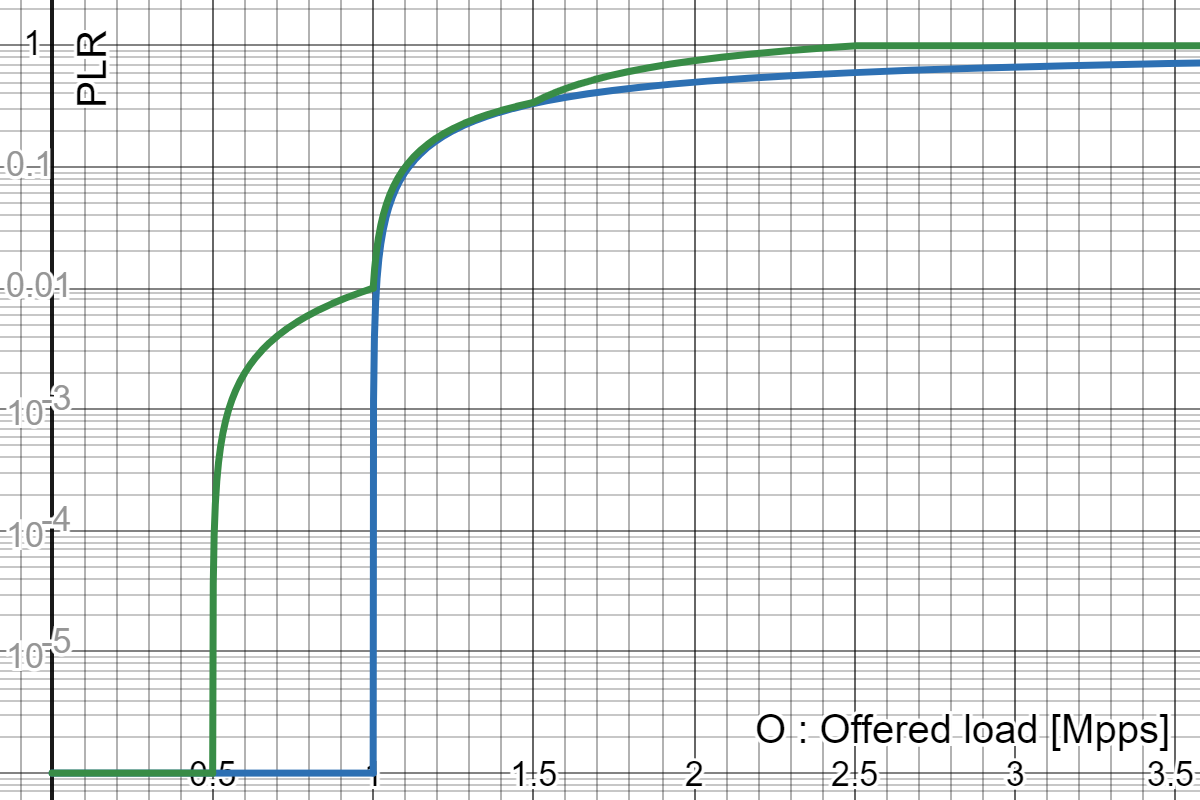}
    \caption{Both the ideal (blue) and the more realistic (green) load-loss curves are plotted in log scale, considering as the bottom of scale a loss ratio of $10^{-6}$.}
    \label{fig:real-and-ideal-log-curve}
\end{figure}

Plotting the load-loss curve in linear scale as in Figs.~\ref{fig:load_loss_curve} and \ref{fig:real_loss_curve} (green line) does not provide useful information for the region of low offered load/low loss represented on the left part of the figure. For this reason, we can use a logarithmic scale to represent the loss ratio on the y axis. On such a scale, it is not possible to represent a zero loss ratio which would correspond to minus infinite, and we have to arbitrarily choose a packet loss ratio at the bottom of the scale. We chose $10^{-6}$ because it corresponds to a value of the packet loss ratio that is not even possible to be evaluated reliably with measurement sessions of reasonable duration. In Fig.~\ref{fig:real-and-ideal-log-curve} we plot both the ideal and the realistic loss ratio curves in log scale. It is clear how we can easily differentiate the performance for the offered load below 1 Mpps between the ideal and the realistic loss-load curves, which are barely indistinguishable in Figs.~\ref{fig:load_loss_curve} and \ref{fig:real_loss_curve}

Fig.~\ref{fig:real_loss_curve} also reports the load-delivery curve $DR(O)$ (in red), where $DR(O)= 1-PDR(O)$. The curve decreases rapidly after the saturation offered load $O^\text{sat}$.

\begin{figure*}[ht!]
    \centering
    \begin{subfigure}{0.485\textwidth}
        \centering
        \includegraphics[width=\linewidth]{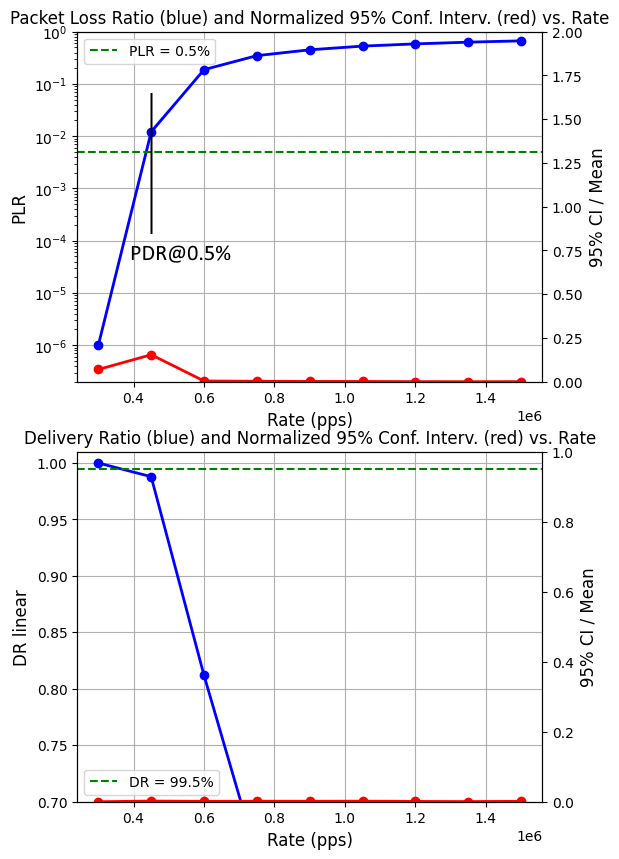}
        \caption{Example of a load-loss curve with a rapid increase in packet loss, indicating a quick transition to saturation after a certain load.}
        \label{fig:kernel-6.8-pin1-cpu-large}
    \end{subfigure}%
    \hfill
    \begin{subfigure}{0.485\textwidth}
        \centering
        \includegraphics[width=\linewidth]{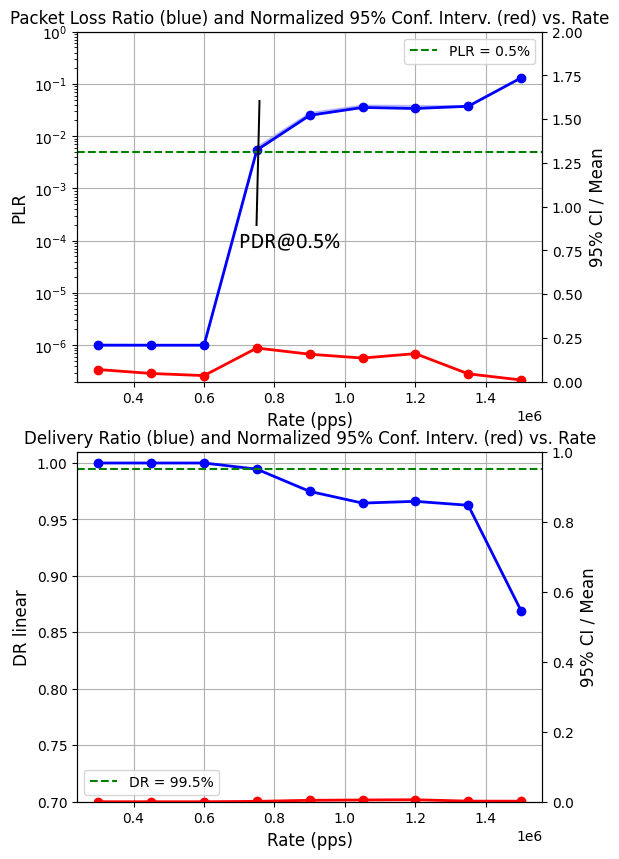}
        \caption{Example of a load-loss curve with a gradual increase in packet loss, suggesting a delayed transition into saturation.}
        \label{fig:kernel-6.5-pin1-cpu-large}
    \end{subfigure}
    \caption{Comparison of two load-loss curves illustrating different behaviors of systems approaching saturation.}
    \label{fig:kernel-comparison}
\end{figure*}

\subsection{Evaluating the representativeness of PDR@0.5\% }

We aim to identify the conditions under which the single PDR@0.5\% metric accurately reflects router performance. Fig.~\ref{fig:kernel-comparison} reports the packet loss ratio $PLR(O)$ and the delivery ratio $DR(O)$ experimentally measured over two Linux software routers with two different kernel versions\footnote{The specific experiment consists in plain routing of IPv6 packets}. In Fig.~\ref{fig:kernel-6.8-pin1-cpu-large} (see top left) the system is operating with low loss until the PDR@0.5\% is reached. Beyond the PDR@0.5\% threshold, the system enters the saturation region. Any further increase in load results in a significant increase in packet loss, indicating that the router is operating at its capacity limit. Hence the PDR@0.5\% scalar metric provides an accurate characterization of the software router performance.

In contrast, the curves shown in Fig.~\ref{fig:kernel-6.5-pin1-cpu-large} (right side) depict a situation where the PDR@0.5\% metric fails to provide a comprehensive evaluation of router performance. Here, when the load exceeds the PDR@0.5\% threshold, the system experiences only a slight increase in packet loss and continues to operate without reaching saturation immediately. This indicates that the router retains some capacity for additional load before fully entering the saturation region, making PDR@0.5\% less indicative of its overall performance.

In other words, while the left curves show a rapid transition to saturation for load > PDR@0.5\%, the right curves demonstrate a slow increase of packet loss beyond PDR@0.5\%before reaching saturation throughput. This is a suboptimal performance, as the system shows non-negligible loss even when it is far from reaching the saturation.

\section{Assessment Methodology}
\label{sec:meth_stability_assess}

In this section, we first outline a systematic approach for detecting saturation conditions based on the relationship between the increase of offered load and the increasy of throughput. Then we use this approach to evaluate the relation between the value of PDR@0.5\% and the saturation offered load, which in turn is used to assess the representativeness and accuracy of the PDR@0.5\% metric. Finally, we present the detailed design and Python implementation of the methodology. 

\subsection{Identification of saturation conditions}

When a system is in a non-saturated state, increases in the offered load \( O \) result in proportional increases in throughput \( T(O) \). However, as the system approaches its limits, it reaches a point where further increases in the offered load primarily result in packet loss rather than increased throughput.  

To quantify the transition to saturation, we analyze the change in throughput relative to an increase in offered load. Let us consider \( O_1 \) and \( O_2 = O_1 + \Delta O \), where \(\Delta O\) represents a small increment in the offered load. The corresponding throughputs are \( T(O_1) \) and \( T(O_2) \). The change in throughput between these two points is defined as \(\Delta T = T(O_2) - T(O_1)\).

To determine how efficiently the system converts the additional load offered into throughput, we define the \textit{Throughput Increase Ratio (TIR)} as:

\[
\text{TIR} = \frac{\Delta T}{\Delta O} = \frac{T(O_2) - T(O_1)}{O_2 - O_1}
\]

The \textit{TIR} measures the fraction of the additional offered load that is converted into increased throughput. In a non-saturated state, \textit{TIR} remains relatively high, indicating that most of the increase in offered load is effectively handled by the system. However, as the system reaches saturation, \textit{TIR} decreases, reflecting that a larger fraction of the additional load results in packet loss. 

To detect saturation, we define a threshold \(\eta\), which is a small value (e.g., \(\eta = 0.1\)), representing the point at which less than 10\% of the additional load results in increased throughput. The system is classified as entering saturation the first time the \textit{TIR} falls below this threshold, i.e., if \(\text{TIR} < \eta\), then classify the system as entering saturation.

Once the system is classified as entering saturation, we verify that it remains in this state as the offered load continues to increase by checking that the \textit{TIR} remains below or equal to the threshold \(\eta\) for all subsequent increases in the offered load.


\subsection{How to assess the accuracy of PDR@0.5\%}

The PDR@0.5\% metric is an accurate measure of a system's packet processing capabilities if, starting from the offered load at which the PLR reaches 0.5\%, the system is in saturation state or near to enter this state. This means that as soon as the packet loss ratio reaches 0.5\%, further increases in offered load primarily results in packet loss rather than a significant increase in throughput.

When the system behaves in this manner, the PDR@0.5\% effectively characterizes the system's maximum achievable throughput. Since the transition to saturation is accompanied by a steep increase in the loss ratio, we can precisely identify this point using the \textit{TIR} methodology outlined above. If the \textit{TIR} is smaller than the threshold \(\eta\) as the offered load increases beyond the PDR@0.5\% point, the system has reached its capacity for processing packets. In this case, the PDR@0.5\% is a reliable indicator of the system's performance limit and for example it can be reliably used to compare the performance of two different implementations of a given networking function.

By using the \textit{TIR} as a detection mechanism, we have a precise methodology to assess the effectiveness of the PDR@0.5\% metric in evaluating a system's packet processing capabilities. It allows us to ensure that the reported PDR value is not just a snapshot but is closely related to the onset of saturation, thus providing a meaningful and accurate measurement for comparing different systems or configurations.

\begin{figure*}[ht!]
    \centering
    \begin{algorithmic}[1]
        \Function{detect\_saturation}{PLR values, offered load values}
            \State Identify the point where PLR exceeds the threshold (PDR).
            \If{no such point is found}
                \State \Return "No saturation detected"
            \EndIf

            \State Set this point as the potential saturation point.
            \State Calculate the Throughput Increase Ratio (TIR) for subsequent offered loads.
            
            \For{each offered load after the potential saturation point}
                \If{TIR is below a specified threshold}
                    \State Mark the system as entering saturation.
                    \State Verify that TIR remains low for all further increases in offered load.
                \Else
                    \State Mark the system as no longer in saturation and clear previous detections.
                \EndIf
            \EndFor

            \State Decide if the experiment is \textit{good} based on consistent detection of saturation starting from the potential saturation point.
            \State \Return the detected saturation point and evaluation of the experiment's quality.
        \EndFunction
    \end{algorithmic}
    \caption{High-Level Pseudocode for Detecting Saturation in a Networked System}
    \label{fig:saturation-pseudocode}
\end{figure*}

\subsection{Detailed design and implementation}

The proposed algorithm is listed in \ref{fig:saturation-pseudocode}. It detects the transition to a saturated state in a system by analyzing the behavior of the Packet Loss Ratio (PLR) as the offered load increases. The algorithm begins by identifying the point at which the PLR exceeds a specified threshold, designating this as the potential saturation point. From this point onward, it calculates the Throughput Increase Ratio (TIR), which quantifies the system's response to further increments in load. If the TIR drops below a defined threshold, the system is considered to have entered a state of saturation. The algorithm then verifies that this condition persists as the offered load continues to increase, ensuring stable saturation behavior. The experiment is deemed \textit{good} if the saturation behavior starts at the potential point and remains consistent thereafter. This systematic approach allows for a thorough evaluation of the system's capacity and stability in handling network traffic. 

We used Python and Google Colab \cite{colab} to implement the proposed methodology. Colab is ideal for data analysis, machine learning, and coding. It allows us to run algorithms efficiently without local setup.
The source code, experiment result data, documentation, and usage examples are on GitHub at \cite{github-pastrami}. This repository is for replicating, reviewing, and developing our methodology. We encourage the research community to examine the repository and offer feedback. Researchers can access and duplicate the public Colab, linked in \cite{github-pastrami} so that they can easily replicate the processing of the proposed methodology on the provided experiment result data.

\section{Description of first round of experiments}
\label{sec:experiment_desc}

\begin{figure}[ht!]
    \centering
    \includegraphics[width=\columnwidth]{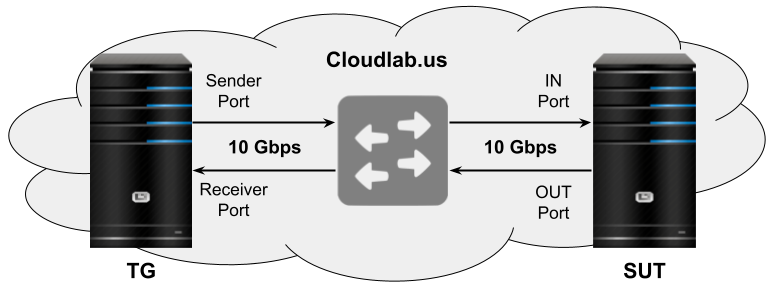}
    \caption{Cloudlab Testbed architecture}
    \label{fig:cloudlab_testbed}
\end{figure}

To validate our research work, a testbed was set up to facilitate the execution of experiments, data collection, and result analysis. The testbed has been deployed on CloudLab \cite{cloudlabWebsite,ricci2014introducing,stoller2019cloudlab}, a cloud-based research infrastructure for network and distributed systems. 
The two testbed nodes are powered by bare metal servers comprising an Intel Xeon E5-2630 v3 processor and 128 GB of RAM. Each bare metal server is equipped with two Intel 82599ES 10-Gigabit network cards, thereby ensuring back-to-back connectivity with the other node through a switch as illustrated in Fig.~\ref{fig:cloudlab_testbed}. We have also a local deployment of a testbed with two servers interconnected with 10G Ethernet cards, but the experiments described in this paper have been executed on Cloudlab as it represents a replicable infrastructure.  
The Traffic Generator (TG) runs the TRex \cite{trex-cisco} open-source traffic generator, which is built on the DPDK \cite{dpdk} framework and is capable of sending traffic at full NIC line-rate. The System Under Test (SUT) runs different Linux kernel releases, according to the specific experiments being conducted.

The space of the experiments is defined by the aspects (or variables) reported in Table~\ref{tab:exp_configuration}. These aspects can be modified in order to explore different configurations and to study the impact of such changes on performance.
By varying parameters such as the number of runs, duration, testbed (Cloudlab or local server), execution environment (bare-metal vs. virtualized), Linux kernel versions, CPU pinning, and the dimension of the NIC ring buffer, the influence of each factor on the overall performance of the system can be analyzed. The experiment class represent the actual networking function that is tested in an experiment (most of our experiments concerned plain IPv6 routing) along with the execution platform for the networking function (e.g. Linux kernel networking, eBPF...).



A considerable number of experiments were conducted (more than 100 at the time of writing) and their results are detailed \shortver{in the extended version of the paper \cite{lungaroni2024pastramiarxiv}.} \extended{in Section \ref{sec:appendix} of this extended version \cite{lungaroni2024pastramiarxiv}.} The reported experiment results include the graphs of the PLR and DR metrics, conveniently plotted using our Python scripts by the Colab. The Python scripts mark the graphs by explicitly adding \textit{GOOD} and \textit{BAD} visual label following the methodology described in Section \ref{sec:meth_stability_assess}.

The evidence collected in this first round of experiments made it clear that packet-level metrics alone were not always sufficient to interpret the observed behavior, and motivated a new development activity aimed at adding CPU-load observability inside the Linux networking stack.

\begin{table}[ht!]
\centering
\renewcommand{\arraystretch}{1.5}
\begin{tabular}{|>{\raggedright\arraybackslash}m{0.28\columnwidth}|>{\raggedright\arraybackslash}m{0.61\columnwidth}|}
\hline
\textbf{Aspect}                & \textbf{Options}                                 \\ \hline
\textbf{Number of Runs}        & 50 (default), 30, configurable                       \\ \hline
\textbf{Duration (seconds)}    & 20, 10 (default), 6, configurable                       \\ \hline
\textbf{Experiment class}      & IPv6 routing, eBPF routing...                        \\ \hline
\textbf{Testbed}               & Cloudlab, local server                             \\ \hline
\textbf{Exec. environment}     & \begin{minipage}[t]{\linewidth}
                                  \begin{itemize}[leftmargin=*]
                                      \item Bare metal server
                                      \item Virtual Machine
                                      \item Containers
                                  \end{itemize}
                                \end{minipage}                                      \\ \hline
\textbf{Linux Kernel Versions} & 5.6, 5.12, 5.13, 5.15, 5.18, 6.5, 6.8 \\ \hline
\textbf{CPU Pinning}           & \begin{minipage}[t]{\linewidth}
                                  \begin{itemize}[leftmargin=*]
                                      \item unpinned
                                      \item pin-1-cpu (RX and TX tasks on 1 CPU)
                                      \item pin-2-cpu (RX on 1 CPU, TX on 2nd CPU)
                                  \end{itemize}
                                \end{minipage}                                      \\ \hline
\textbf{NIC Ring Buffer Size}  & \begin{minipage}[t]{\linewidth}
                                  \begin{itemize}[leftmargin=*]
                                      \item SMALL-nic-buf (512 packets)
                                      \item LARGE-nic-buf (4096 packets)
                                  \end{itemize}
                                \end{minipage}                                      \\ \hline
\end{tabular}
\caption{Configuration options for the experimental space.}
\label{tab:exp_configuration}
\end{table}

\section{Lessons learned from the first round of experiments}
\label{sec:lessons_learned_first_round}

The first round of experiments provided a broad characterization of the forwarding behavior across different Linux kernel versions, execution environments, and system configurations. At the same time, it highlighted an important limitation of the initial measurement workflow. While the methodology based on DR, PLR, and the stability assessment of the resulting curves proved effective in identifying \textit{GOOD} and \textit{BAD} behaviors, packet-level metrics alone were not always sufficient to explain \emph{why} a given configuration exhibited instability or reduced efficiency.

This limitation became particularly relevant when comparing experiments that produced similar forwarding-rate trends but showed different levels of variability, or when non-negligible packet loss appeared under conditions that, based on the offered load alone, should have been sustainable. In such cases, the observed behavior suggested that the packet-processing activity inside the networking stack had to be examined more closely, in order to understand whether the system was approaching a genuine processing limit or whether the anomaly originated from less visible effects related to scheduling and kernel activity.

The experience gained during the first campaign therefore made it clear that the original PASTRAMI workflow had to be extended with additional observability capabilities. In particular, it became necessary to complement forwarding performance measurements with a metric able to capture the processing effort required by the System Under Test during each experiment. Such an extension was expected to improve the interpretation of the results and to support a more accurate analysis of anomalous or unstable behaviors.

For this reason, a new activity was initiated after the first round of experiments, with the goal of integrating CPU-load observability into PASTRAMI. The focus was placed on monitoring the packet-processing activity of the Linux networking stack with sufficiently fine granularity and low overhead, so that the additional measurements could be incorporated into the automated experimental workflow without perturbing the experiments themselves. The next section describes this extension and the integration of the \texttt{netrace} eBPF-based monitoring tool into the PASTRAMI framework.

\section{CPU-load observability with \texttt{netrace}}
\label{sec:cpu_load_netrace}

The need for an additional observability dimension emerged directly from the analysis of the first round of experiments. The forwarding metrics collected by PASTRAMI were sufficient to characterize the external behavior of the System Under Test, but they did not provide direct information on the amount of processing effort required inside the Linux networking stack. As a consequence, when instability or unexpected packet loss was observed, the available measurements did not always allow one to determine whether the system was close to a true saturation condition or whether the anomaly was caused by other effects internal to the operating system.

For this reason, the PASTRAMI framework was extended to include CPU-load observability as an additional metric to be correlated with the forwarding results. The objective was not to replace the original DR and PLR metrics, but rather to complement them with information capable of explaining how intensively the system was processing packets during each experiment. In the context of Linux software forwarding, this aspect is particularly relevant because packet reception and transmission involve not only the execution of protocol-stack functions, but also the activity of the interrupt and deferred-processing mechanisms used by the kernel.

A key element in this regard is the handling of software interrupts, or \textit{softirqs}, which are widely used in the Linux networking subsystem to defer part of the packet-processing work outside the hard-interrupt context. Under sustained traffic conditions, softirq processing may become a dominant component of the overall CPU consumption of the forwarding path. Moreover, when the amount of deferred work cannot be completed promptly in the immediate execution context, Linux may offload part of such activity to the \texttt{ksoftirqd} kernel threads. Monitoring the time spent in these processing paths is therefore useful to interpret forwarding behavior and to understand whether a given experiment is approaching a processing bottleneck.

An initial attempt to add CPU-load measurements relied on standard user-level monitoring tools such as \texttt{top}. However, this approach quickly proved inadequate for the purposes of PASTRAMI. First, its sampling granularity was too coarse with respect to the short duration and high dynamics of packet-level experiments. Second, its output was not well suited for accurate automated correlation with the measurements already collected by the framework. Third, the use of generic user-space monitoring tools did not provide direct visibility into the specific kernel execution paths most relevant to packet processing.

To overcome these limitations, a dedicated tool named \texttt{netrace} was developed and adopted within the PASTRAMI workflow. \texttt{netrace} is based on eBPF and is designed to trace the execution time of the softirq processing path with low overhead. More specifically, the tool measures the time spent by the kernel in the relevant deferred packet-processing activity and uses such information to estimate the CPU load associated with networking operations over a fixed sampling interval. In the current implementation, the sampling interval is set to one second, thus obtaining a time series of CPU-load values during the execution of each experiment.

The use of eBPF is instrumental to achieving the required level of observability with limited perturbation of the measurements. By attaching at a low-level point of the kernel execution path, \texttt{netrace} can observe the relevant processing intervals with greater precision than generic user-space tools, while preserving a lightweight instrumentation model suitable for repeated automated experiments. In addition, the tool exports the collected data in JSON format, thereby making it possible to integrate the CPU-load traces into the existing post-processing chain already used by PASTRAMI for result visualization and analysis.

This extension represents a natural evolution of the framework. The original methodology remains centered on the evaluation of forwarding performance through DR, PLR, and stability analysis, but it is now complemented by an additional metric that improves the interpretation of the observed behaviors. In particular, CPU-load observability makes it possible to associate forwarding anomalies with the internal packet-processing effort of the system, thereby providing a stronger basis for understanding unstable conditions and for comparing different kernel versions and execution environments.

\subsection{Integration of \texttt{netrace} into PASTRAMI}
\label{sec:netrace_integration}

The introduction of \texttt{netrace} required an extension of the PASTRAMI workflow both in the experiment-execution phase and in the post-processing phase. The goal was to preserve the automated nature of the framework while augmenting each experiment with CPU-load observability information collected directly on the System Under Test.

During the execution of an experiment, \texttt{netrace} is remotely started on the System Under Test together with the networking function under analysis. While the Traffic Generator injects packets according to the selected offered-load configuration, \texttt{netrace} monitors the execution time associated with the softirq-based packet-processing activity and periodically records the corresponding CPU-load samples. In this way, the framework is able to collect forwarding-performance metrics and CPU-load information within the same experimental run, thus enabling a direct correlation between the external behavior of the system and its internal processing effort.

In order to support automated processing, the output generated by \texttt{netrace} is exported in JSON format. This design choice simplifies the transfer of the collected data from the System Under Test to the controller node and allows the new measurements to be incorporated into the same analysis pipeline already used by PASTRAMI for the other experiment results. As a consequence, CPU-load traces can be archived together with the packet-level measurements and handled in a uniform way during post-processing.

The Colab-based post-processing scripts were correspondingly extended to parse the JSON files generated by \texttt{netrace} and to extract the CPU-load time series associated with each experiment. This made it possible to enrich the analysis phase with an additional metric that can be inspected jointly with DR, PLR, and the stability labels produced by the methodology. The integration was designed so that the new observability dimension could be added without altering the overall organization of the framework and without introducing manual steps in the experimental procedure.

The integration of \texttt{netrace} into PASTRAMI therefore represents an incremental but significant improvement of the framework. The original workflow is preserved, yet each measurement campaign can now provide, in addition to forwarding results, an explicit view of the processing effort sustained by the Linux networking stack during the experiment. This extension lays the groundwork for a more informed interpretation of anomalous behaviors and for future analyses aimed at correlating forwarding instability with internal kernel activity.

\section{Conclusion and Future Work}
\label{sec:conclusion}

This paper presented PASTRAMI, a methodology and an associated toolchain for a more reliable assessment of software forwarding performance. The proposed approach moves beyond the sole use of PDR@0.5\% by combining packet-level measurements with a stability assessment procedure based on the analysis of DR and PLR curves. This makes it possible to distinguish measurement conditions in which a single summary metric is representative from cases in which the observed behavior is unstable and requires a more cautious interpretation.

The first round of experiments confirmed the usefulness of this methodological approach across different Linux kernel versions and execution environments. At the same time, it also highlighted a limitation of the initial workflow: packet-level metrics alone are not always sufficient to explain the causes of anomalous behavior or instability. For this reason, the framework has been extended with a new observability capability aimed at monitoring the CPU load associated with packet processing inside the Linux networking stack.

To this end, the paper has introduced the integration of \texttt{netrace}, an eBPF-based monitoring tool designed to trace softirq processing activity with low overhead and to export CPU-load measurements in a format suitable for automated post-processing. This extension represents an important step in the evolution of PASTRAMI, since it complements forwarding-performance measurements with an additional metric that can support a more informed interpretation of the observed behavior.

The Colab-based workflow remains a central element of the framework, as it provides an open and reproducible environment for experiment control, result collection, and post-processing. By making the methodology and the associated tooling openly available, this work aims to support further experimental validation and to facilitate comparative studies across different platforms and configurations.

Future work will focus on the second round of experiments enabled by this extension, with the objective of correlating forwarding performance with CPU-load measurements and of better understanding the sources of instability observed in some experimental conditions. Further developments may also include a deeper analysis of the internal causes of packet drops and processing inefficiencies in the Linux kernel, as well as the application of the methodology to additional software forwarding scenarios and execution environments.



\section{Acknowledgements}
This work was supported by the European Union - Next Generation EU under the Italian National Recovery and Resilience Plan (NRRP), Mission 4, Component 2, Investment 1.3, CUP E83C22004640001, partnership on “Telecommunications of the Future” (PE00000001 - program “RESTART”) and by the Italian Ministry of University and Research (MUR) under the PRIN 2022 project NEWTON, Project ID 2022ZA8T22.

\bibliographystyle{IEEEtran}
\bibliography{references}

\extended{
\newpage
\section{Appendix}
\label{sec:appendix}
In this section you can find the results of our measurements described in section \ref{sec:experiment_desc}.

\begin{figure}[ht!]
    \centering
    \includegraphics[width=\linewidth]{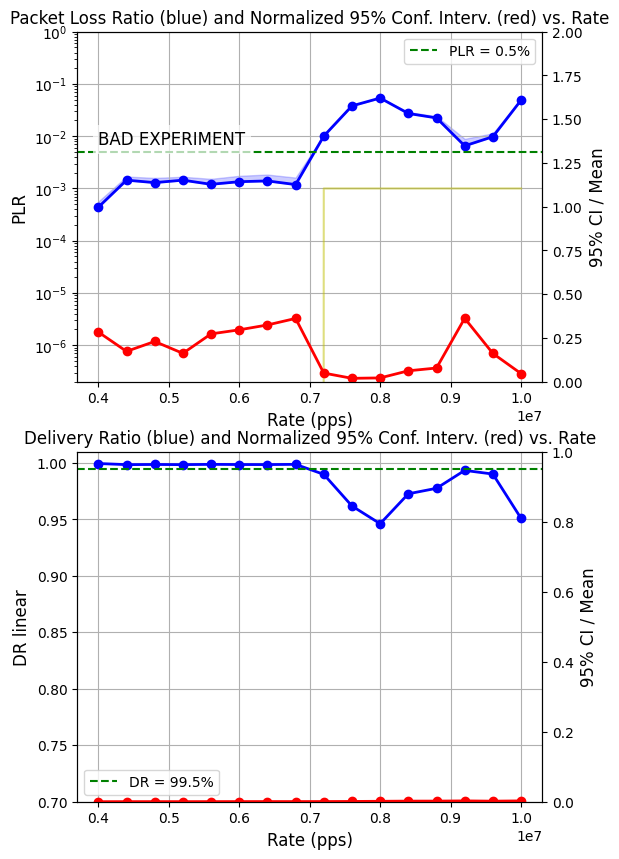}
    \caption{ebpf, CloudLab, bare metal server, Testbed 1, 30 runs, T = 10s, kernel 5.15, unpinned, SMALL nic buffer, date 0000-00-00, version 00}
    \label{fig:clab_tb1_class_ebpf_raw_30_exp_t_10_k5.15_intel_bare-metal_as-is}
\end{figure}

\begin{figure}[ht!]
    \centering
    \includegraphics[width=\linewidth]{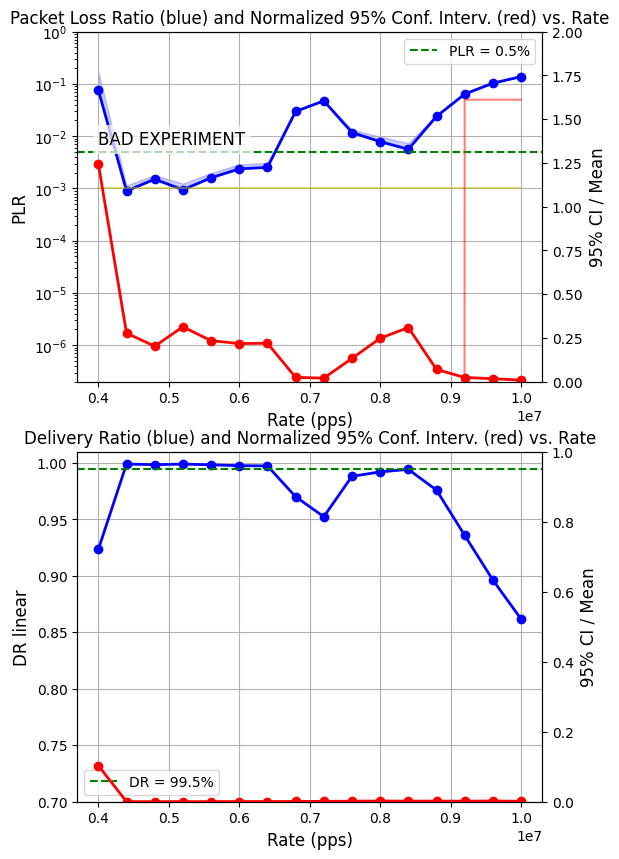}
    \caption{ebpf, CloudLab, bare metal server, Testbed 1, 30 runs, T = 10s, kernel 5.15, unpinned, LARGE nic buffer, date 0000-00-00, version 00}
    \label{fig:clab_tb1_class_ebpf_raw_30_exp_t_10_k5.15_intel_bare-metal_nic_buf_4096}
\end{figure}

\begin{figure}[ht!]
    \centering
    \includegraphics[width=\linewidth]{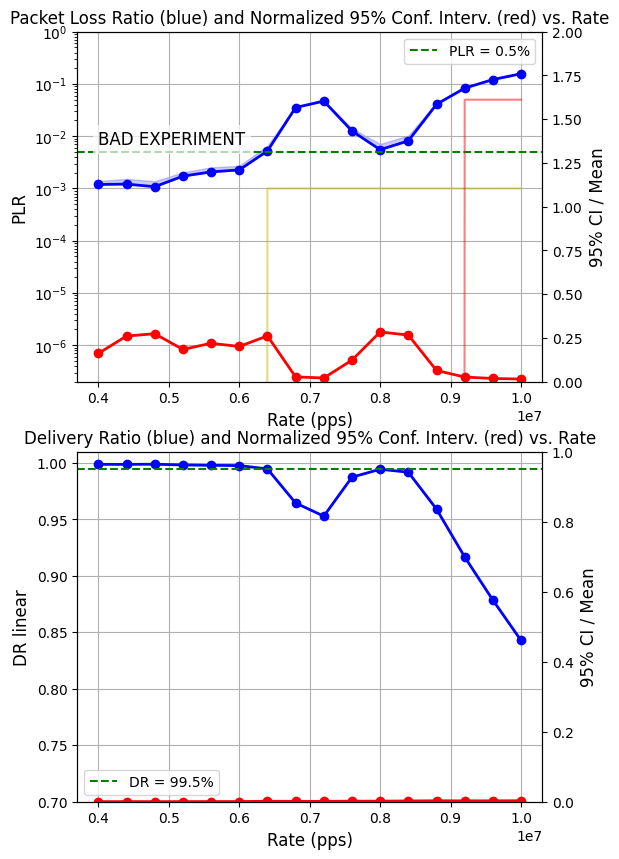}
    \caption{ebpf, CloudLab, bare metal server, Testbed 1, 30 runs, T = 10s, kernel 5.15, pin-1cpu, LARGE nic buffer, date 0000-00-00, version 00}
    \label{fig:clab_tb1_class_ebpf_raw_30_exp_t_10_k5.15_intel_bare-metal_nic_buf_4096_irq_pf0-4_pf1-4}
\end{figure}

\begin{figure}[ht!]
    \centering
    \includegraphics[width=\linewidth]{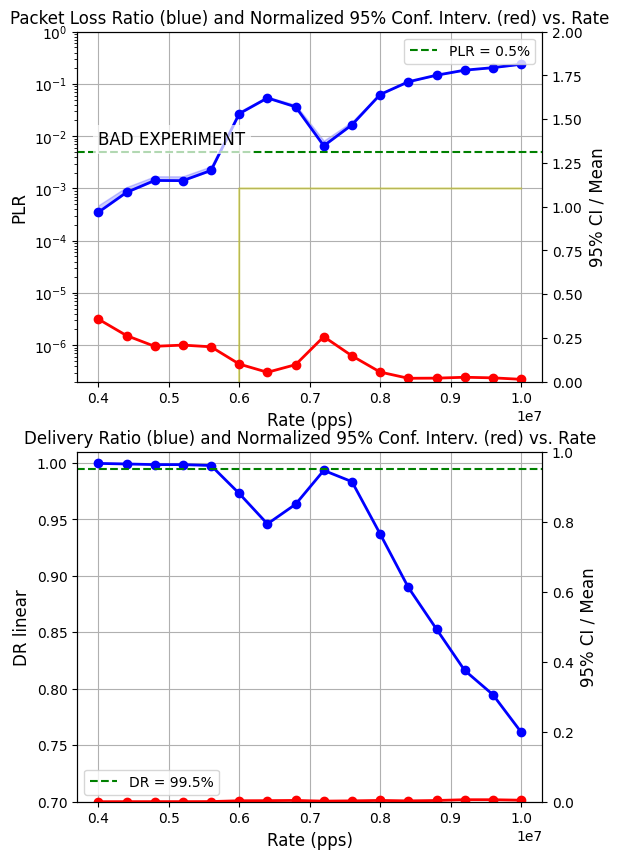}
    \caption{ebpf, CloudLab, bare metal server, Testbed 1, 30 runs, T = 10s, kernel 5.15, pin-2cpu, LARGE nic buffer, date 0000-00-00, version 00}
    \label{fig:clab_tb1_class_ebpf_raw_30_exp_t_10_k5.15_intel_bare-metal_nic_buf_4096_irq_pf0-4_pf1-6}
\end{figure}

\begin{figure}[ht!]
    \centering
    \includegraphics[width=\linewidth]{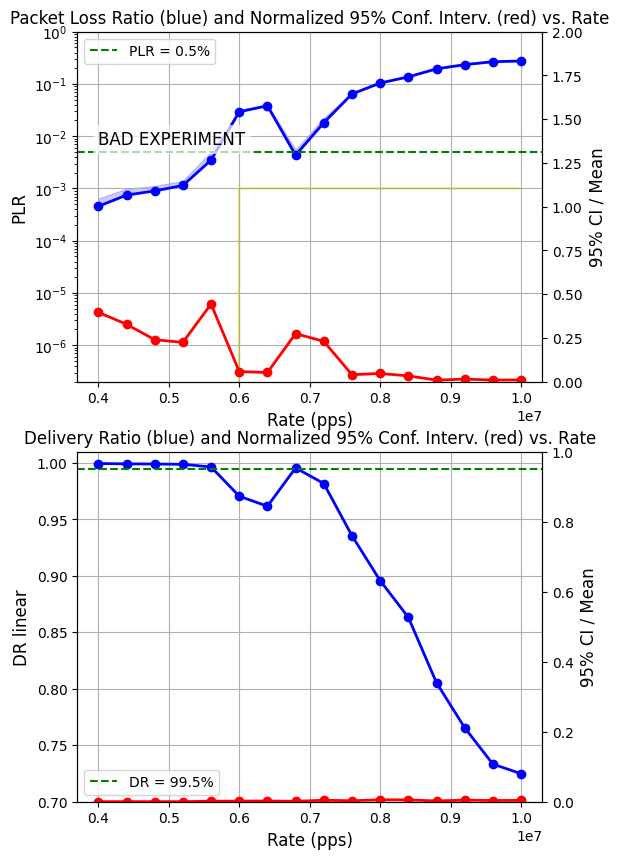}
    \caption{ebpf, CloudLab, bare metal server, Testbed 1, 30 runs, T = 10s, kernel 6.5, unpinned, SMALL nic buffer, date 0000-00-00, version 00}
    \label{fig:clab_tb1_class_ebpf_raw_30_exp_t_10_k6.5_intel_bare-metal_as-is}
\end{figure}

\begin{figure}[ht!]
    \centering
    \includegraphics[width=\linewidth]{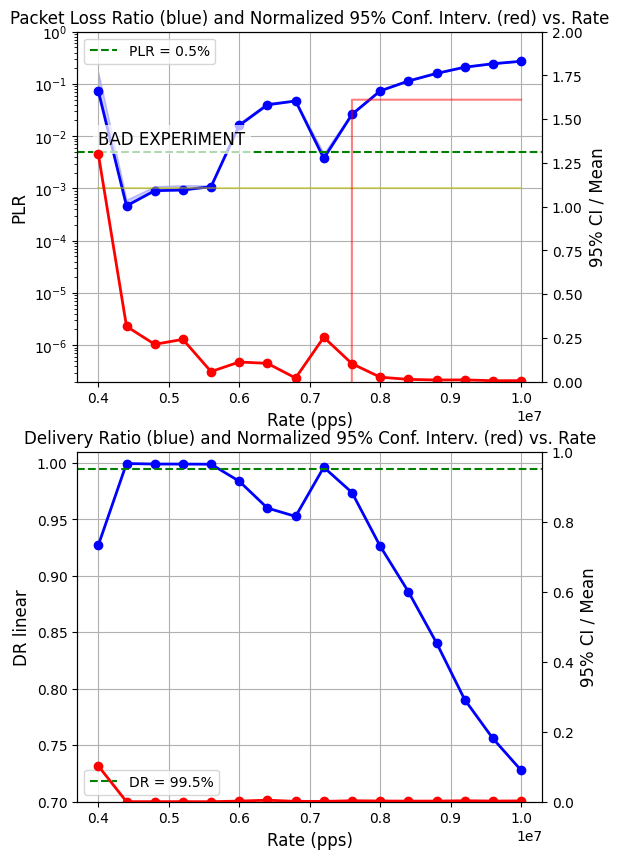}
    \caption{ebpf, CloudLab, bare metal server, Testbed 1, 30 runs, T = 10s, kernel 6.5, unpinned, LARGE nic buffer, date 0000-00-00, version 00}
    \label{fig:clab_tb1_class_ebpf_raw_30_exp_t_10_k6.5_intel_bare-metal_nic_buf_4096}
\end{figure}

\begin{figure}[ht!]
    \centering
    \includegraphics[width=\linewidth]{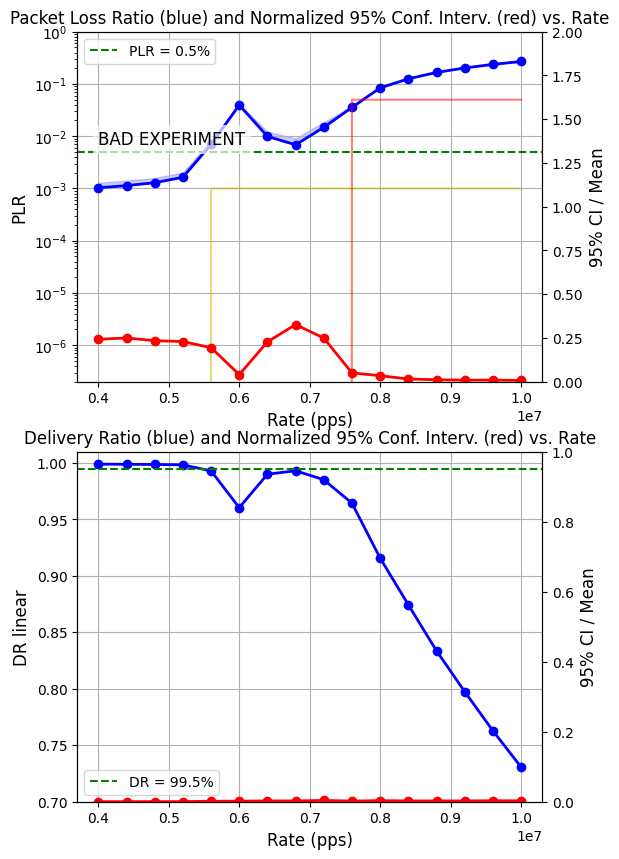}
    \caption{ebpf, CloudLab, bare metal server, Testbed 1, 30 runs, T = 10s, kernel 6.5, pin-1cpu, LARGE nic buffer, date 0000-00-00, version 00}
    \label{fig:clab_tb1_class_ebpf_raw_30_exp_t_10_k6.5_intel_bare-metal_nic_buf_4096_irq_pf0-4_pf1-4}
\end{figure}

\begin{figure}[ht!]
    \centering
    \includegraphics[width=\linewidth]{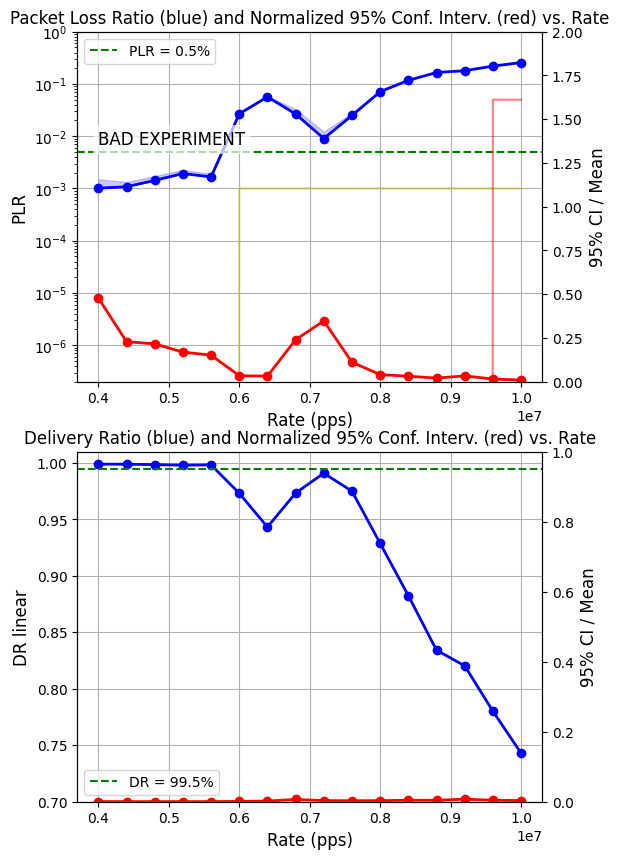}
    \caption{ebpf, CloudLab, bare metal server, Testbed 1, 30 runs, T = 10s, kernel 6.5, pin-2cpu, LARGE nic buffer, date 0000-00-00, version 00}
    \label{fig:clab_tb1_class_ebpf_raw_30_exp_t_10_k6.5_intel_bare-metal_nic_buf_4096_irq_pf0-4_pf1-6}
\end{figure}

\begin{figure}[ht!]
    \centering
    \includegraphics[width=\linewidth]{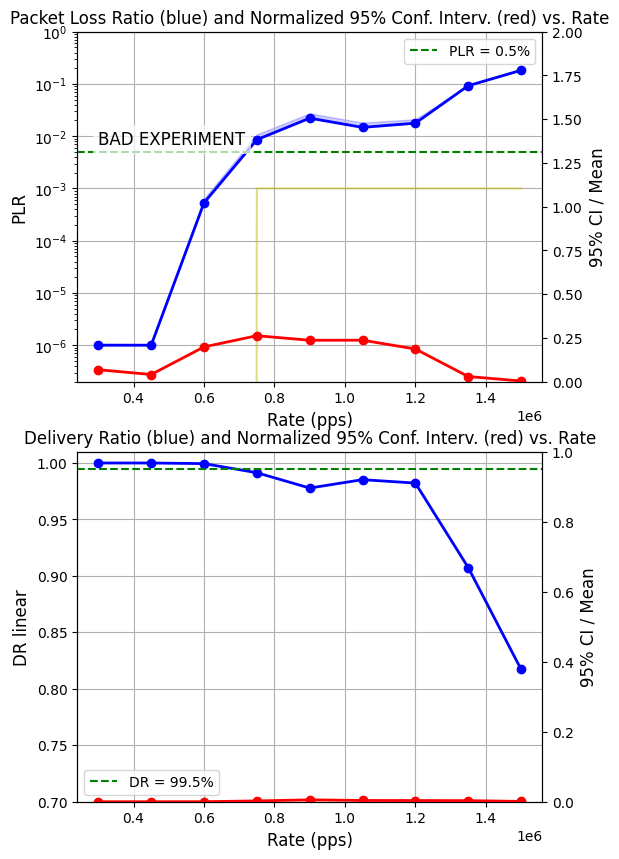}
    \caption{routing, CloudLab, bare metal server, Testbed 0, 50 runs, T = 10s, kernel 5.15, unpinned, SMALL nic buffer, date 0000-00-00, version 00}
    \label{fig:clab_raw_50_k5.15_intel_bare-metal_as-is_2nd_exec}
\end{figure}

\begin{figure}[ht!]
    \centering
    \includegraphics[width=\linewidth]{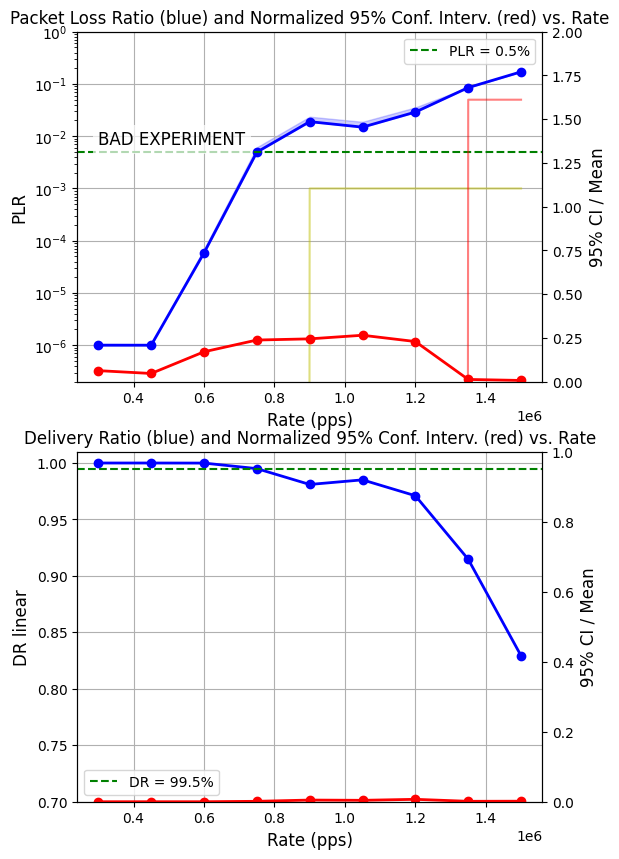}
    \caption{routing, CloudLab, bare metal server, Testbed 0, 50 runs, T = 10s, kernel 5.15, unpinned, SMALL nic buffer, date 0000-00-00, version 00}
    \label{fig:clab_raw_50_k5.15_intel_bare-metal_as-is}
\end{figure}

\begin{figure}[ht!]
    \centering
    \includegraphics[width=\linewidth]{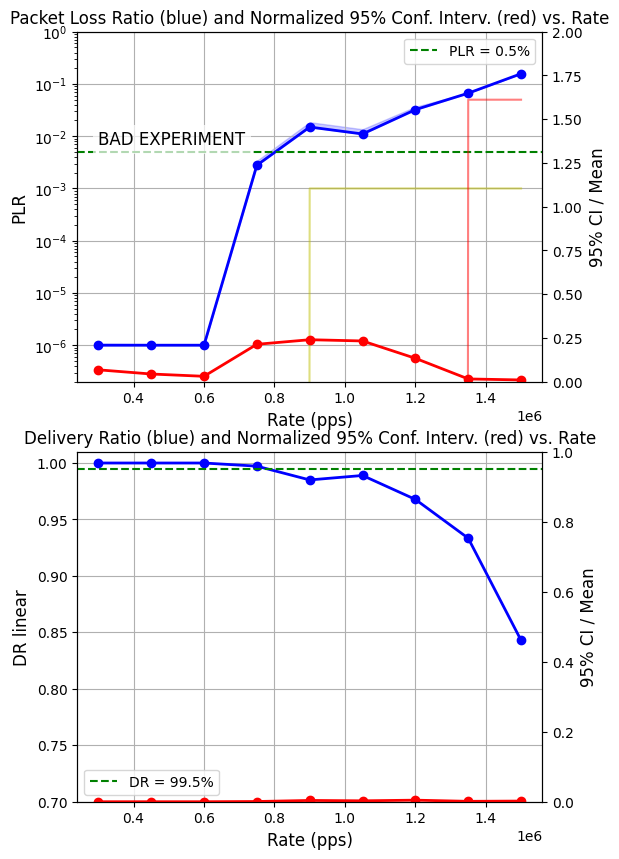}
    \caption{routing, CloudLab, bare metal server, Testbed 0, 50 runs, T = 10s, kernel 5.15, unpinned, LARGE nic buffer, date 0000-00-00, version 00}
    \label{fig:clab_raw_50_k5.15_intel_bare-metal_nic_buf_4096}
\end{figure}

\begin{figure}[ht!]
    \centering
    \includegraphics[width=\linewidth]{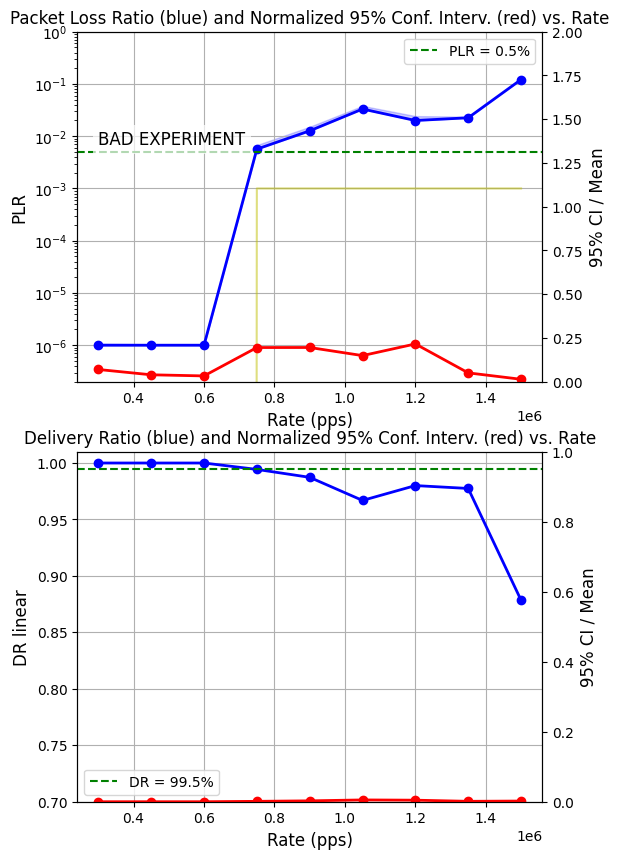}
    \caption{routing, CloudLab, bare metal server, Testbed 0, 50 runs, T = 10s, kernel 5.15, pin-1cpu, LARGE nic buffer, date 0000-00-00, version 00}
    \label{fig:clab_raw_50_k5.15_intel_bare-metal_nic_buf_4096_irq_pf0-4_pf1-4}
\end{figure}

\begin{figure}[ht!]
    \centering
    \includegraphics[width=\linewidth]{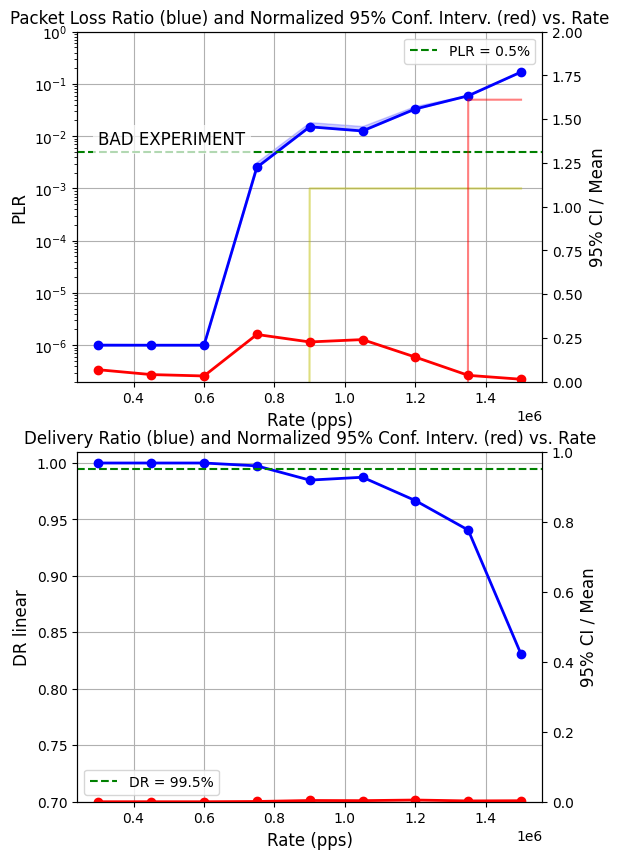}
    \caption{routing, CloudLab, bare metal server, Testbed 0, 50 runs, T = 10s, kernel 5.15, pin-2cpu, LARGE nic buffer, date 0000-00-00, version 00}
    \label{fig:clab_raw_50_k5.15_intel_bare-metal_nic_buf_4096_irq_pf0-4_pf1-6}
\end{figure}

\begin{figure}[ht!]
    \centering
    \includegraphics[width=\linewidth]{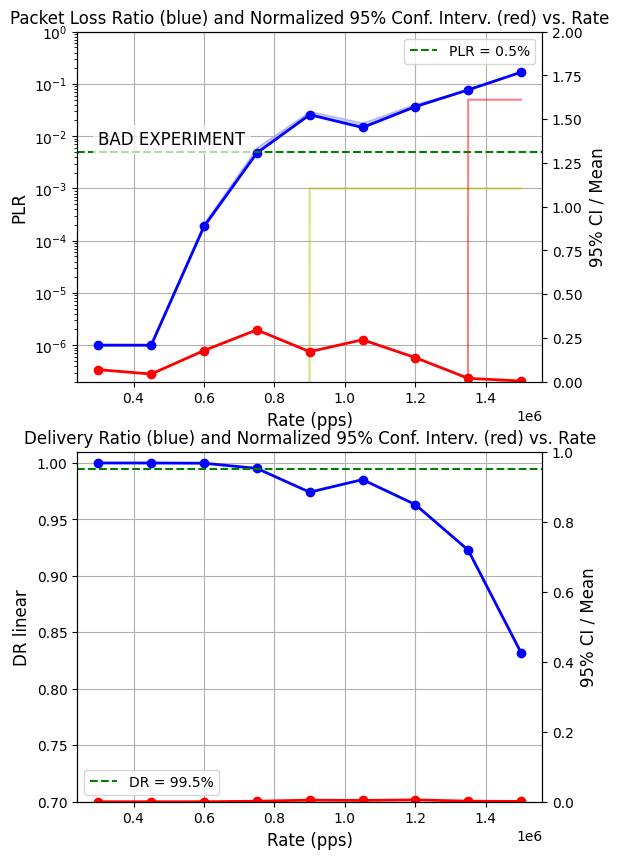}
    \caption{routing, CloudLab, bare metal server, Testbed 0, 50 runs, T = 10s, kernel 6.5, unpinned, SMALL nic buffer, date 0000-00-00, version 00}
    \label{fig:clab_raw_50_k6.5_intel_bare-metal_as-is}
\end{figure}

\begin{figure}[ht!]
    \centering
    \includegraphics[width=\linewidth]{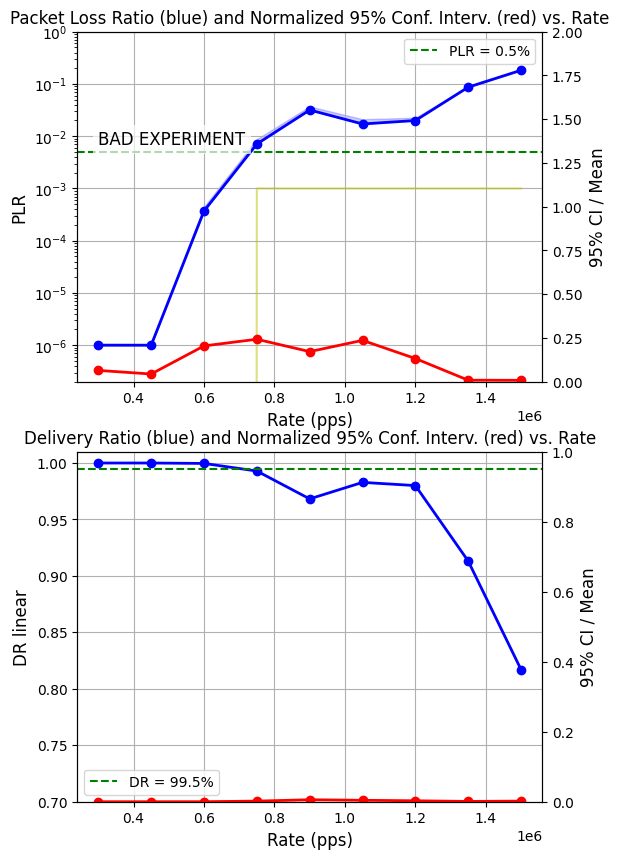}
    \caption{routing, CloudLab, bare metal server, Testbed 0, 50 runs, T = 10s, kernel 6.5, unpinned, SMALL nic buffer, date 0000-00-00, version 00}
    \label{fig:clab_raw_50_k6.5_intel_bare-metal_as-is_2nd_exec}
\end{figure}

\begin{figure}[ht!]
    \centering
    \includegraphics[width=\linewidth]{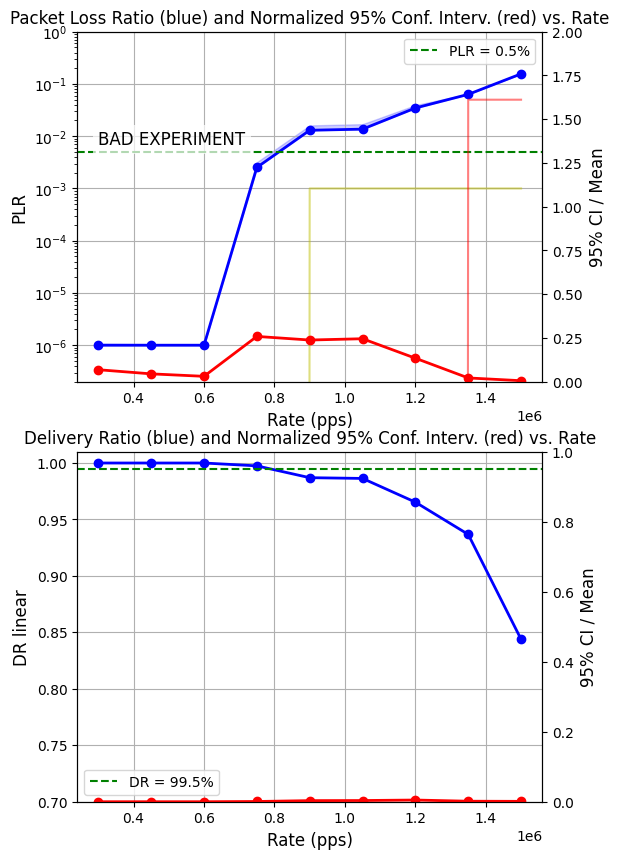}
    \caption{routing, CloudLab, bare metal server, Testbed 0, 50 runs, T = 10s, kernel 6.5, unpinned, LARGE nic buffer, date 0000-00-00, version 00}
    \label{fig:clab_raw_50_k6.5_intel_bare-metal_nic_buf_4096}
\end{figure}

\begin{figure}[ht!]
    \centering
    \includegraphics[width=\linewidth]{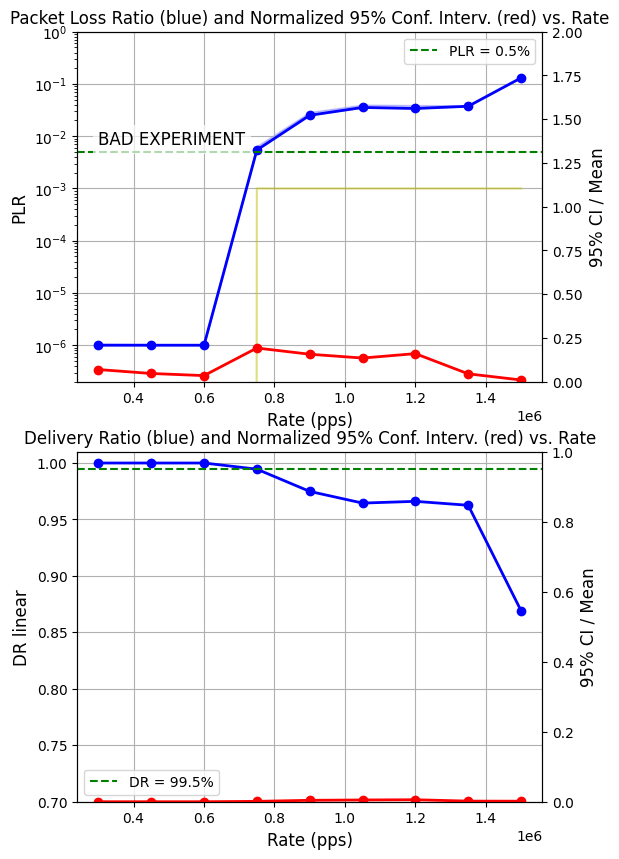}
    \caption{routing, CloudLab, bare metal server, Testbed 0, 50 runs, T = 10s, kernel 6.5, pin-1cpu, LARGE nic buffer, date 0000-00-00, version 00}
    \label{fig:clab_raw_50_k6.5_intel_bare-metal_nic_buf_4096_irq_pf0-4_pf1-4}
\end{figure}

\begin{figure}[ht!]
    \centering
    \includegraphics[width=\linewidth]{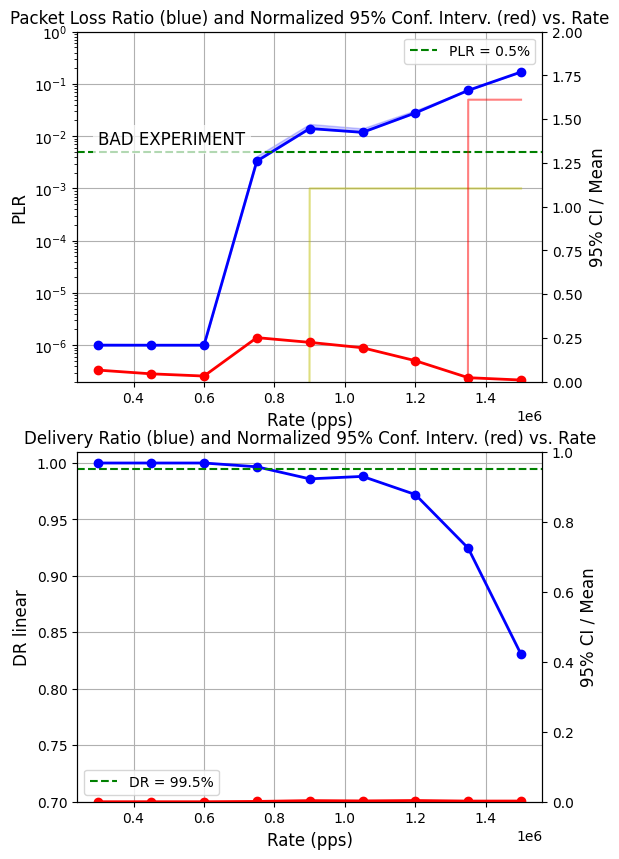}
    \caption{routing, CloudLab, bare metal server, Testbed 0, 50 runs, T = 10s, kernel 6.5, pin-2cpu, LARGE nic buffer, date 0000-00-00, version 00}
    \label{fig:clab_raw_50_k6.5_intel_bare-metal_nic_buf_4096_irq_pf0-4_pf1-6}
\end{figure}

\begin{figure}[ht!]
    \centering
    \includegraphics[width=\linewidth]{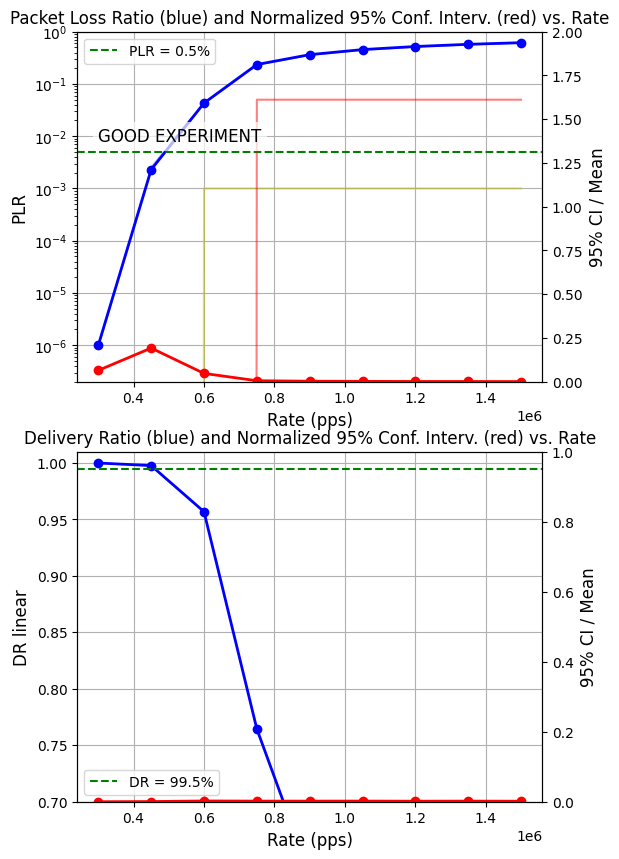}
    \caption{routing, CloudLab, bare metal server, Testbed 0, 50 runs, T = 10s, kernel 6.8, unpinned, SMALL nic buffer, date 0000-00-00, version 00}
    \label{fig:clab_raw_50_k6.8_intel_bare-metal_as-is}
\end{figure}

\begin{figure}[ht!]
    \centering
    \includegraphics[width=\linewidth]{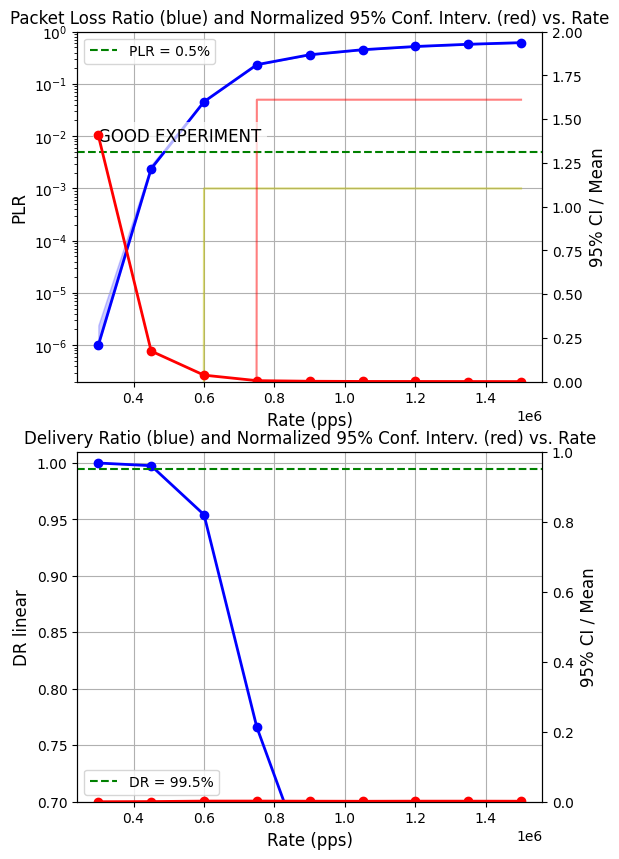}
    \caption{routing, CloudLab, bare metal server, Testbed 0, 50 runs, T = 10s, kernel 6.8, unpinned, SMALL nic buffer, date 0000-00-00, version 00}
    \label{fig:clab_raw_50_k6.8_intel_bare-metal_as-is_3rd_exec}
\end{figure}

\begin{figure}[ht!]
    \centering
    \includegraphics[width=\linewidth]{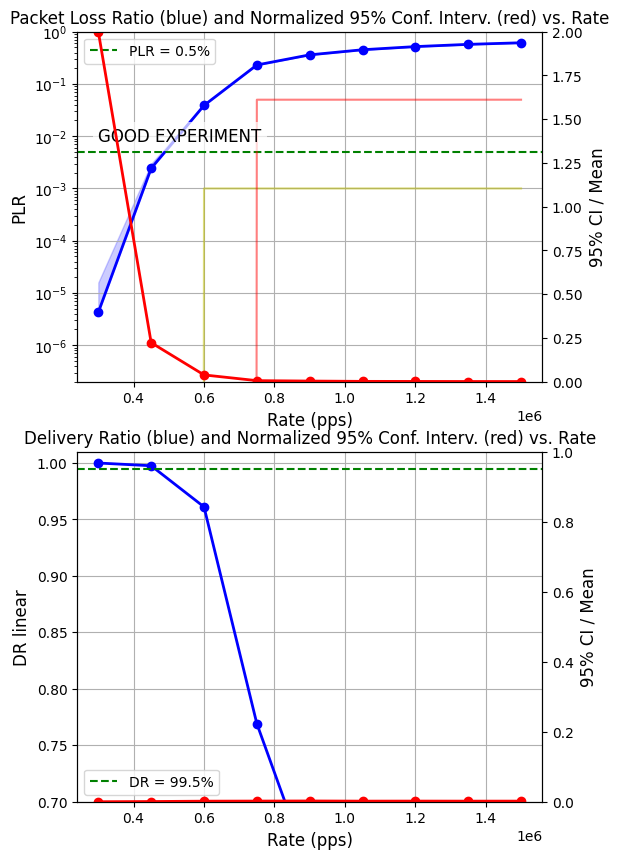}
    \caption{routing, CloudLab, bare metal server, Testbed 0, 50 runs, T = 10s, kernel 6.8, unpinned, SMALL nic buffer, date 0000-00-00, version 00}
    \label{fig:clab_raw_50_k6.8_intel_bare-metal_as-is_2nd_exec}
\end{figure}

\begin{figure}[ht!]
    \centering
    \includegraphics[width=\linewidth]{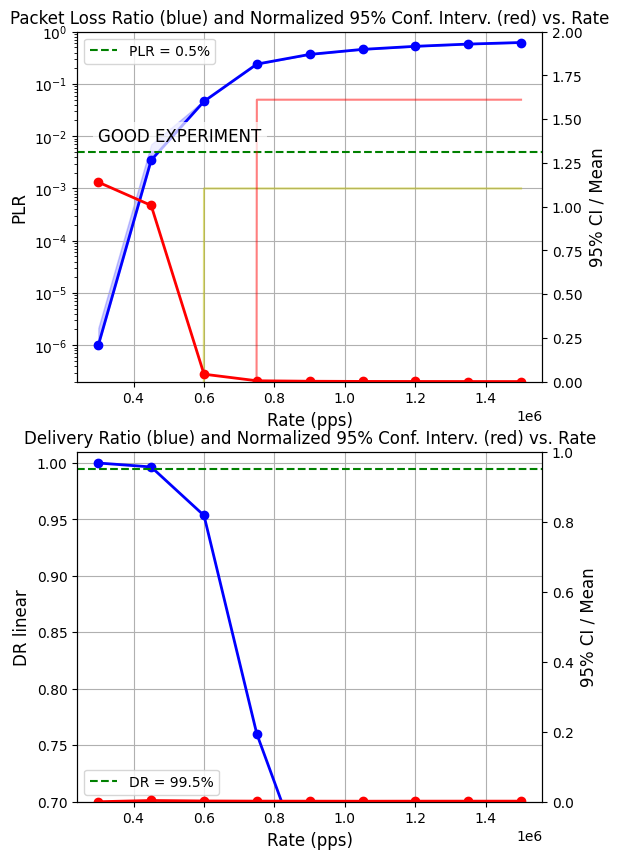}
    \caption{routing, CloudLab, bare metal server, Testbed 0, 50 runs, T = 10s, kernel 6.8, unpinned, SMALL nic buffer, date 0000-00-00, version 00}
    \label{fig:clab_raw_50_k6.8_intel_bare-metal_as-is_4th_exec_after-reboot}
\end{figure}

\begin{figure}[ht!]
    \centering
    \includegraphics[width=\linewidth]{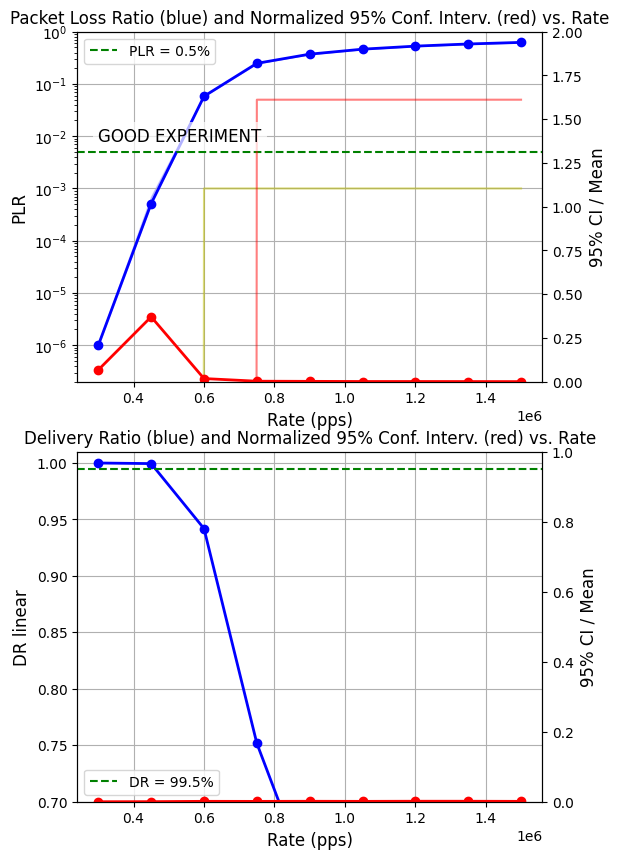}
    \caption{routing, CloudLab, bare metal server, Testbed 0, 50 runs, T = 10s, kernel 6.8, unpinned, LARGE nic buffer, date 0000-00-00, version 00}
    \label{fig:clab_raw_50_k6.8_intel_bare-metal_nic_buf_4096}
\end{figure}

\begin{figure}[ht!]
    \centering
    \includegraphics[width=\linewidth]{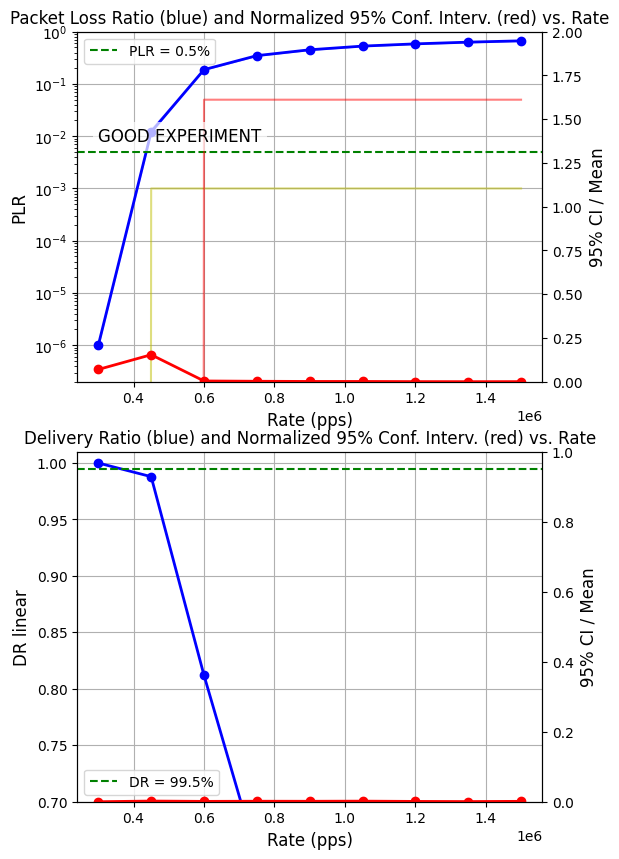}
    \caption{routing, CloudLab, bare metal server, Testbed 0, 50 runs, T = 10s, kernel 6.8, pin-1cpu, LARGE nic buffer, date 0000-00-00, version 00}
    \label{fig:clab_raw_50_k6.8_intel_bare-metal_nic_buf_4096_irq_pf0-4_pf1-4}
\end{figure}

\begin{figure}[ht!]
    \centering
    \includegraphics[width=\linewidth]{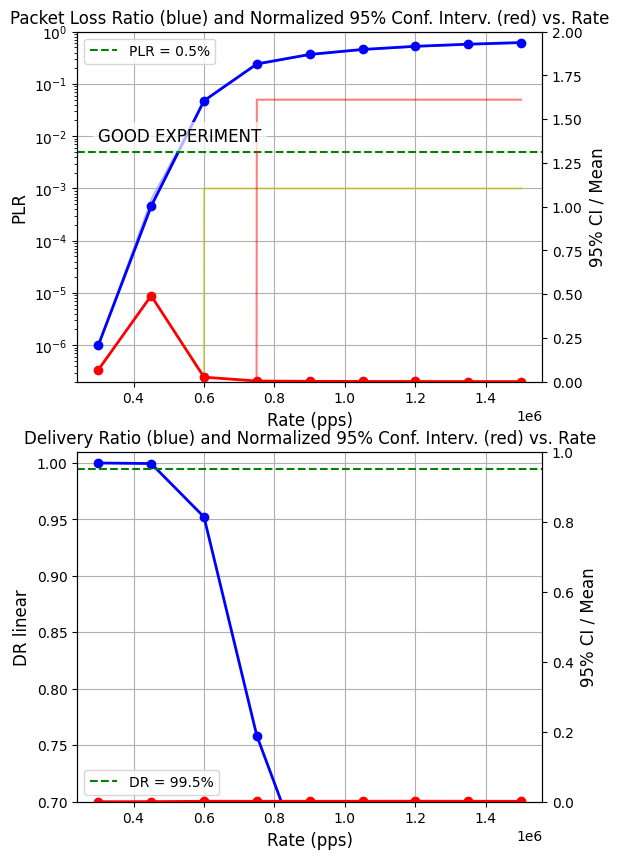}
    \caption{routing, CloudLab, bare metal server, Testbed 0, 50 runs, T = 10s, kernel 6.8, pin-2cpu, LARGE nic buffer, date 0000-00-00, version 00}
    \label{fig:clab_raw_50_k6.8_intel_bare-metal_nic_buf_4096_irq_pf0-4_pf1-6}
\end{figure}

\begin{figure}[ht!]
    \centering
    \includegraphics[width=\linewidth]{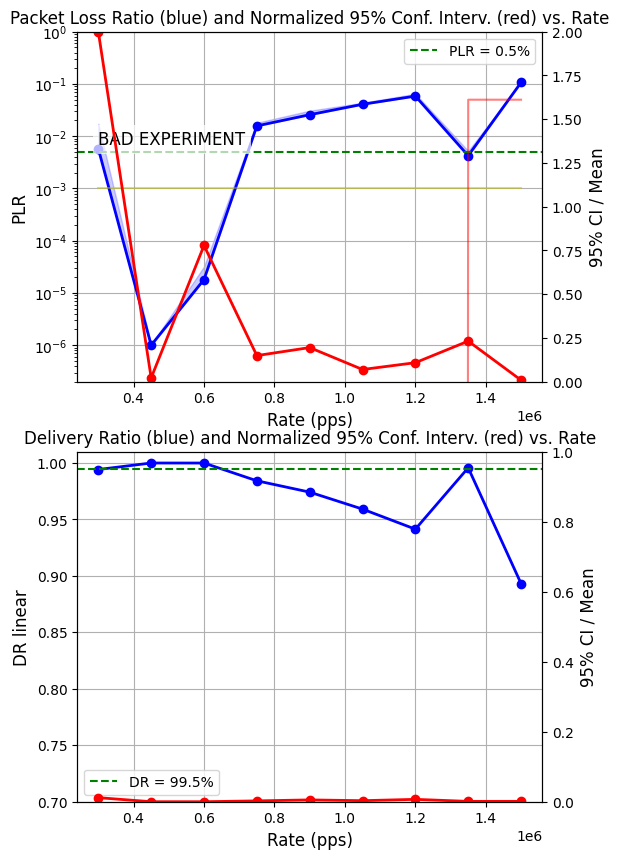}
    \caption{routing, CloudLab, bare metal server, Testbed 0, 50 runs, T = 20s, kernel 5.15, unpinned, SMALL nic buffer, date 0000-00-00, version 00}
    \label{fig:clab_raw_50_exp_t_20_k5.15_intel_bare-metal_as-is}
\end{figure}

\begin{figure}[ht!]
    \centering
    \includegraphics[width=\linewidth]{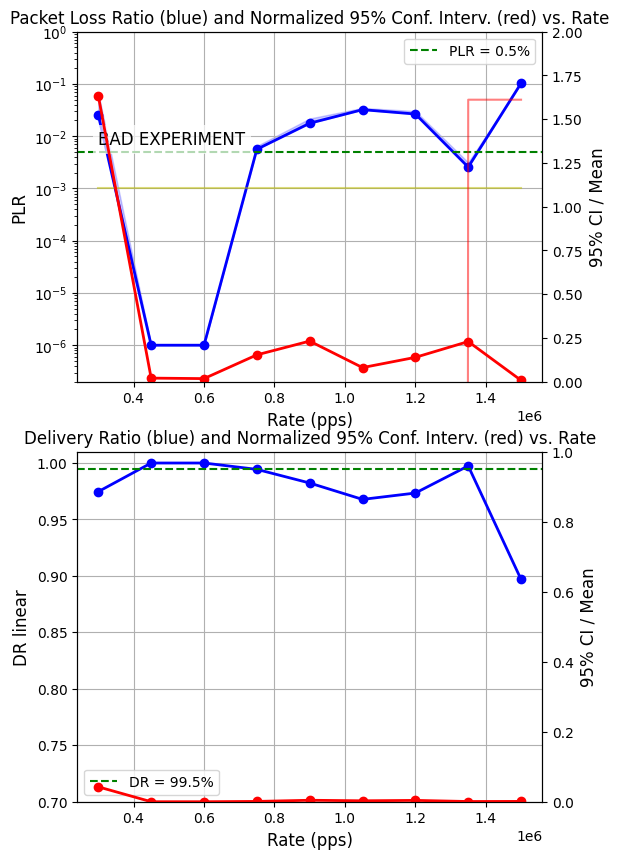}
    \caption{routing, CloudLab, bare metal server, Testbed 0, 50 runs, T = 20s, kernel 5.15, unpinned, LARGE nic buffer, date 0000-00-00, version 00}
    \label{fig:clab_raw_50_exp_t_20_k5.15_intel_bare-metal_nic_buf_4096}
\end{figure}

\begin{figure}[ht!]
    \centering
    \includegraphics[width=\linewidth]{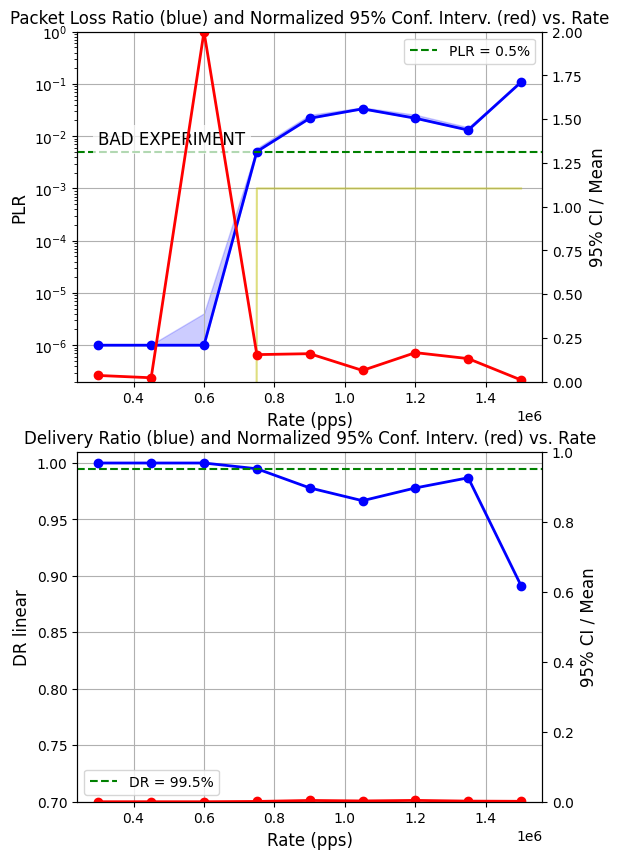}
    \caption{routing, CloudLab, bare metal server, Testbed 0, 50 runs, T = 20s, kernel 5.15, pin-1cpu, LARGE nic buffer, date 0000-00-00, version 00}
    \label{fig:clab_raw_50_exp_t_20_k5.15_intel_bare-metal_nic_buf_4096_irq_pf0-4_pf1-4}
\end{figure}

\begin{figure}[ht!]
    \centering
    \includegraphics[width=\linewidth]{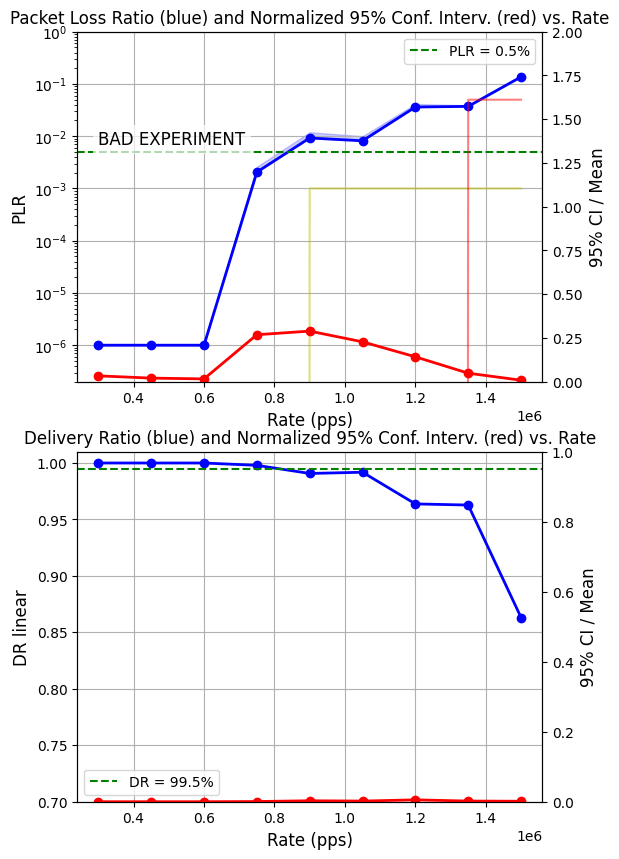}
    \caption{routing, CloudLab, bare metal server, Testbed 0, 50 runs, T = 20s, kernel 5.15, pin-2cpu, LARGE nic buffer, date 0000-00-00, version 00}
    \label{fig:clab_raw_50_exp_t_20_k5.15_intel_bare-metal_nic_buf_4096_irq_pf0-4_pf1-6}
\end{figure}

\begin{figure}[ht!]
    \centering
    \includegraphics[width=\linewidth]{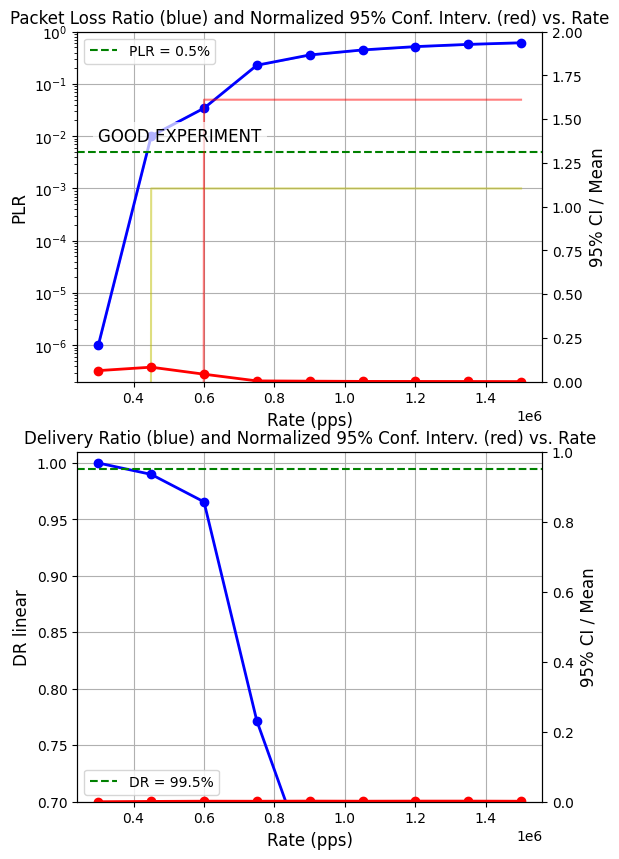}
    \caption{routing, CloudLab, bare metal server, Testbed 1, 50 runs, T = 10s, kernel 5.15, unpinned, SMALL nic buffer, date 0000-00-00, version 00}
    \label{fig:clab_tb1_raw_50_exp_t_10_k5.15_intel_bare-metal_as-is}
\end{figure}

\begin{figure}[ht!]
    \centering
    \includegraphics[width=\linewidth]{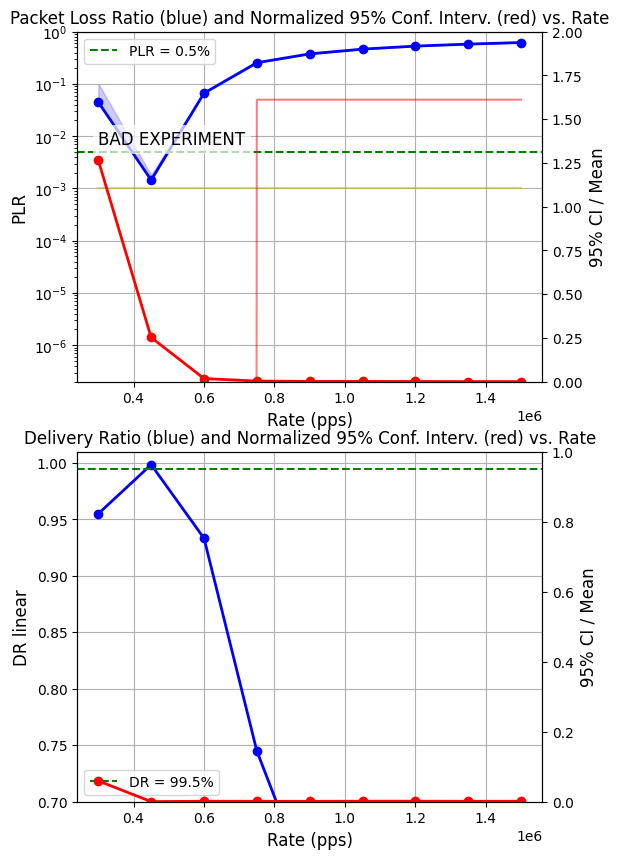}
    \caption{routing, CloudLab, bare metal server, Testbed 1, 50 runs, T = 10s, kernel 5.15, unpinned, LARGE nic buffer, date 0000-00-00, version 00}
    \label{fig:clab_tb1_raw_50_exp_t_10_k5.15_intel_bare-metal_nic_buf_4096}
\end{figure}

\begin{figure}[ht!]
    \centering
    \includegraphics[width=\linewidth]{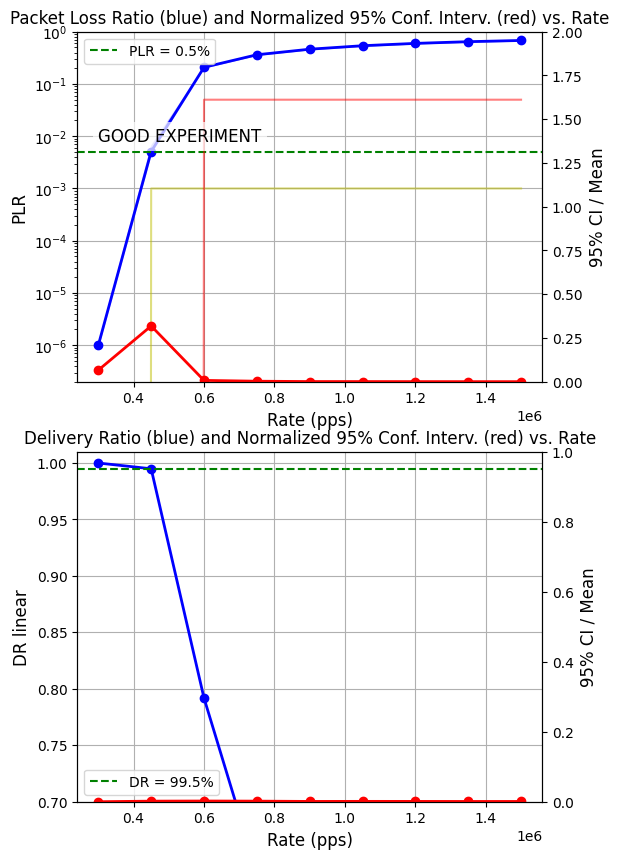}
    \caption{routing, CloudLab, bare metal server, Testbed 1, 50 runs, T = 10s, kernel 5.15, pin-1cpu, LARGE nic buffer, date 0000-00-00, version 00}
    \label{fig:clab_tb1_raw_50_exp_t_10_k5.15_intel_bare-metal_nic_buf_4096_irq_pf0-4_pf1-4}
\end{figure}

\begin{figure}[ht!]
    \centering
    \includegraphics[width=\linewidth]{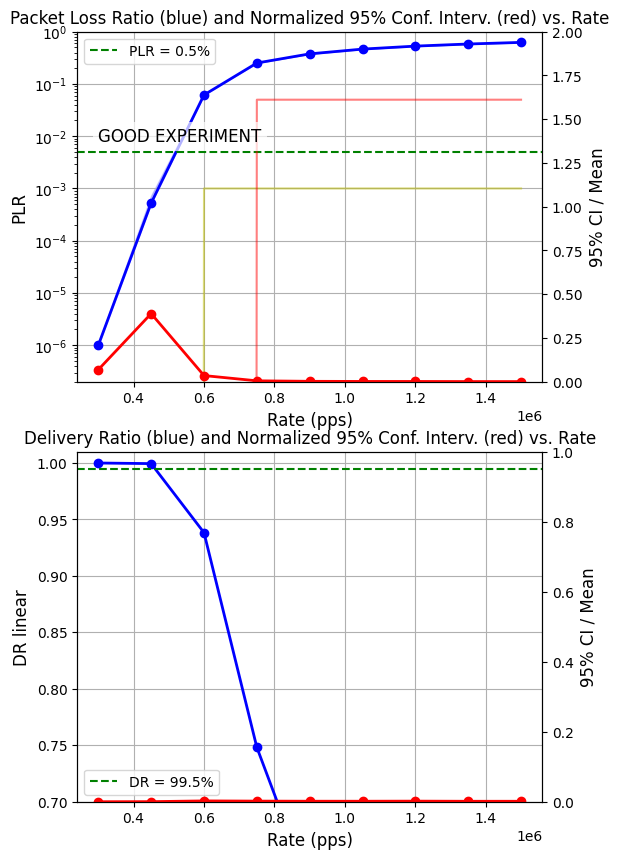}
    \caption{routing, CloudLab, bare metal server, Testbed 1, 50 runs, T = 10s, kernel 5.15, pin-2cpu, LARGE nic buffer, date 0000-00-00, version 00}
    \label{fig:clab_tb1_raw_50_exp_t_10_k5.15_intel_bare-metal_nic_buf_4096_irq_pf0-4_pf1-6}
\end{figure}

\begin{figure}[ht!]
    \centering
    \includegraphics[width=\linewidth]{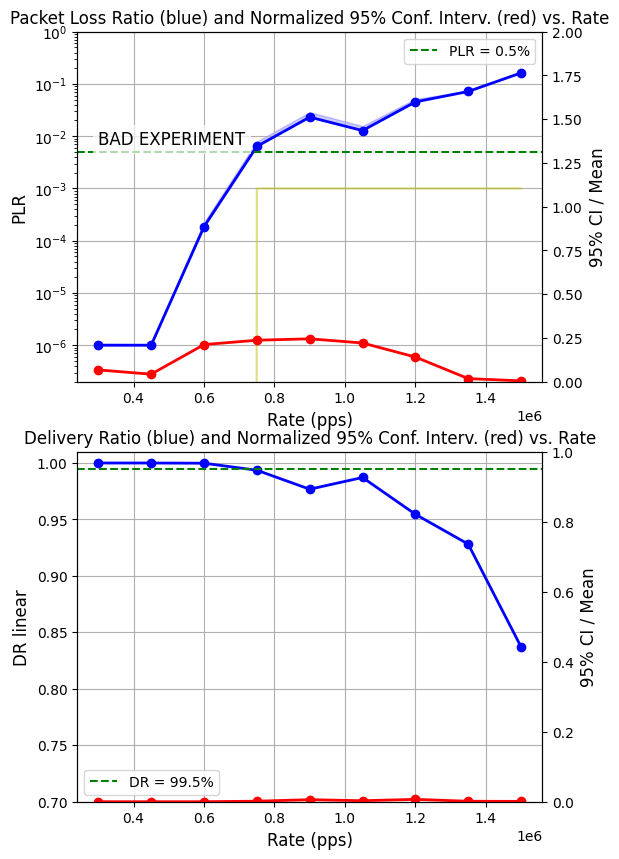}
    \caption{routing, CloudLab, bare metal server, Testbed 1, 50 runs, T = 10s, kernel 6.5, unpinned, SMALL nic buffer, date 0000-00-00, version 00}
    \label{fig:clab_tb1_raw_50_exp_t_10_k6.5_intel_bare-metal_as-is}
\end{figure}

\begin{figure}[ht!]
    \centering
    \includegraphics[width=\linewidth]{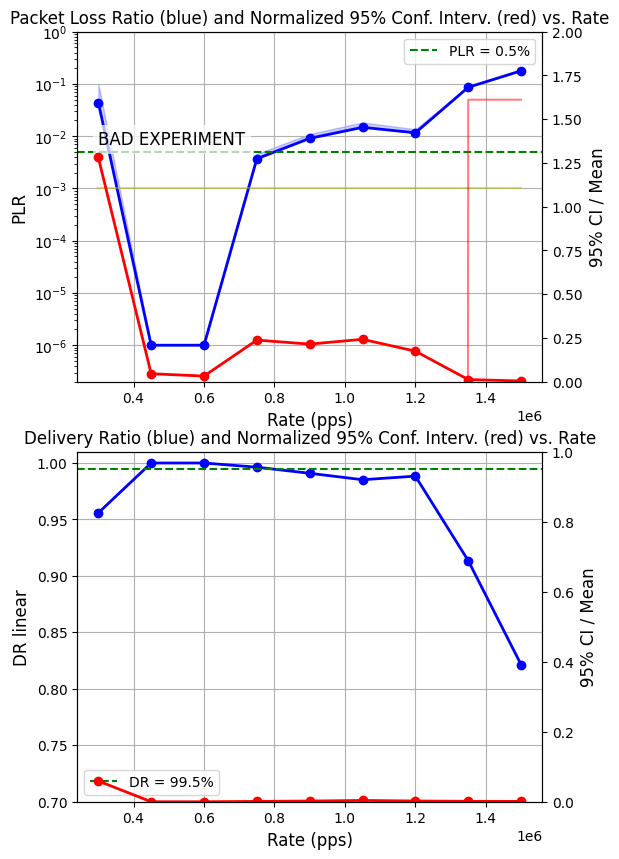}
    \caption{routing, CloudLab, bare metal server, Testbed 1, 50 runs, T = 10s, kernel 6.5, unpinned, LARGE nic buffer, date 0000-00-00, version 00}
    \label{fig:clab_tb1_raw_50_exp_t_10_k6.5_intel_bare-metal_nic_buf_4096}
\end{figure}

\begin{figure}[ht!]
    \centering
    \includegraphics[width=\linewidth]{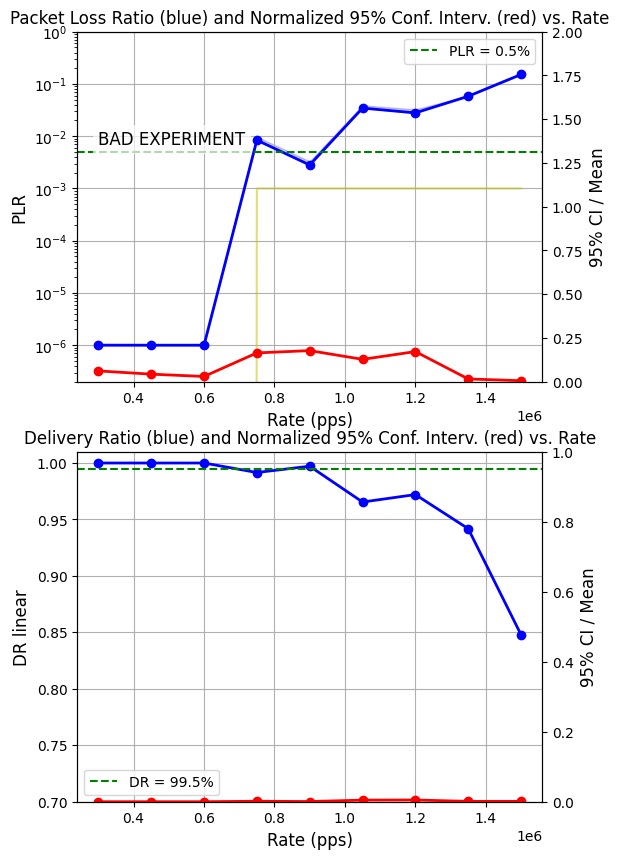}
    \caption{routing, CloudLab, bare metal server, Testbed 1, 50 runs, T = 10s, kernel 6.5, pin-1cpu, LARGE nic buffer, date 0000-00-00, version 00}
    \label{fig:clab_tb1_raw_50_exp_t_10_k6.5_intel_bare-metal_nic_buf_4096_irq_pf0-4_pf1-4}
\end{figure}

\begin{figure}[ht!]
    \centering
    \includegraphics[width=\linewidth]{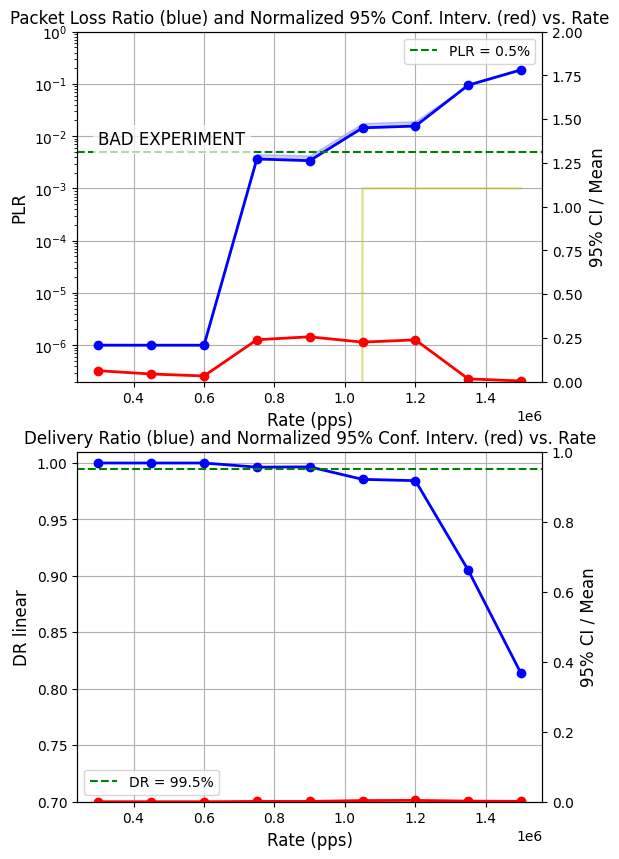}
    \caption{routing, CloudLab, bare metal server, Testbed 1, 50 runs, T = 10s, kernel 6.5, pin-2cpu, LARGE nic buffer, date 0000-00-00, version 00}
    \label{fig:clab_tb1_raw_50_exp_t_10_k6.5_intel_bare-metal_nic_buf_4096_irq_pf0-4_pf1-6}
\end{figure}

\begin{figure}[ht!]
    \centering
    \includegraphics[width=\linewidth]{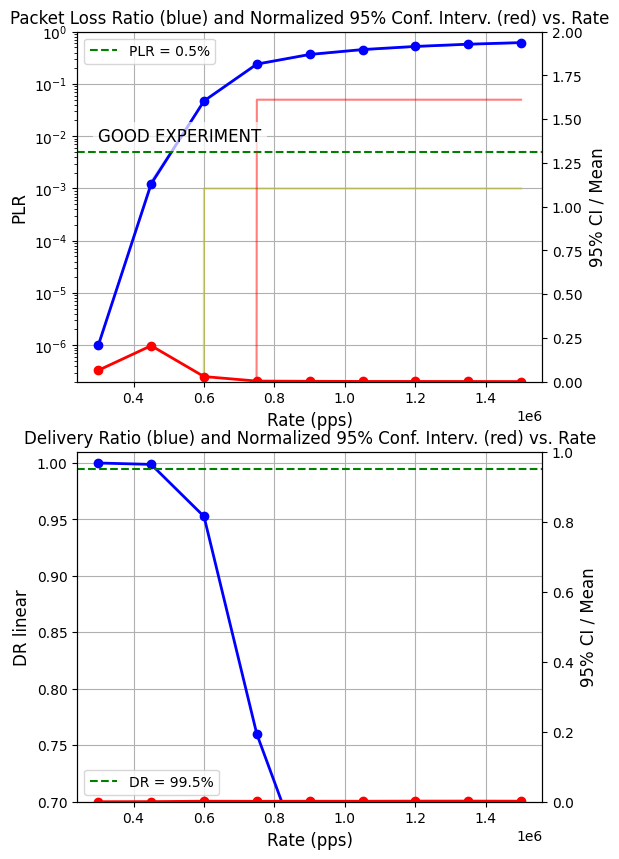}
    \caption{routing, CloudLab, bare metal server, Testbed 1, 50 runs, T = 10s, kernel 6.8, unpinned, SMALL nic buffer, date 0000-00-00, version 00}
    \label{fig:clab_tb1_raw_50_exp_t_10_k6.8_intel_bare-metal_as-is}
\end{figure}

\begin{figure}[ht!]
    \centering
    \includegraphics[width=\linewidth]{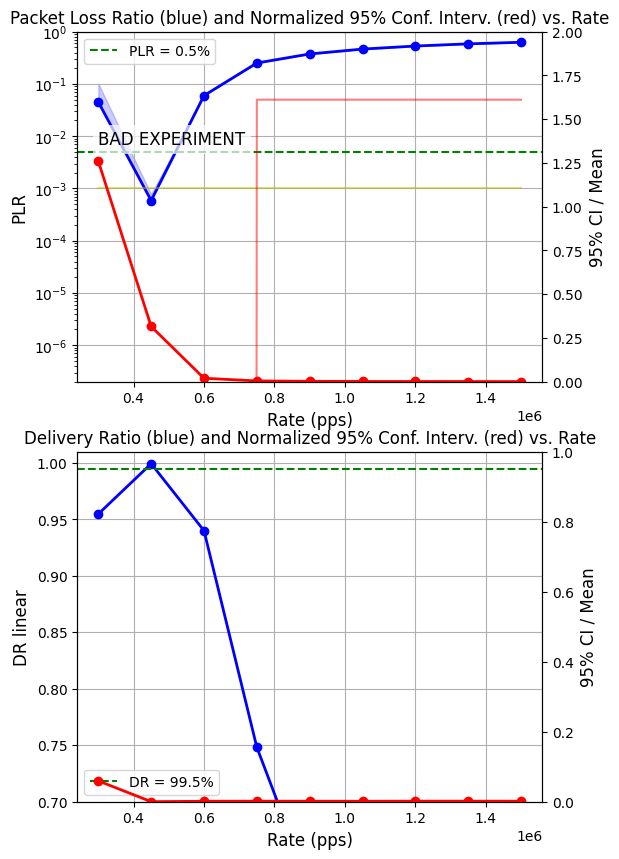}
    \caption{routing, CloudLab, bare metal server, Testbed 1, 50 runs, T = 10s, kernel 6.8, unpinned, LARGE nic buffer, date 0000-00-00, version 00}
    \label{fig:clab_tb1_raw_50_exp_t_10_k6.8_intel_bare-metal_nic_buf_4096}
\end{figure}

\begin{figure}[ht!]
    \centering
    \includegraphics[width=\linewidth]{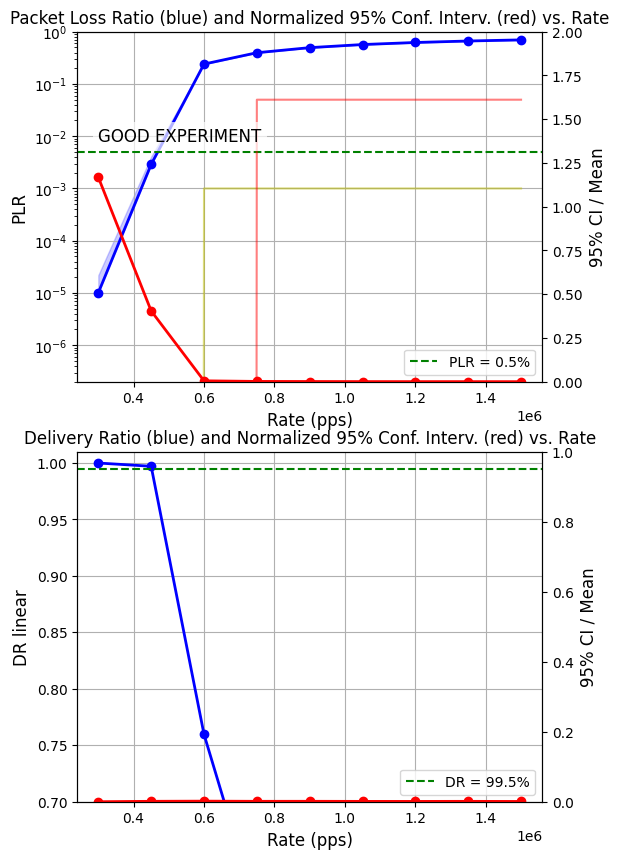}
    \caption{routing, CloudLab, bare metal server, Testbed 1, 50 runs, T = 10s, kernel 6.8, pin-1cpu, LARGE nic buffer, date 0000-00-00, version 00}
    \label{fig:clab_tb1_raw_50_exp_t_10_k6.8_intel_bare-metal_nic_buf_4096_irq_pf0-4_pf1-4}
\end{figure}

\begin{figure}[ht!]
    \centering
    \includegraphics[width=\linewidth]{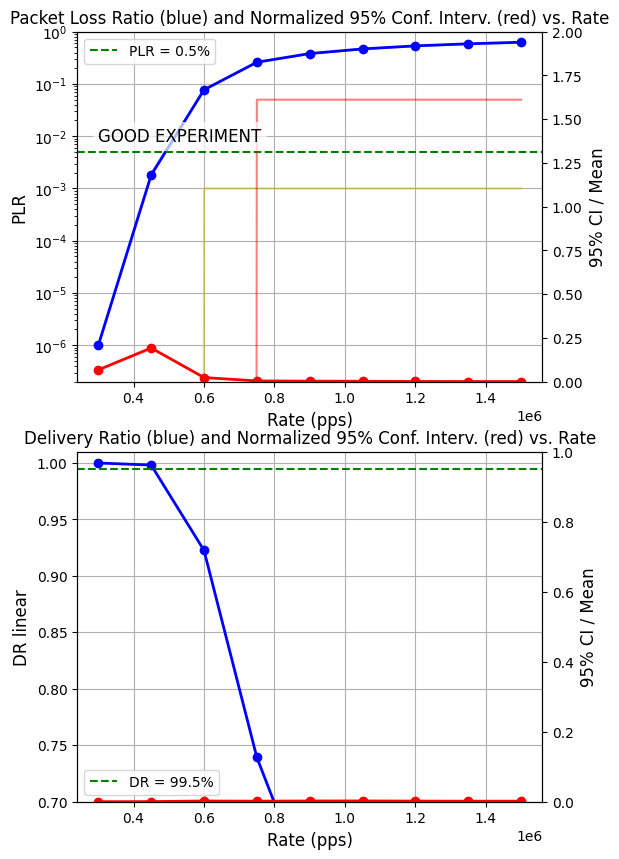}
    \caption{routing, CloudLab, bare metal server, Testbed 1, 50 runs, T = 10s, kernel 6.8, pin-2cpu, LARGE nic buffer, date 0000-00-00, version 00}
    \label{fig:clab_tb1_raw_50_exp_t_10_k6.8_intel_bare-metal_nic_buf_4096_irq_pf0-4_pf1-6}
\end{figure}

\begin{figure}[ht!]
    \centering
    \includegraphics[width=\linewidth]{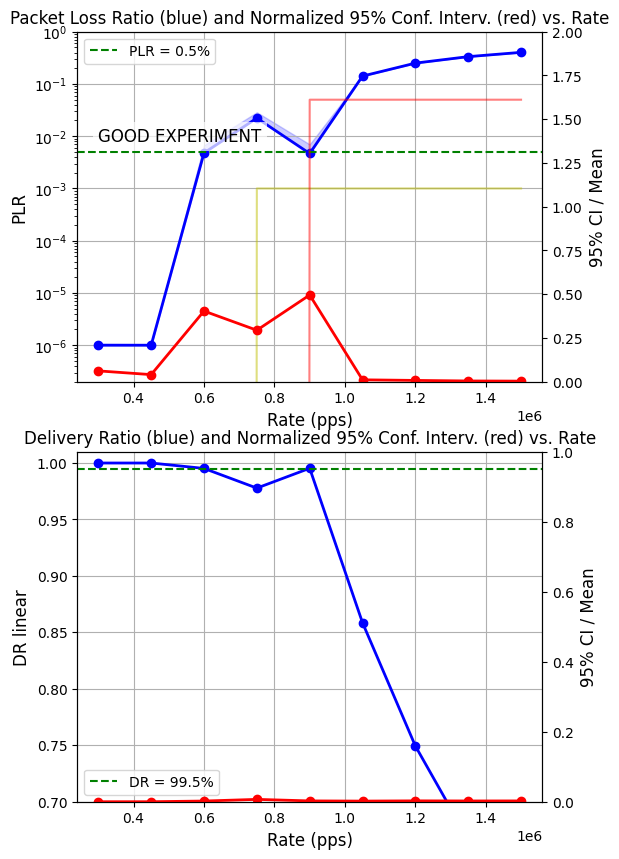}
    \caption{routing, CloudLab, bare metal server, Testbed 2, 30 runs, T = 10s, kernel 6.5, unpinned, LARGE nic buffer, date 2024-10-24, version 02}
    \label{fig:clab_tb2_raw_30_exp_t_10_k6.5_intel_bare-metal_nic_buf_4096_date_2024-10-24_ver_02}
\end{figure}

\begin{figure}[ht!]
    \centering
    \includegraphics[width=\linewidth]{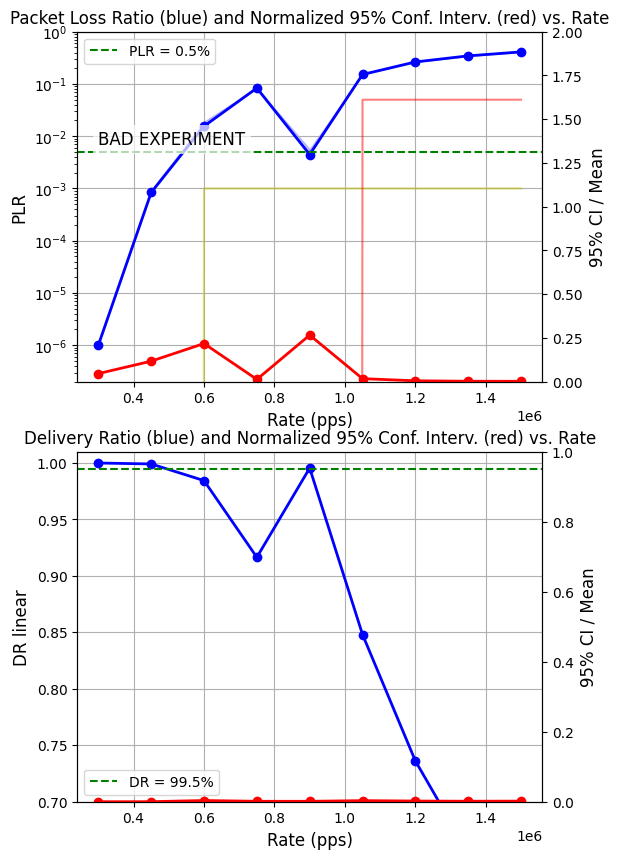}
    \caption{routing, CloudLab, bare metal server, Testbed 2, 50 runs, T = 10s, kernel 5.15, unpinned, SMALL nic buffer, date 0000-00-00, version 00}
    \label{fig:clab_tb2_raw_50_exp_t_10_k5.15_intel_bare-metal_as-is}
\end{figure}

\begin{figure}[ht!]
    \centering
    \includegraphics[width=\linewidth]{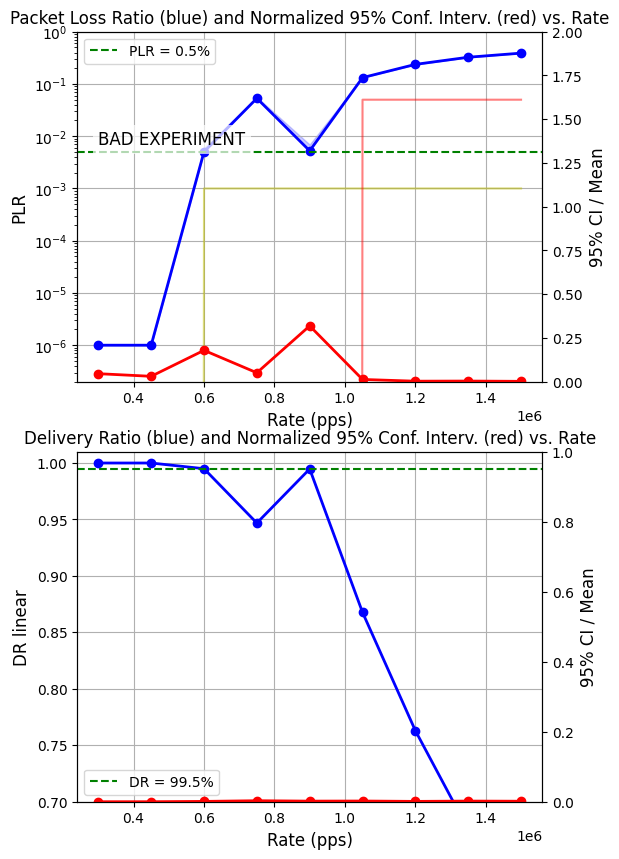}
    \caption{routing, CloudLab, bare metal server, Testbed 2, 50 runs, T = 10s, kernel 5.15, unpinned, LARGE nic buffer, date 0000-00-00, version 00}
    \label{fig:clab_tb2_raw_50_exp_t_10_k5.15_intel_bare-metal_nic_buf_4096}
\end{figure}

\begin{figure}[ht!]
    \centering
    \includegraphics[width=\linewidth]{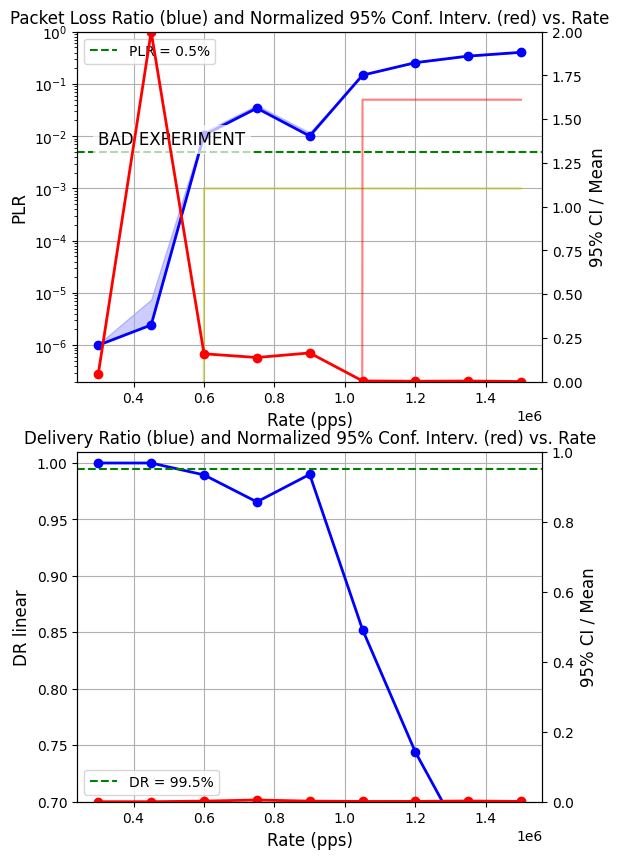}
    \caption{routing, CloudLab, bare metal server, Testbed 2, 50 runs, T = 10s, kernel 5.15, unpinned, LARGE nic buffer, date 2024-10-24, version 02}
    \label{fig:clab_tb2_raw_50_exp_t_10_k5.15_intel_bare-metal_nic_buf_4096_date_2024-10-24_ver_02}
\end{figure}

\begin{figure}[ht!]
    \centering
    \includegraphics[width=\linewidth]{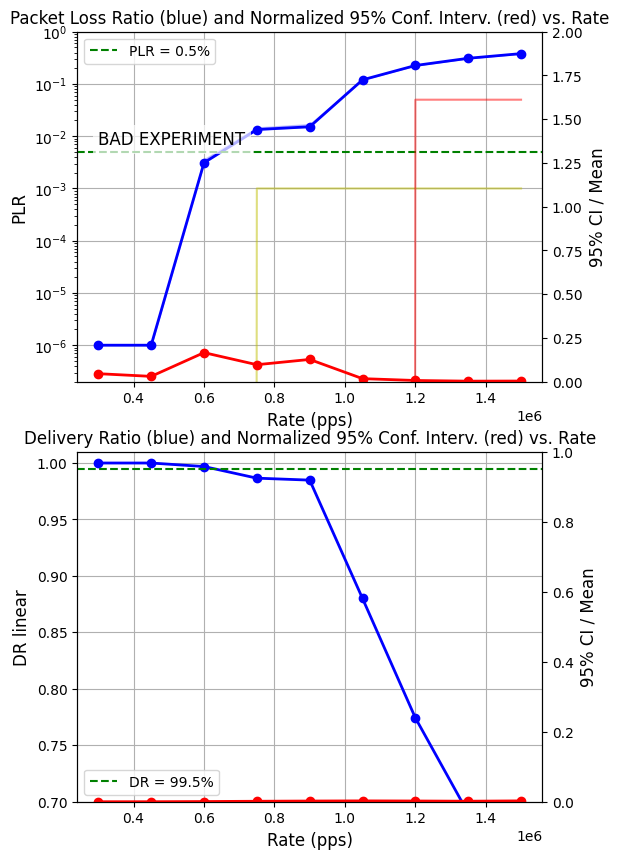}
    \caption{routing, CloudLab, bare metal server, Testbed 2, 50 runs, T = 10s, kernel 5.15, pin-1cpu, LARGE nic buffer, date 0000-00-00, version 00}
    \label{fig:clab_tb2_raw_50_exp_t_10_k5.15_intel_bare-metal_nic_buf_4096_irq_pf0-4_pf1-4}
\end{figure}

\begin{figure}[ht!]
    \centering
    \includegraphics[width=\linewidth]{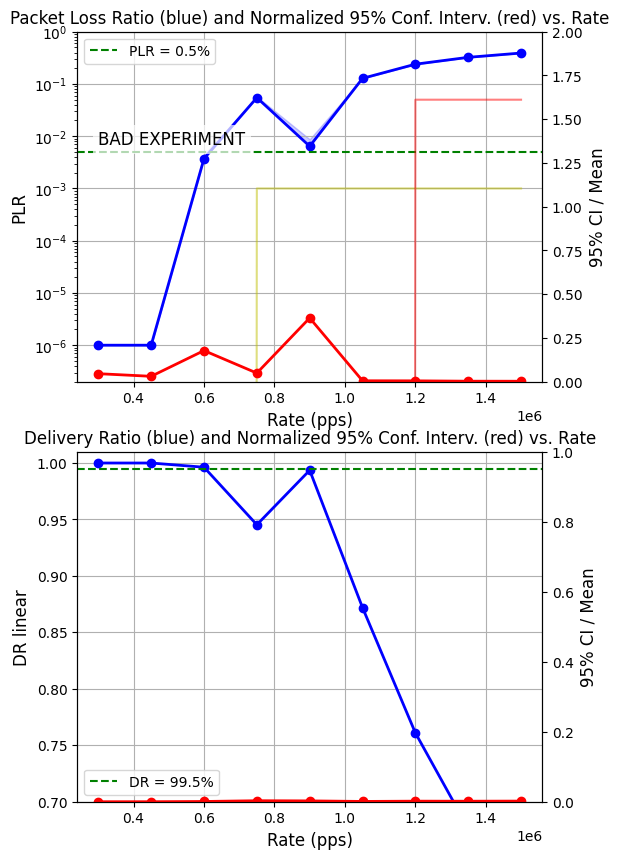}
    \caption{routing, CloudLab, bare metal server, Testbed 2, 50 runs, T = 10s, kernel 5.15, pin-2cpu, LARGE nic buffer, date 0000-00-00, version 00}
    \label{fig:clab_tb2_raw_50_exp_t_10_k5.15_intel_bare-metal_nic_buf_4096_irq_pf0-4_pf1-6}
\end{figure}

\begin{figure}[ht!]
    \centering
    \includegraphics[width=\linewidth]{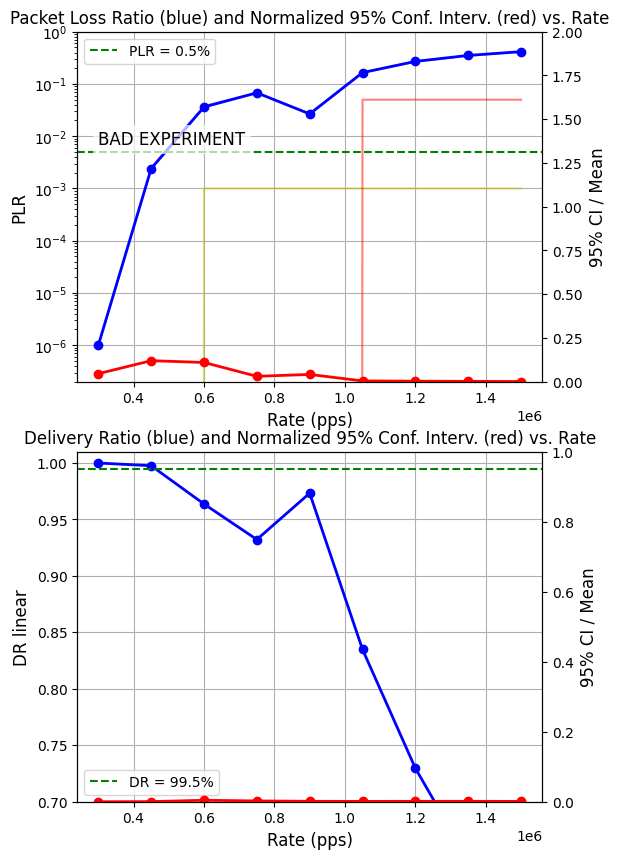}
    \caption{routing, CloudLab, bare metal server, Testbed 2, 50 runs, T = 10s, kernel 6.5, unpinned, SMALL nic buffer, date 0000-00-00, version 00}
    \label{fig:clab_tb2_raw_50_exp_t_10_k6.5_intel_bare-metal_as-is}
\end{figure}

\begin{figure}[ht!]
    \centering
    \includegraphics[width=\linewidth]{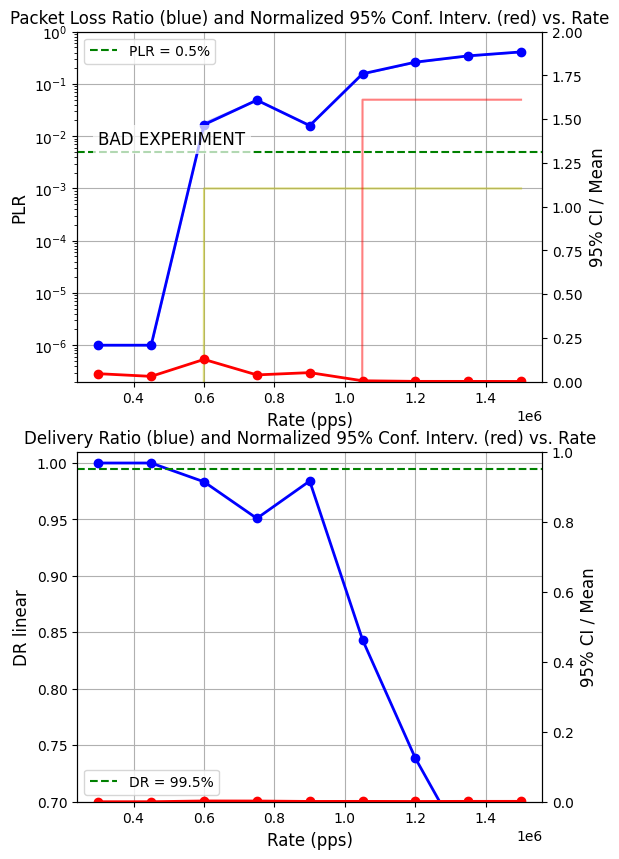}
    \caption{routing, CloudLab, bare metal server, Testbed 2, 50 runs, T = 10s, kernel 6.5, unpinned, LARGE nic buffer, date 0000-00-00, version 00}
    \label{fig:clab_tb2_raw_50_exp_t_10_k6.5_intel_bare-metal_nic_buf_4096}
\end{figure}

\begin{figure}[ht!]
    \centering
    \includegraphics[width=\linewidth]{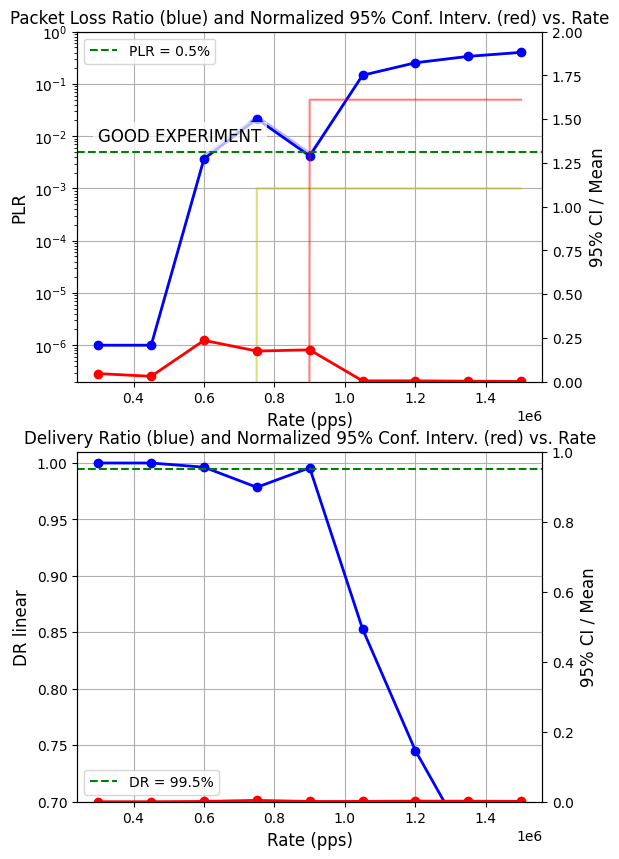}
    \caption{routing, CloudLab, bare metal server, Testbed 2, 50 runs, T = 10s, kernel 6.5, unpinned, LARGE nic buffer, date 2024-10-24, version 02}
    \label{fig:clab_tb2_raw_50_exp_t_10_k6.5_intel_bare-metal_nic_buf_4096_date_2024-10-24_ver_02}
\end{figure}

\begin{figure}[ht!]
    \centering
    \includegraphics[width=\linewidth]{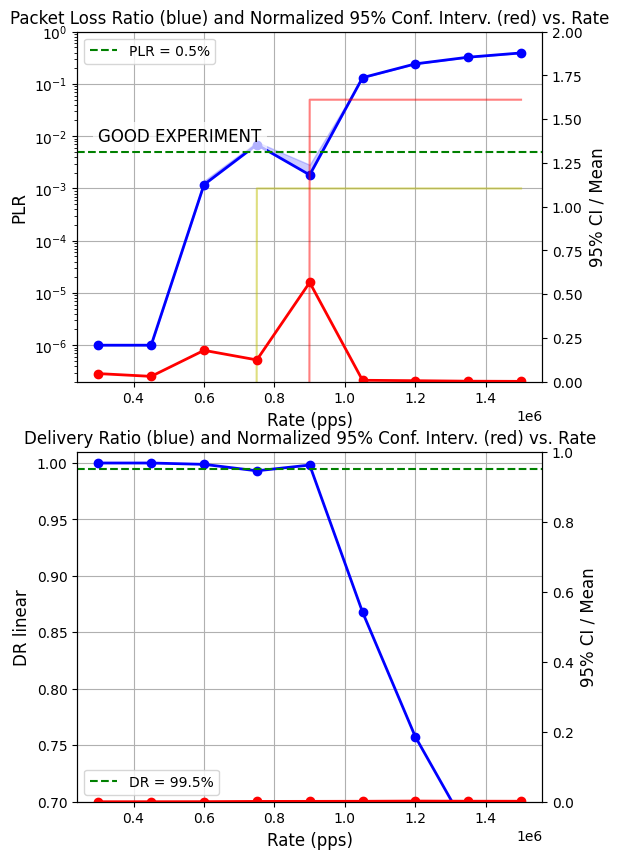}
    \caption{routing, CloudLab, bare metal server, Testbed 2, 50 runs, T = 10s, kernel 6.5, pin-1cpu, LARGE nic buffer, date 0000-00-00, version 00}
    \label{fig:clab_tb2_raw_50_exp_t_10_k6.5_intel_bare-metal_nic_buf_4096_irq_pf0-4_pf1-4}
\end{figure}

\begin{figure}[ht!]
    \centering
    \includegraphics[width=\linewidth]{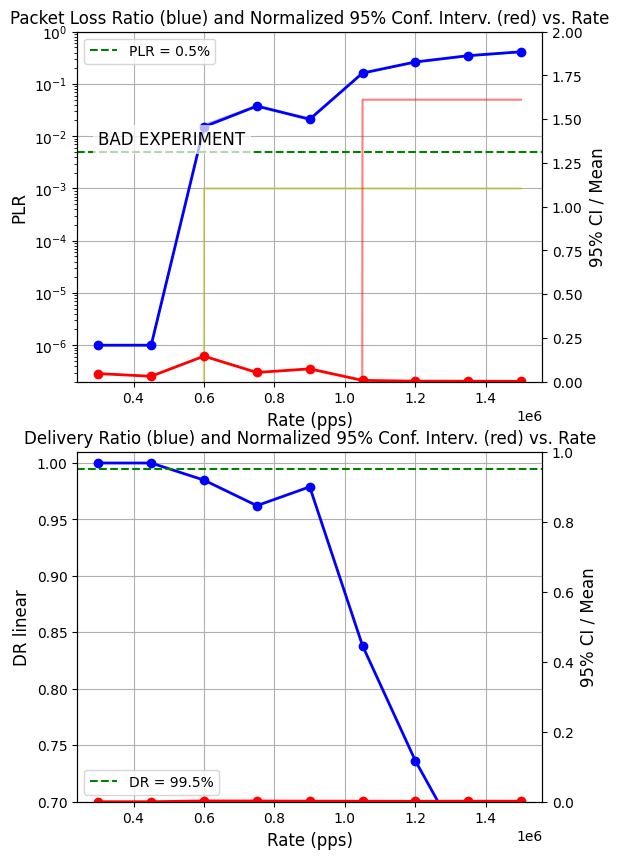}
    \caption{routing, CloudLab, bare metal server, Testbed 2, 50 runs, T = 10s, kernel 6.5, pin-2cpu, LARGE nic buffer, date 0000-00-00, version 00}
    \label{fig:clab_tb2_raw_50_exp_t_10_k6.5_intel_bare-metal_nic_buf_4096_irq_pf0-4_pf1-6}
\end{figure}

\begin{figure}[ht!]
    \centering
    \includegraphics[width=\linewidth]{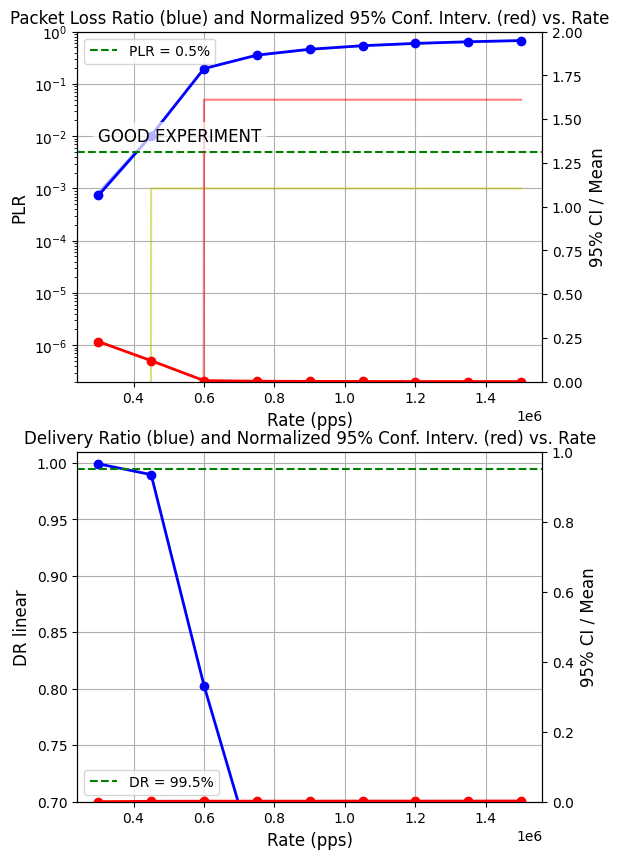}
    \caption{routing, CloudLab, bare metal server, Testbed 2, 50 runs, T = 10s, kernel 6.8, unpinned, SMALL nic buffer, date 0000-00-00, version 00}
    \label{fig:clab_tb2_raw_50_exp_t_10_k6.8_intel_bare-metal_as-is}
\end{figure}

\begin{figure}[ht!]
    \centering
    \includegraphics[width=\linewidth]{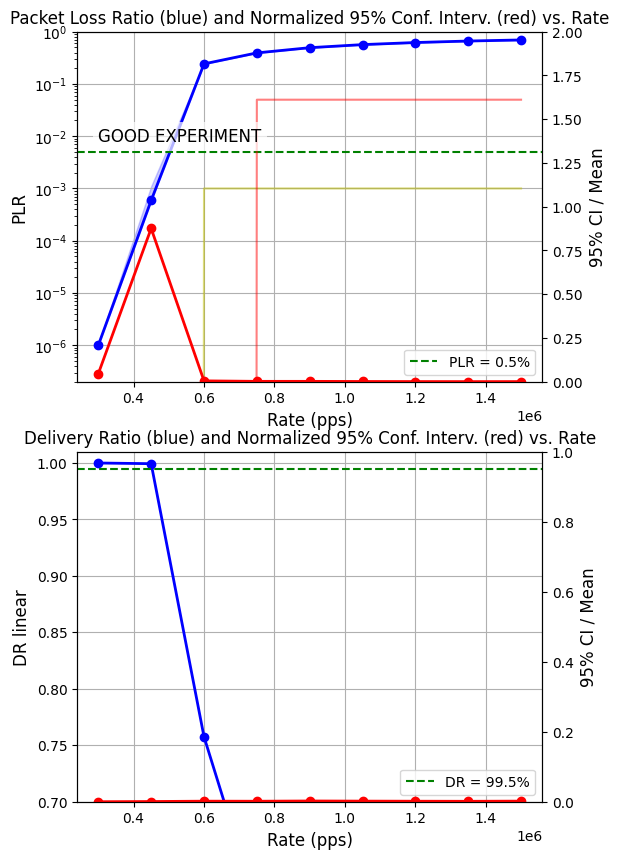}
    \caption{routing, CloudLab, bare metal server, Testbed 2, 50 runs, T = 10s, kernel 6.8, unpinned, LARGE nic buffer, date 0000-00-00, version 00}
    \label{fig:clab_tb2_raw_50_exp_t_10_k6.8_intel_bare-metal_nic_buf_4096}
\end{figure}

\begin{figure}[ht!]
    \centering
    \includegraphics[width=\linewidth]{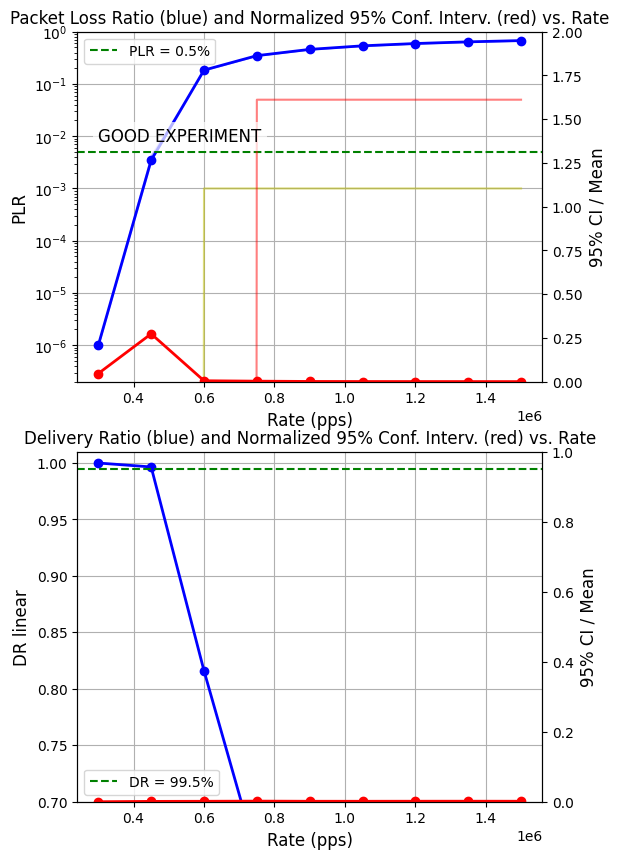}
    \caption{routing, CloudLab, bare metal server, Testbed 2, 50 runs, T = 10s, kernel 6.8, unpinned, LARGE nic buffer, date 2024-10-24, version 02}
    \label{fig:clab_tb2_raw_50_exp_t_10_k6.8_intel_bare-metal_nic_buf_4096_date_2024-10-24_ver_02}
\end{figure}

\begin{figure}[ht!]
    \centering
    \includegraphics[width=\linewidth]{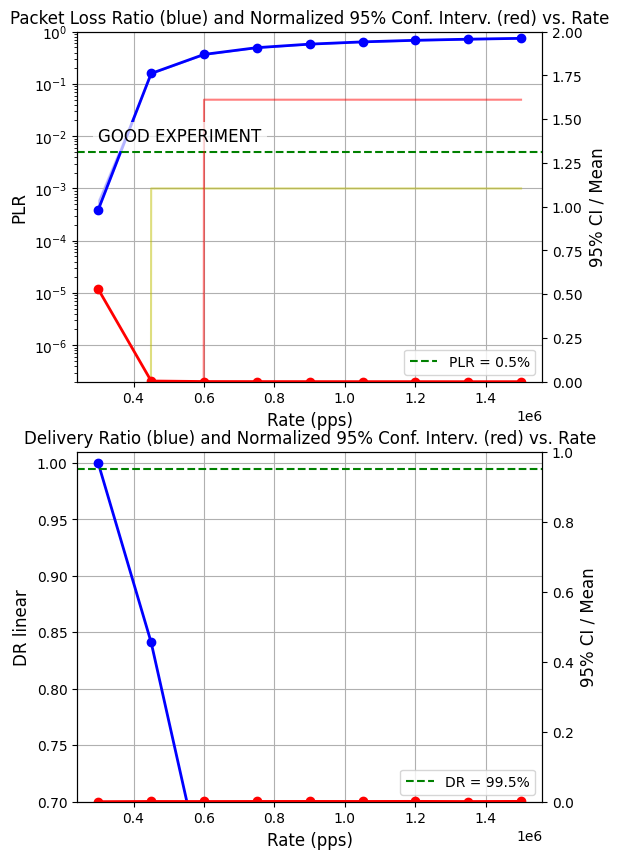}
    \caption{routing, CloudLab, bare metal server, Testbed 2, 50 runs, T = 10s, kernel 6.8, pin-1cpu, LARGE nic buffer, date 0000-00-00, version 00}
    \label{fig:clab_tb2_raw_50_exp_t_10_k6.8_intel_bare-metal_nic_buf_4096_irq_pf0-4_pf1-4}
\end{figure}

\begin{figure}[ht!]
    \centering
    \includegraphics[width=\linewidth]{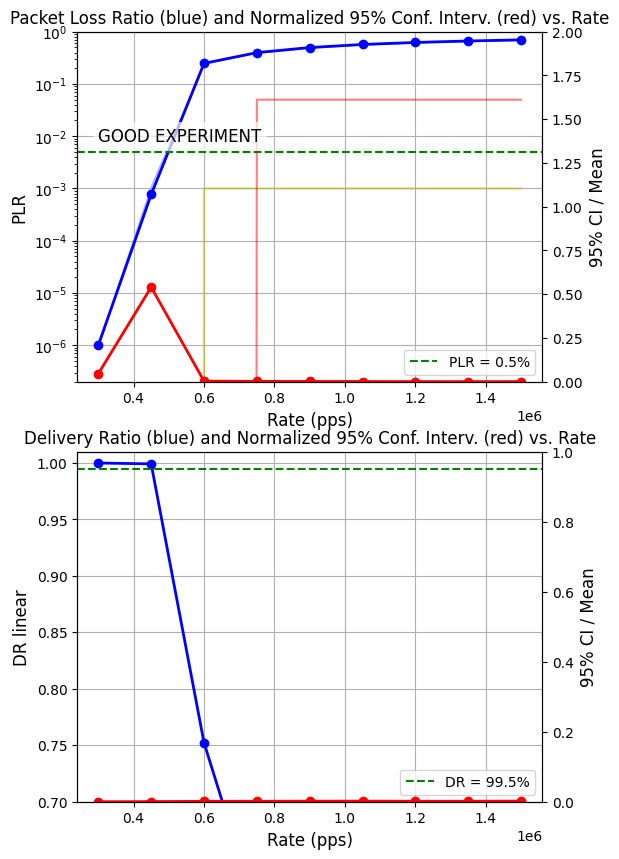}
    \caption{routing, CloudLab, bare metal server, Testbed 2, 50 runs, T = 10s, kernel 6.8, pin-2cpu, LARGE nic buffer, date 0000-00-00, version 00}
    \label{fig:clab_tb2_raw_50_exp_t_10_k6.8_intel_bare-metal_nic_buf_4096_irq_pf0-4_pf1-6}
\end{figure}

\begin{figure}[ht!]
    \centering
    \includegraphics[width=\linewidth]{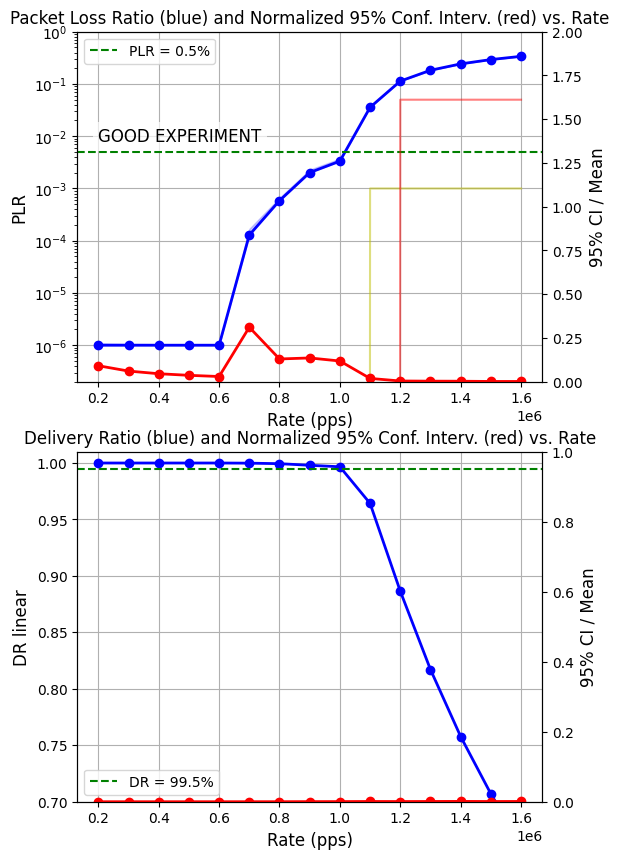}
    \caption{routing, CloudLab, bare metal server, Testbed 3, 30 runs, T = 10s, kernel 5.13, unpinned, SMALL nic buffer, date 0000-00-00, version 00}
    \label{fig:clab_tb3_raw_30_exp_t_10_k5.13_intel_bare-metal_as-is}
\end{figure}

\begin{figure}[ht!]
    \centering
    \includegraphics[width=\linewidth]{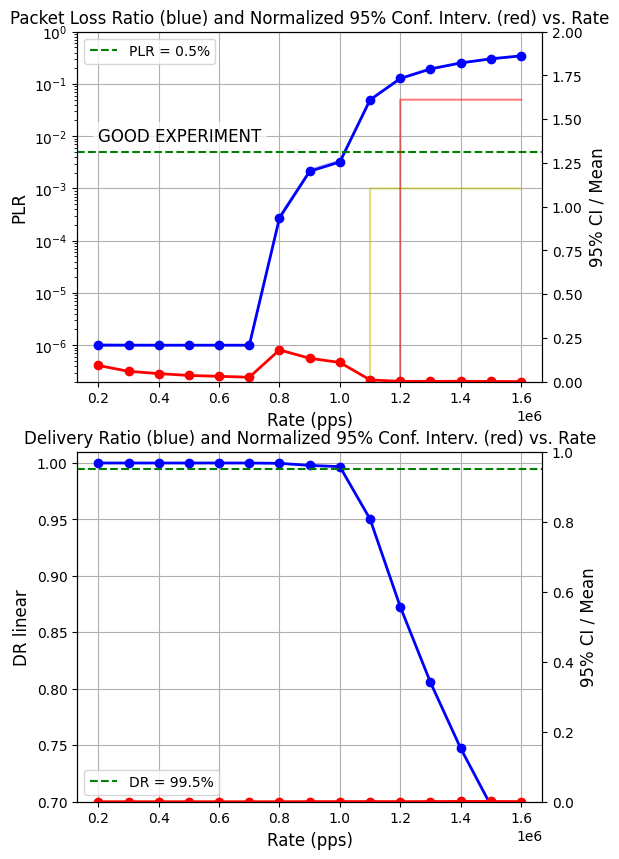}
    \caption{routing, CloudLab, bare metal server, Testbed 3, 30 runs, T = 10s, kernel 5.13, unpinned, LARGE nic buffer, date 0000-00-00, version 00}
    \label{fig:clab_tb3_raw_30_exp_t_10_k5.13_intel_bare-metal_nic_buf_4096}
\end{figure}

\begin{figure}[ht!]
    \centering
    \includegraphics[width=\linewidth]{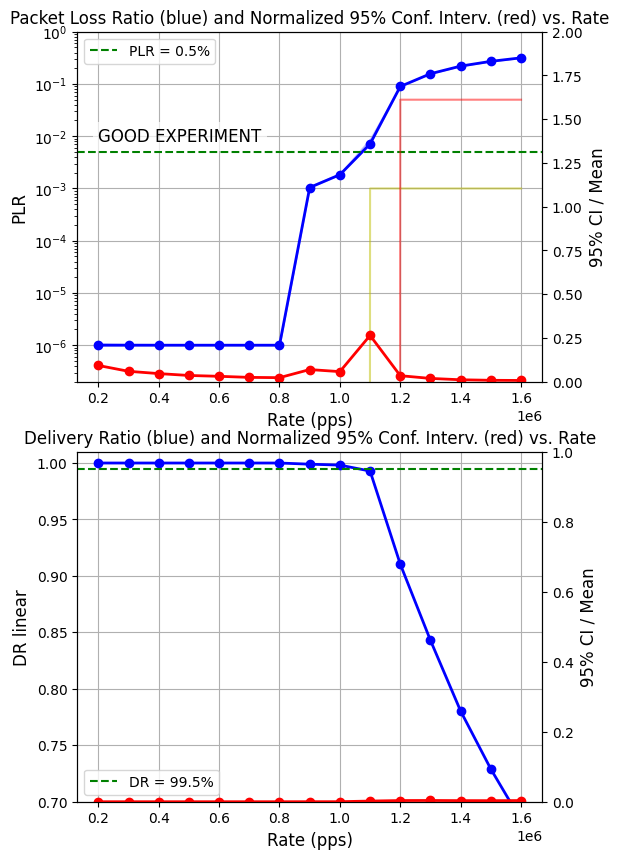}
    \caption{routing, CloudLab, bare metal server, Testbed 3, 30 runs, T = 10s, kernel 5.13, pin-1cpu, LARGE nic buffer, date 0000-00-00, version 00}
    \label{fig:clab_tb3_raw_30_exp_t_10_k5.13_intel_bare-metal_nic_buf_4096_irq_pf0-4_pf1-4}
\end{figure}

\begin{figure}[ht!]
    \centering
    \includegraphics[width=\linewidth]{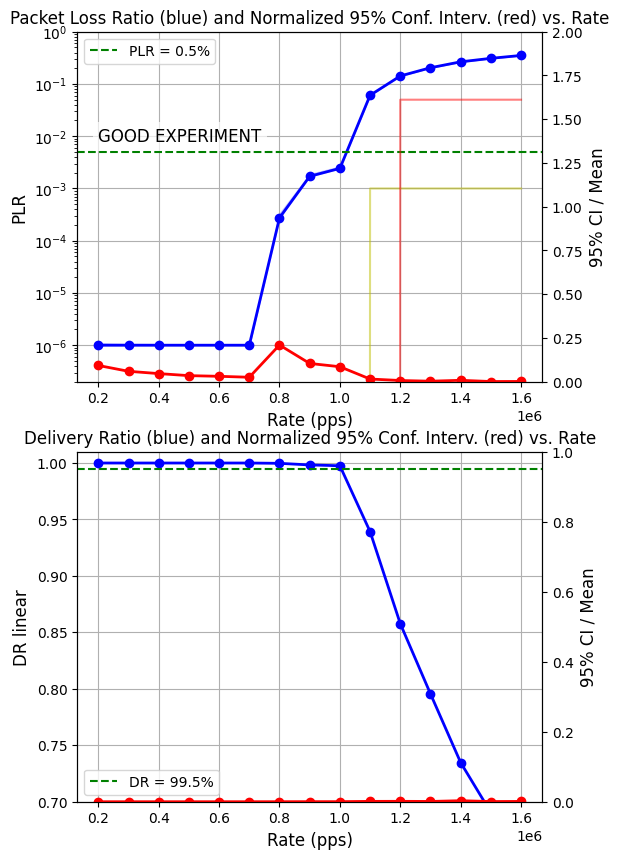}
    \caption{routing, CloudLab, bare metal server, Testbed 3, 30 runs, T = 10s, kernel 5.13, pin-2cpu, LARGE nic buffer, date 0000-00-00, version 00}
    \label{fig:clab_tb3_raw_30_exp_t_10_k5.13_intel_bare-metal_nic_buf_4096_irq_pf0-4_pf1-6}
\end{figure}

\begin{figure}[ht!]
    \centering
    \includegraphics[width=\linewidth]{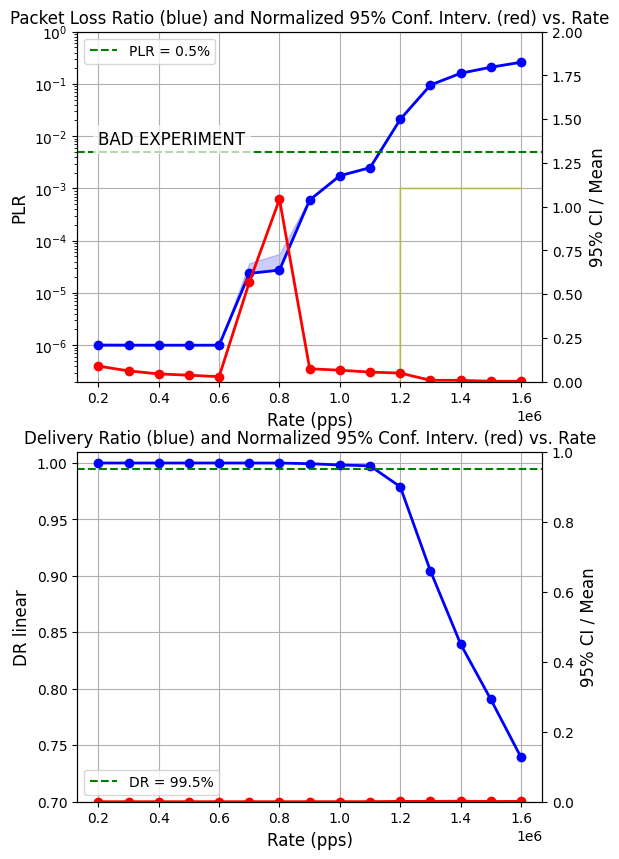}
    \caption{routing, CloudLab, bare metal server, Testbed 3, 30 runs, T = 10s, kernel 5.18, unpinned, SMALL nic buffer, date 0000-00-00, version 00}
    \label{fig:clab_tb3_raw_30_exp_t_10_k5.18_intel_bare-metal_as-is}
\end{figure}

\begin{figure}[ht!]
    \centering
    \includegraphics[width=\linewidth]{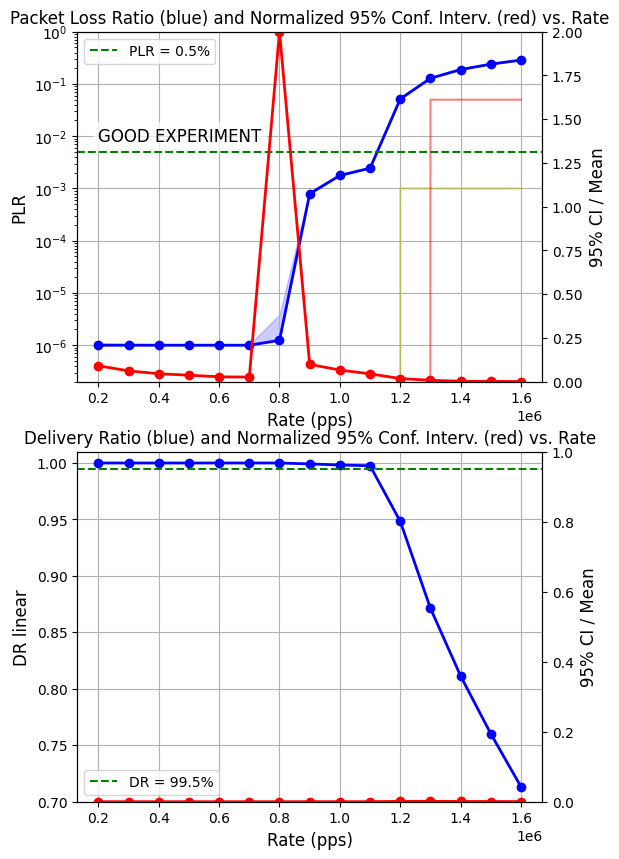}
    \caption{routing, CloudLab, bare metal server, Testbed 3, 30 runs, T = 10s, kernel 5.18, unpinned, LARGE nic buffer, date 0000-00-00, version 00}
    \label{fig:clab_tb3_raw_30_exp_t_10_k5.18_intel_bare-metal_nic_buf_4096}
\end{figure}

\begin{figure}[ht!]
    \centering
    \includegraphics[width=\linewidth]{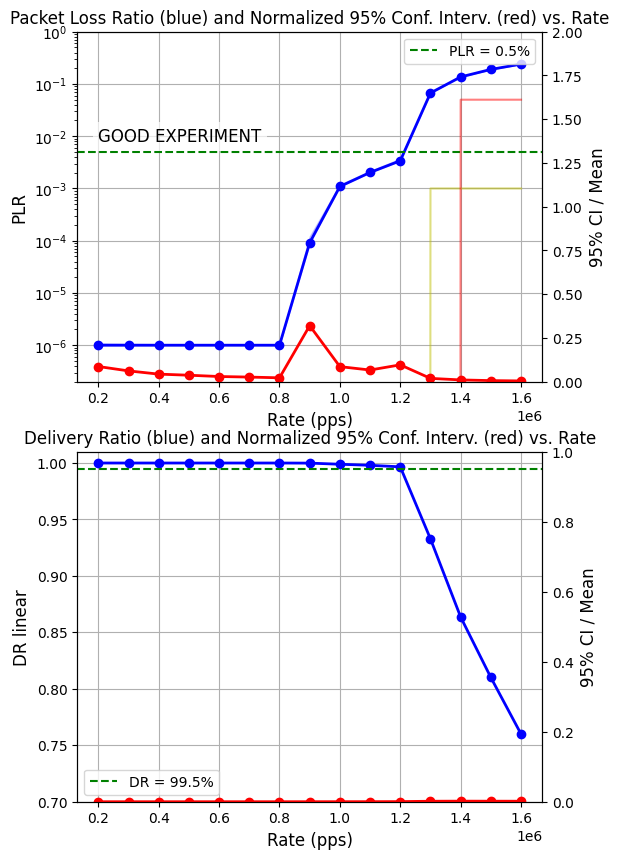}
    \caption{routing, CloudLab, bare metal server, Testbed 3, 30 runs, T = 10s, kernel 5.18, pin-1cpu, LARGE nic buffer, date 0000-00-00, version 00}
    \label{fig:clab_tb3_raw_30_exp_t_10_k5.18_intel_bare-metal_nic_buf_4096_irq_pf0-4_pf1-4}
\end{figure}

\begin{figure}[ht!]
    \centering
    \includegraphics[width=\linewidth]{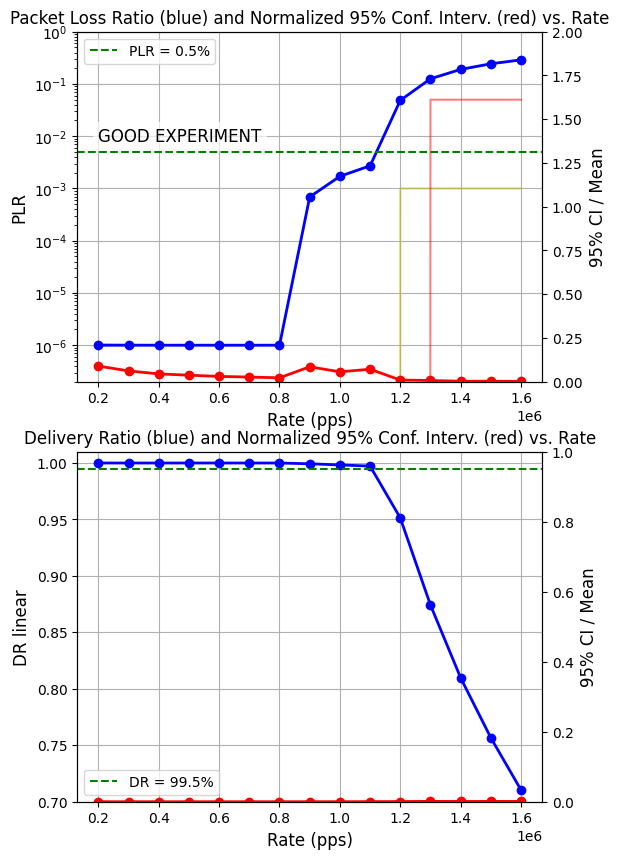}
    \caption{routing, CloudLab, bare metal server, Testbed 3, 30 runs, T = 10s, kernel 5.18, pin-2cpu, LARGE nic buffer, date 0000-00-00, version 00}
    \label{fig:clab_tb3_raw_30_exp_t_10_k5.18_intel_bare-metal_nic_buf_4096_irq_pf0-4_pf1-6}
\end{figure}

\begin{figure}[ht!]
    \centering
    \includegraphics[width=\linewidth]{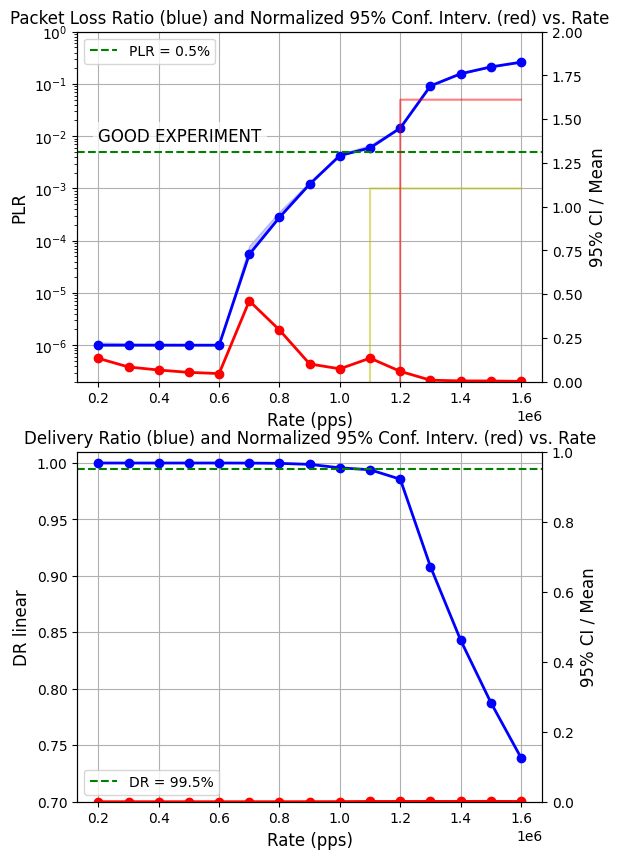}
    \caption{routing, CloudLab, bare metal server, Testbed 3, 30 runs, T = 6s, kernel 5.13, unpinned, SMALL nic buffer, date 0000-00-00, version 00}
    \label{fig:clab_tb3_raw_30_exp_t_6_k5.13_intel_bare-metal_as-is}
\end{figure}

\begin{figure}[ht!]
    \centering
    \includegraphics[width=\linewidth]{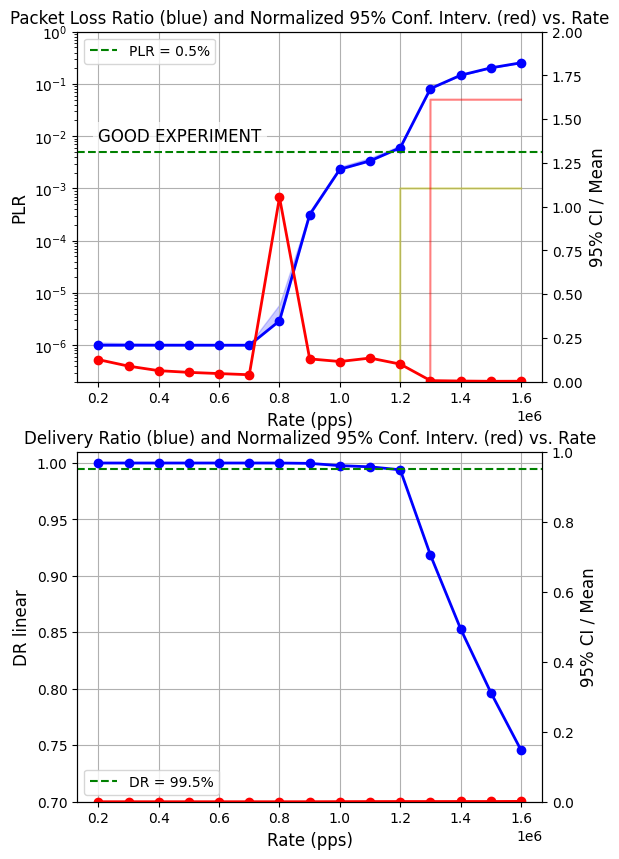}
    \caption{routing, CloudLab, bare metal server, Testbed 3, 30 runs, T = 6s, kernel 5.13, unpinned, LARGE nic buffer, date 0000-00-00, version 00}
    \label{fig:clab_tb3_raw_30_exp_t_6_k5.13_intel_bare-metal_nic_buf_4096}
\end{figure}

\begin{figure}[ht!]
    \centering
    \includegraphics[width=\linewidth]{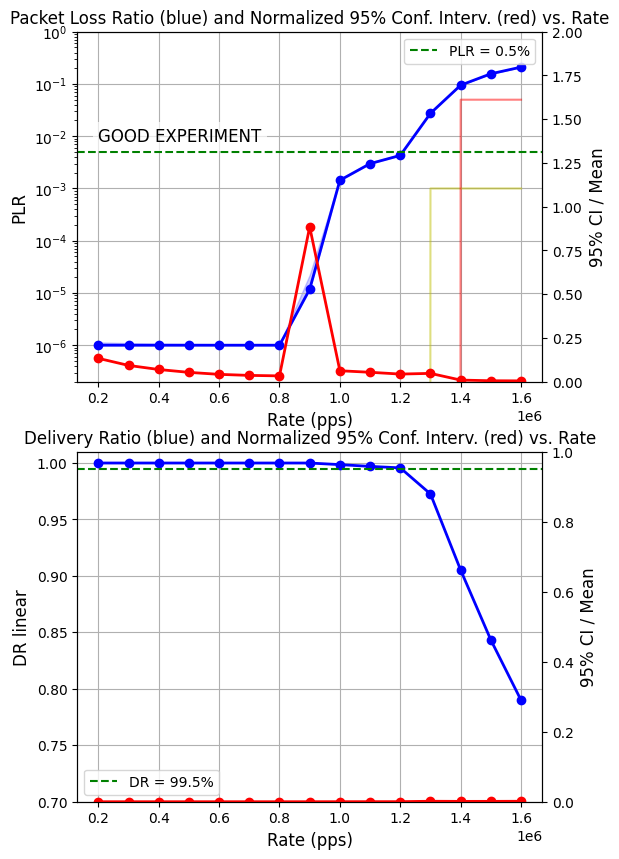}
    \caption{routing, CloudLab, bare metal server, Testbed 3, 30 runs, T = 6s, kernel 5.13, pin-1cpu, LARGE nic buffer, date 0000-00-00, version 00}
    \label{fig:clab_tb3_raw_30_exp_t_6_k5.13_intel_bare-metal_nic_buf_4096_irq_pf0-4_pf1-4}
\end{figure}

\begin{figure}[ht!]
    \centering
    \includegraphics[width=\linewidth]{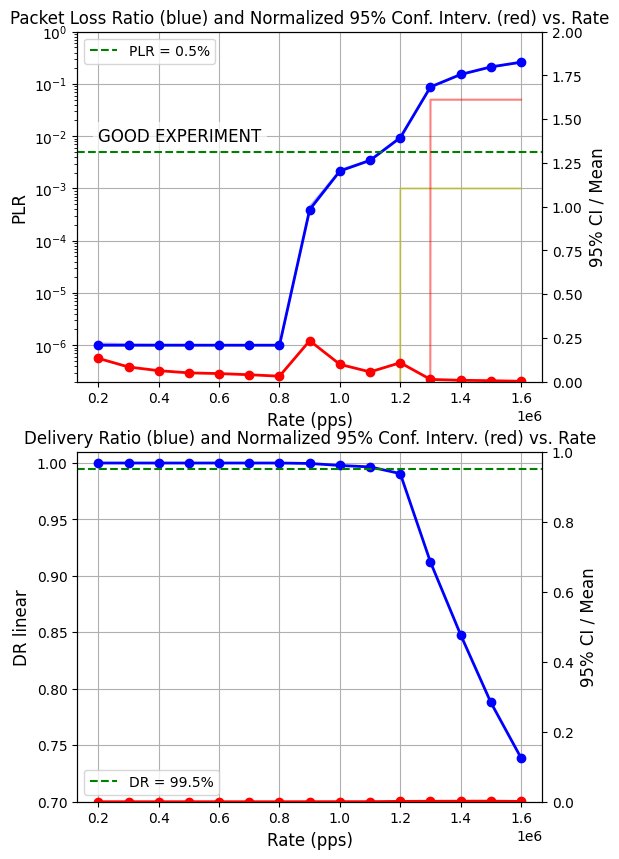}
    \caption{routing, CloudLab, bare metal server, Testbed 3, 30 runs, T = 6s, kernel 5.13, pin-2cpu, LARGE nic buffer, date 0000-00-00, version 00}
    \label{fig:clab_tb3_raw_30_exp_t_6_k5.13_intel_bare-metal_nic_buf_4096_irq_pf0-4_pf1-6}
\end{figure}

\begin{figure}[ht!]
    \centering
    \includegraphics[width=\linewidth]{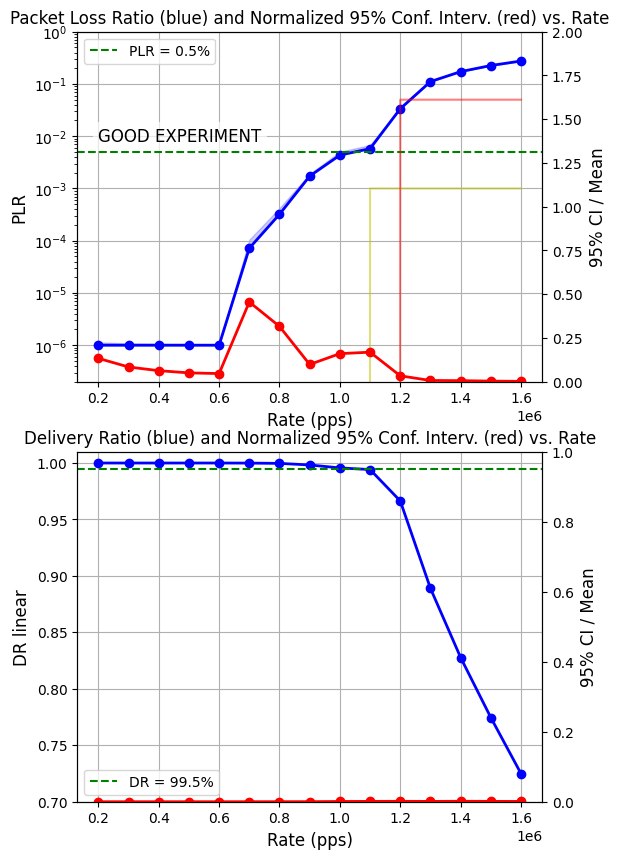}
    \caption{routing, CloudLab, bare metal server, Testbed 3, 30 runs, T = 6s, kernel 5.18, unpinned, SMALL nic buffer, date 0000-00-00, version 00}
    \label{fig:clab_tb3_raw_30_exp_t_6_k5.18_intel_bare-metal_as-is}
\end{figure}

\begin{figure}[ht!]
    \centering
    \includegraphics[width=\linewidth]{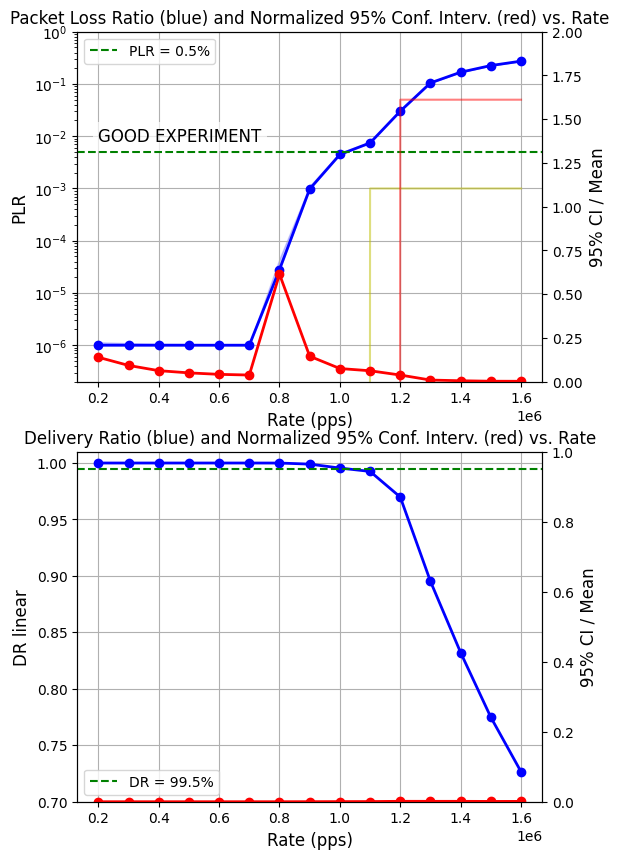}
    \caption{routing, CloudLab, bare metal server, Testbed 3, 30 runs, T = 6s, kernel 5.18, unpinned, LARGE nic buffer, date 0000-00-00, version 00}
    \label{fig:clab_tb3_raw_30_exp_t_6_k5.18_intel_bare-metal_nic_buf_4096}
\end{figure}

\begin{figure}[ht!]
    \centering
    \includegraphics[width=\linewidth]{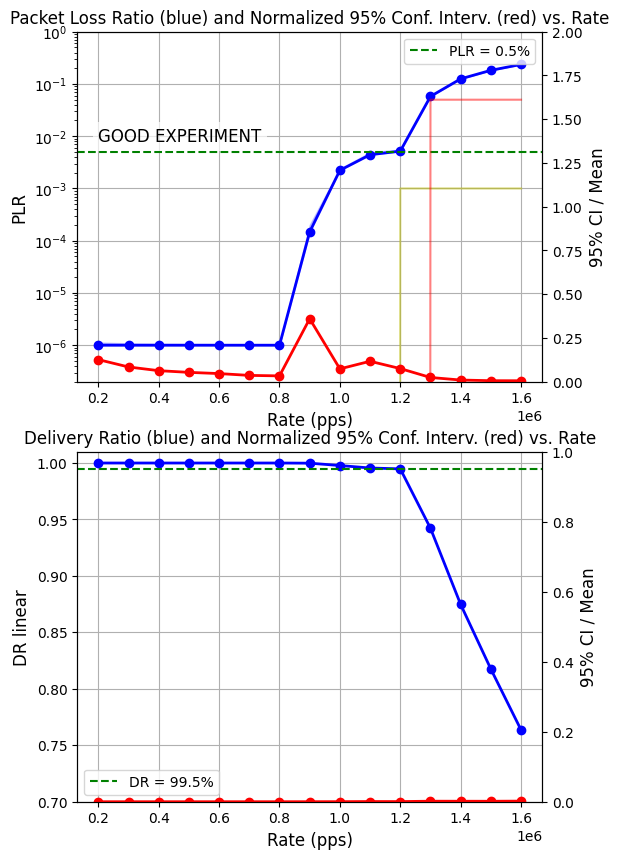}
    \caption{routing, CloudLab, bare metal server, Testbed 3, 30 runs, T = 6s, kernel 5.18, pin-1cpu, LARGE nic buffer, date 0000-00-00, version 00}
    \label{fig:clab_tb3_raw_30_exp_t_6_k5.18_intel_bare-metal_nic_buf_4096_irq_pf0-4_pf1-4}
\end{figure}

\begin{figure}[ht!]
    \centering
    \includegraphics[width=\linewidth]{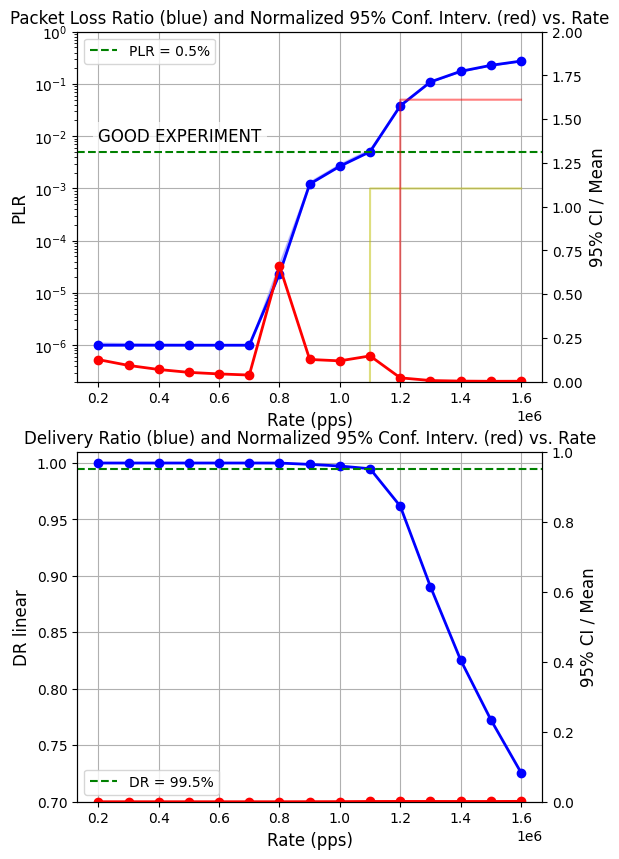}
    \caption{routing, CloudLab, bare metal server, Testbed 3, 30 runs, T = 6s, kernel 5.18, pin-2cpu, LARGE nic buffer, date 0000-00-00, version 00}
    \label{fig:clab_tb3_raw_30_exp_t_6_k5.18_intel_bare-metal_nic_buf_4096_irq_pf0-4_pf1-6}
\end{figure}

\begin{figure}[ht!]
    \centering
    \includegraphics[width=\linewidth]{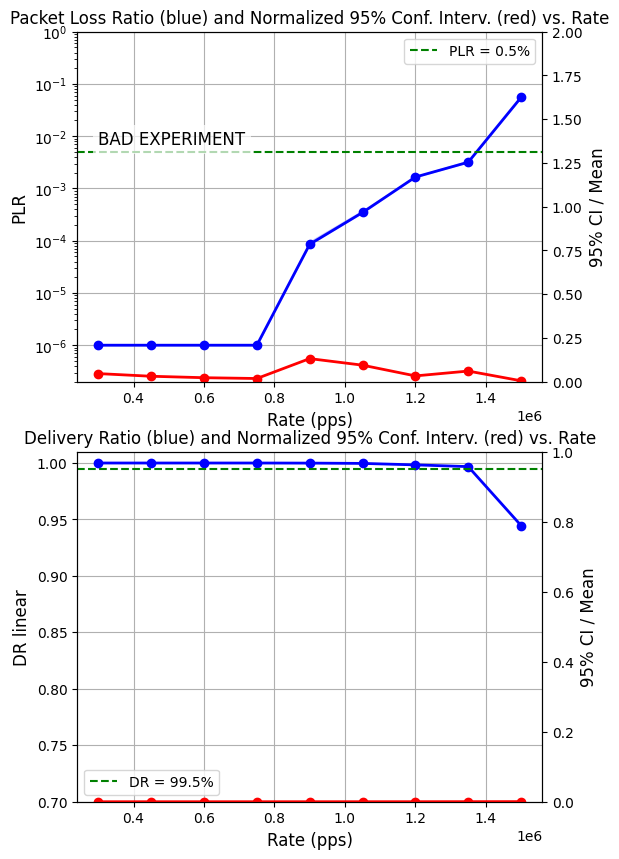}
    \caption{routing, CloudLab, bare metal server, Testbed 3, 50 runs, T = 10s, kernel 5.6, unpinned, SMALL nic buffer, date 0000-00-00, version 00}
    \label{fig:clab_tb3_raw_50_exp_t_10_k5.6_intel_bare-metal_as-is}
\end{figure}

\begin{figure}[ht!]
    \centering
    \includegraphics[width=\linewidth]{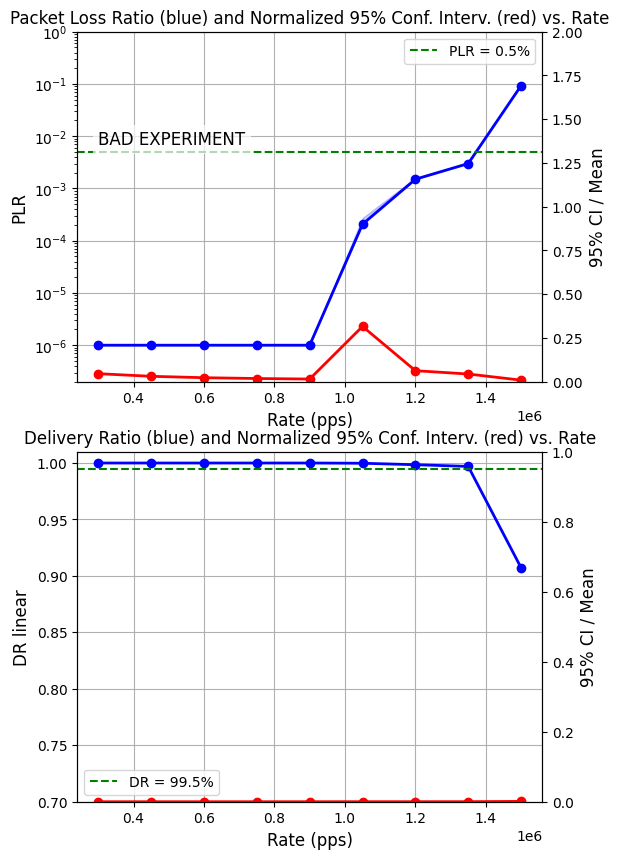}
    \caption{routing, CloudLab, bare metal server, Testbed 3, 50 runs, T = 10s, kernel 5.6, unpinned, LARGE nic buffer, date 0000-00-00, version 00}
    \label{fig:clab_tb3_raw_50_exp_t_10_k5.6_intel_bare-metal_nic_buf_4096}
\end{figure}

\begin{figure}[ht!]
    \centering
    \includegraphics[width=\linewidth]{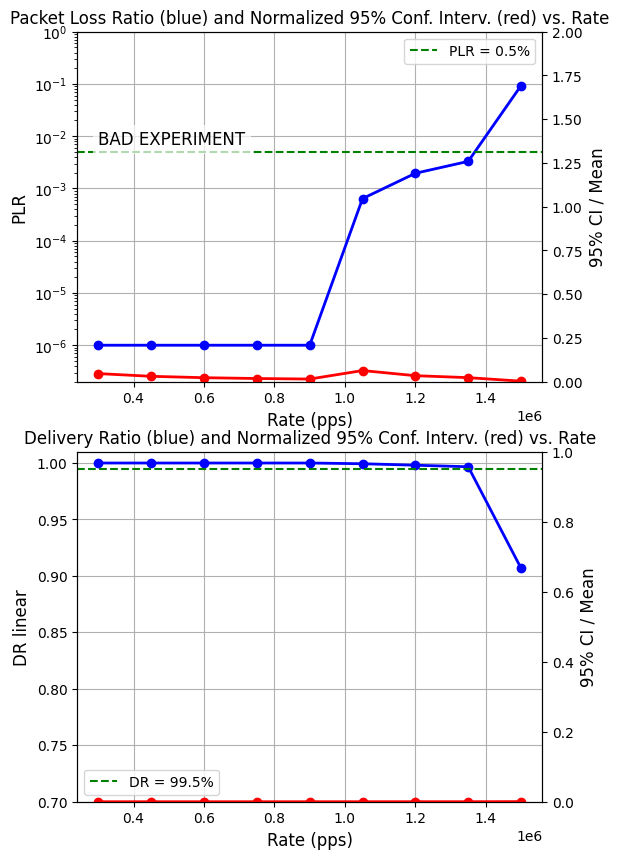}
    \caption{routing, CloudLab, bare metal server, Testbed 3, 50 runs, T = 10s, kernel 5.6, pin-1cpu, LARGE nic buffer, date 0000-00-00, version 00}
    \label{fig:clab_tb3_raw_50_exp_t_10_k5.6_intel_bare-metal_nic_buf_4096_irq_pf0-4_pf1-4}
\end{figure}

\begin{figure}[ht!]
    \centering
    \includegraphics[width=\linewidth]{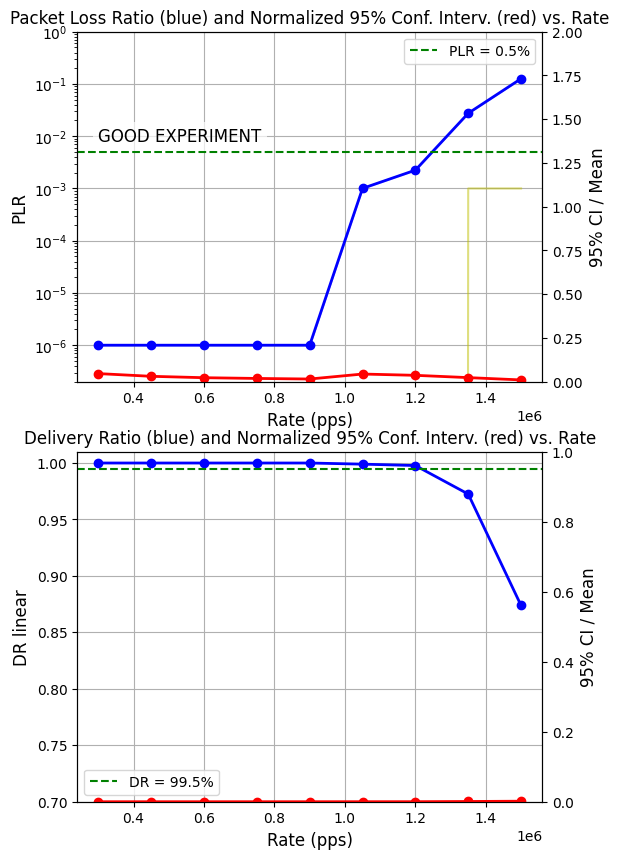}
    \caption{routing, CloudLab, bare metal server, Testbed 3, 50 runs, T = 10s, kernel 5.6, pin-2cpu, LARGE nic buffer, date 0000-00-00, version 00}
    \label{fig:clab_tb3_raw_50_exp_t_10_k5.6_intel_bare-metal_nic_buf_4096_irq_pf0-4_pf1-6}
\end{figure}

\begin{figure}[ht!]
    \centering
    \includegraphics[width=\linewidth]{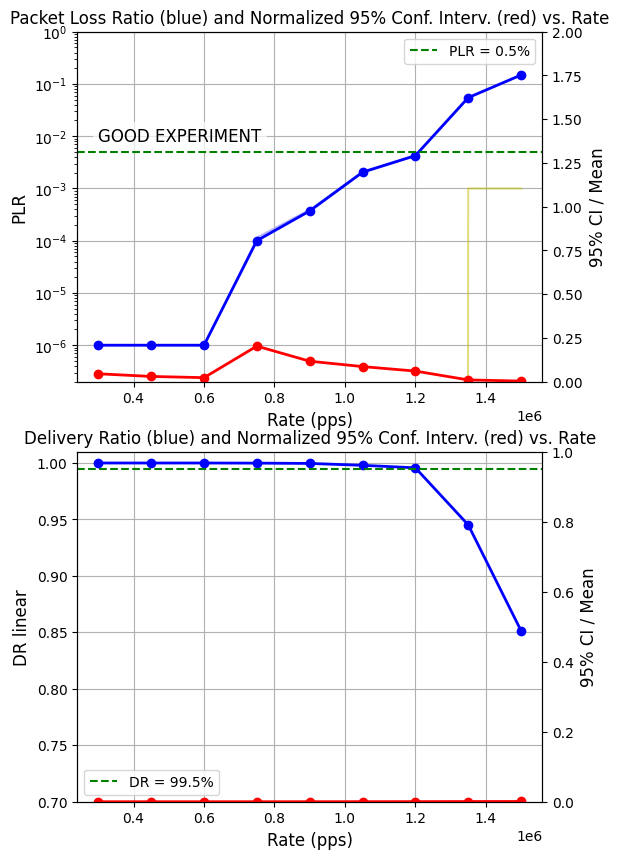}
    \caption{routing, CloudLab, bare metal server, Testbed 3, 50 runs, T = 10s, kernel 5.12, unpinned, SMALL nic buffer, date 0000-00-00, version 00}
    \label{fig:clab_tb3_raw_50_exp_t_10_k5.12_intel_bare-metal_as-is}
\end{figure}

\begin{figure}[ht!]
    \centering
    \includegraphics[width=\linewidth]{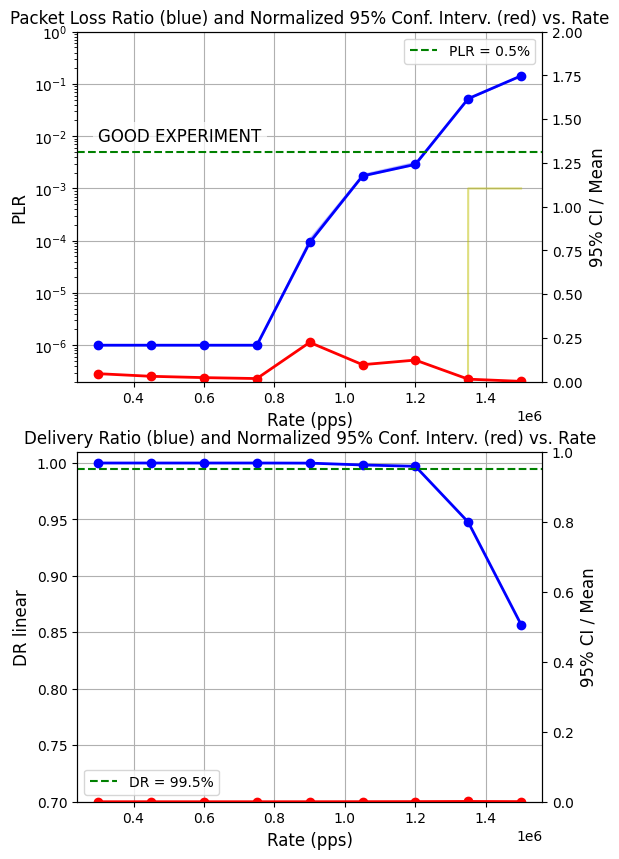}
    \caption{routing, CloudLab, bare metal server, Testbed 3, 50 runs, T = 10s, kernel 5.12, unpinned, LARGE nic buffer, date 0000-00-00, version 00}
    \label{fig:clab_tb3_raw_50_exp_t_10_k5.12_intel_bare-metal_nic_buf_4096}
\end{figure}

\begin{figure}[ht!]
    \centering
    \includegraphics[width=\linewidth]{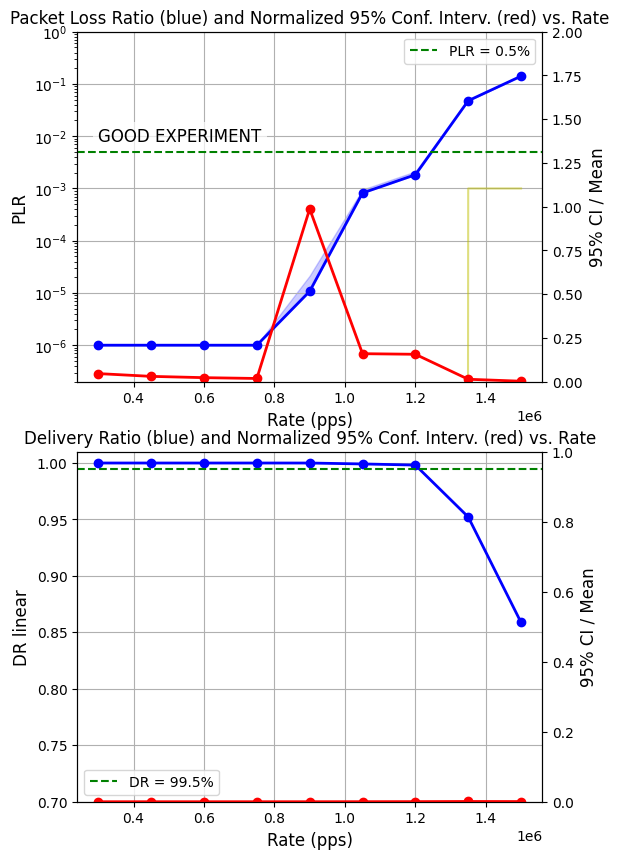}
    \caption{routing, CloudLab, bare metal server, Testbed 3, 50 runs, T = 10s, kernel 5.12, pin-1cpu, LARGE nic buffer, date 0000-00-00, version 00}
    \label{fig:clab_tb3_raw_50_exp_t_10_k5.12_intel_bare-metal_nic_buf_4096_irq_pf0-4_pf1-4}
\end{figure}

\begin{figure}[ht!]
    \centering
    \includegraphics[width=\linewidth]{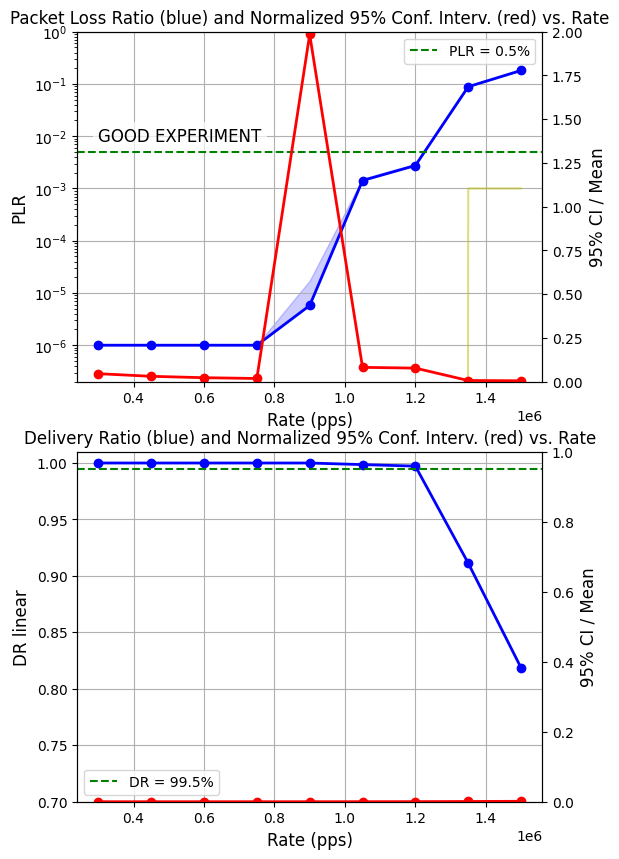}
    \caption{routing, CloudLab, bare metal server, Testbed 3, 50 runs, T = 10s, kernel 5.12, pin-2cpu, LARGE nic buffer, date 0000-00-00, version 00}
    \label{fig:clab_tb3_raw_50_exp_t_10_k5.12_intel_bare-metal_nic_buf_4096_irq_pf0-4_pf1-6}
\end{figure}

\begin{figure}[ht!]
    \centering
    \includegraphics[width=\linewidth]{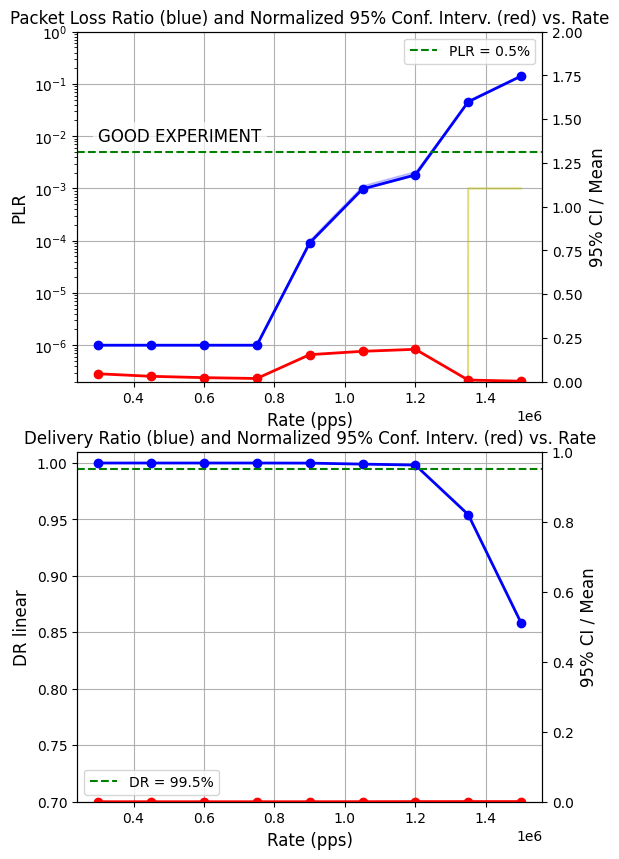}
    \caption{routing, CloudLab, bare metal server, Testbed 3, 50 runs, T = 10s, kernel 5.13, unpinned, SMALL nic buffer, date 0000-00-00, version 00}
    \label{fig:clab_tb3_raw_50_exp_t_10_k5.13_intel_bare-metal_as-is}
\end{figure}

\begin{figure}[ht!]
    \centering
    \includegraphics[width=\linewidth]{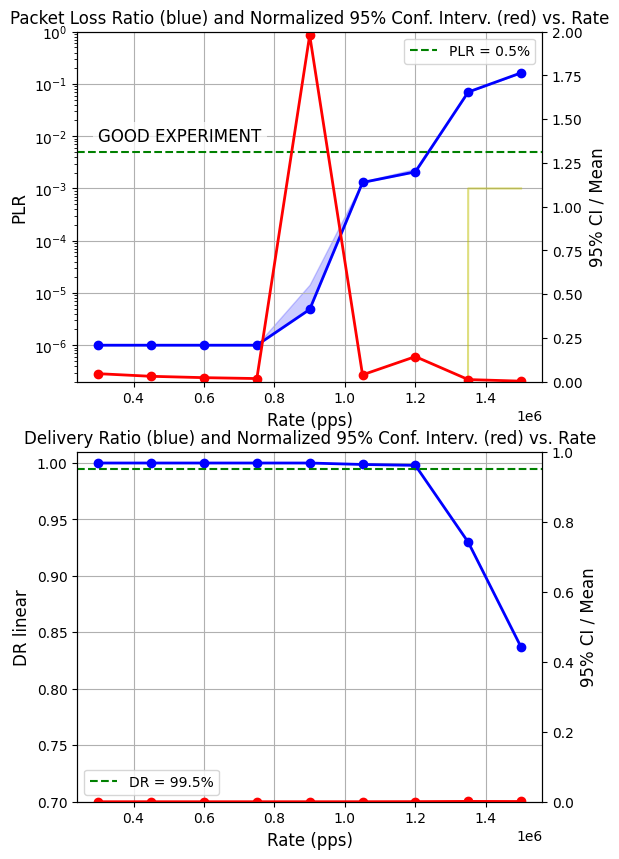}
    \caption{routing, CloudLab, bare metal server, Testbed 3, 50 runs, T = 10s, kernel 5.13, unpinned, LARGE nic buffer, date 0000-00-00, version 00}
    \label{fig:clab_tb3_raw_50_exp_t_10_k5.13_intel_bare-metal_nic_buf_4096}
\end{figure}

\begin{figure}[ht!]
    \centering
    \includegraphics[width=\linewidth]{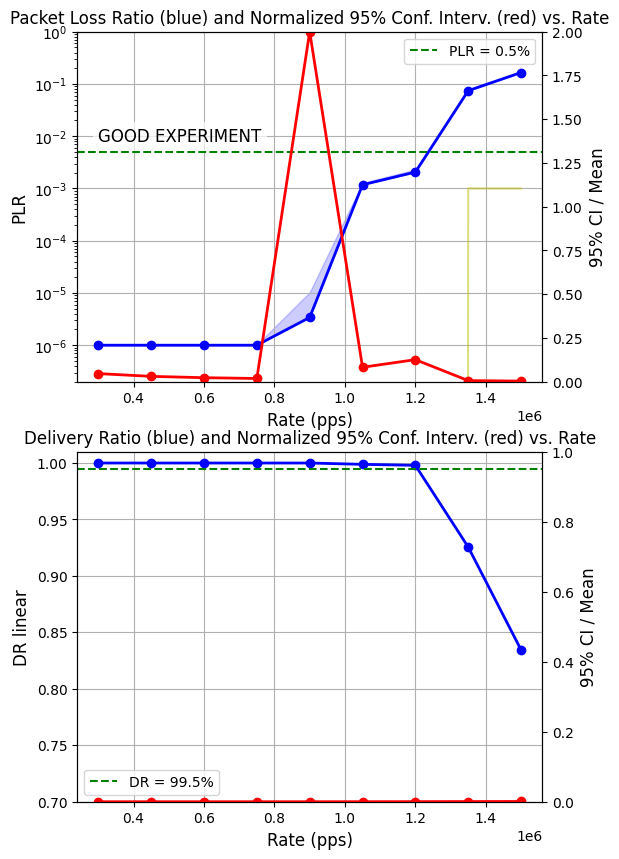}
    \caption{routing, CloudLab, bare metal server, Testbed 3, 50 runs, T = 10s, kernel 5.13, pin-1cpu, LARGE nic buffer, date 0000-00-00, version 00}
    \label{fig:clab_tb3_raw_50_exp_t_10_k5.13_intel_bare-metal_nic_buf_4096_irq_pf0-4_pf1-4}
\end{figure}

\begin{figure}[ht!]
    \centering
    \includegraphics[width=\linewidth]{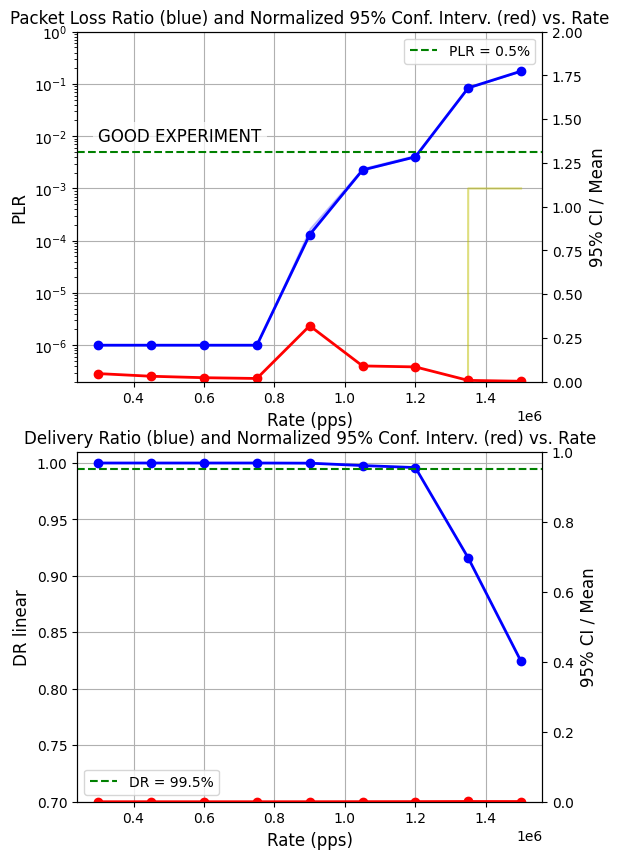}
    \caption{routing, CloudLab, bare metal server, Testbed 3, 50 runs, T = 10s, kernel 5.13, pin-2cpu, LARGE nic buffer, date 0000-00-00, version 00}
    \label{fig:clab_tb3_raw_50_exp_t_10_k5.13_intel_bare-metal_nic_buf_4096_irq_pf0-4_pf1-6}
\end{figure}

\begin{figure}[ht!]
    \centering
    \includegraphics[width=\linewidth]{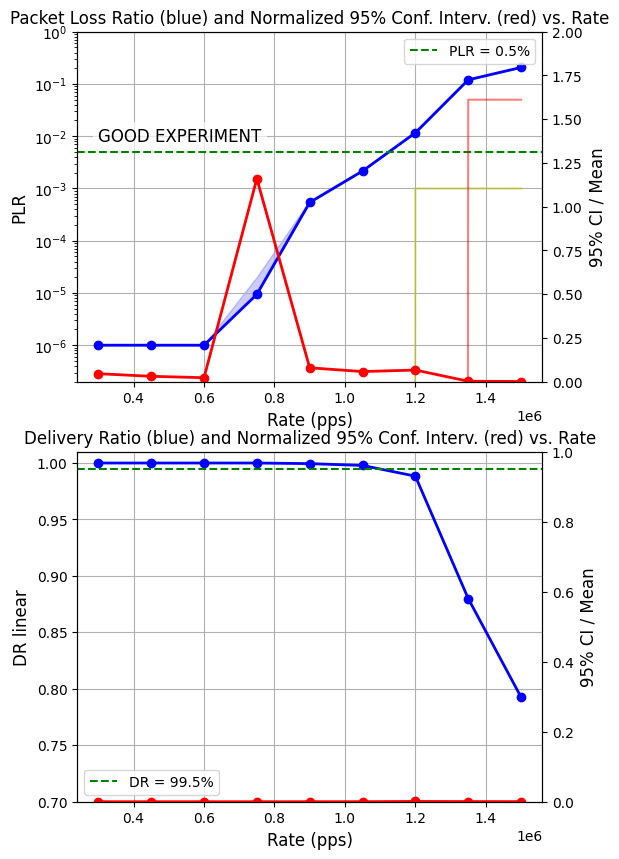}
    \caption{routing, CloudLab, bare metal server, Testbed 3, 50 runs, T = 10s, kernel 5.18, unpinned, SMALL nic buffer, date 0000-00-00, version 00}
    \label{fig:clab_tb3_raw_50_exp_t_10_k5.18_intel_bare-metal_as-is}
\end{figure}

\begin{figure}[ht!]
    \centering
    \includegraphics[width=\linewidth]{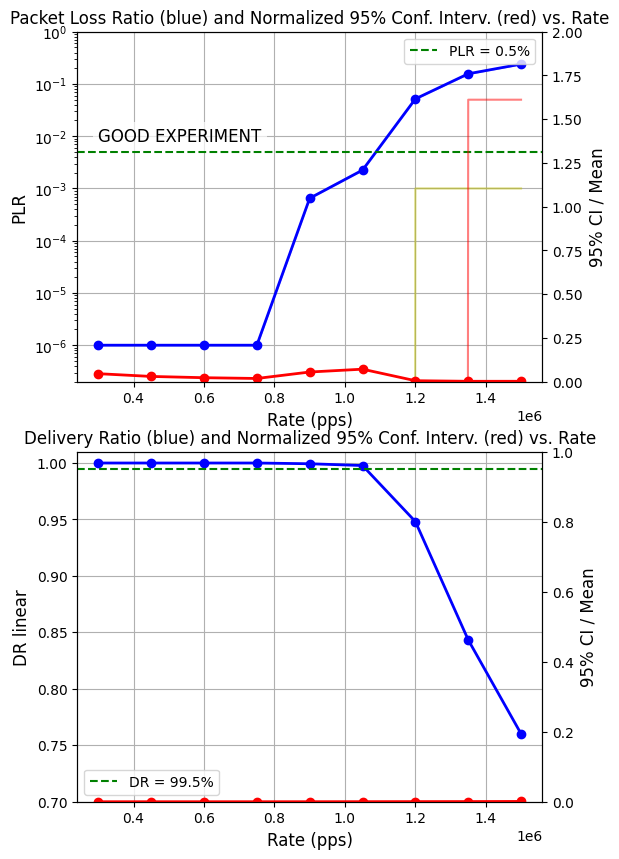}
    \caption{routing, CloudLab, bare metal server, Testbed 3, 50 runs, T = 10s, kernel 5.18, unpinned, LARGE nic buffer, date 0000-00-00, version 00}
    \label{fig:clab_tb3_raw_50_exp_t_10_k5.18_intel_bare-metal_nic_buf_4096}
\end{figure}

\begin{figure}[ht!]
    \centering
    \includegraphics[width=\linewidth]{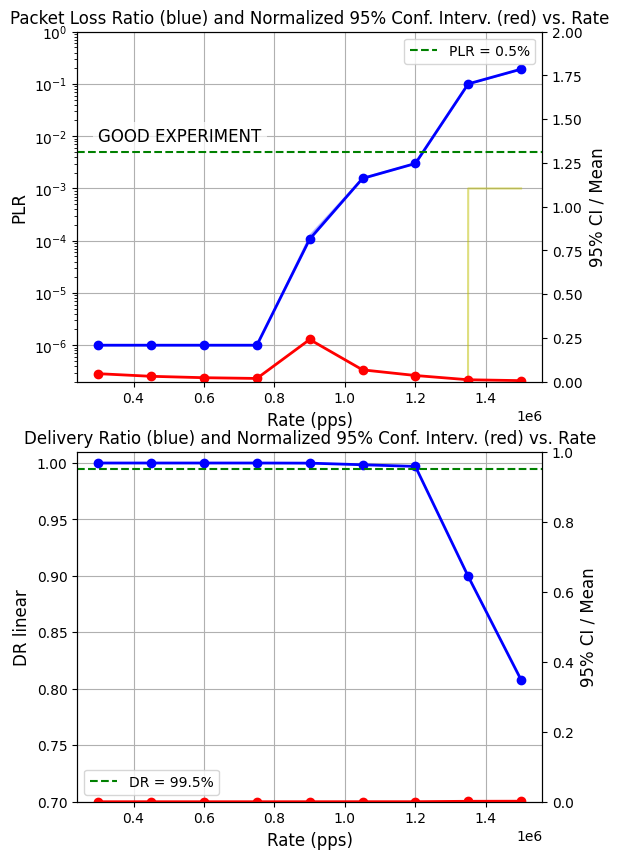}
    \caption{routing, CloudLab, bare metal server, Testbed 3, 50 runs, T = 10s, kernel 5.18, pin-1cpu, LARGE nic buffer, date 0000-00-00, version 00}
    \label{fig:clab_tb3_raw_50_exp_t_10_k5.18_intel_bare-metal_nic_buf_4096_irq_pf0-4_pf1-4}
\end{figure}

\begin{figure}[ht!]
    \centering
    \includegraphics[width=\linewidth]{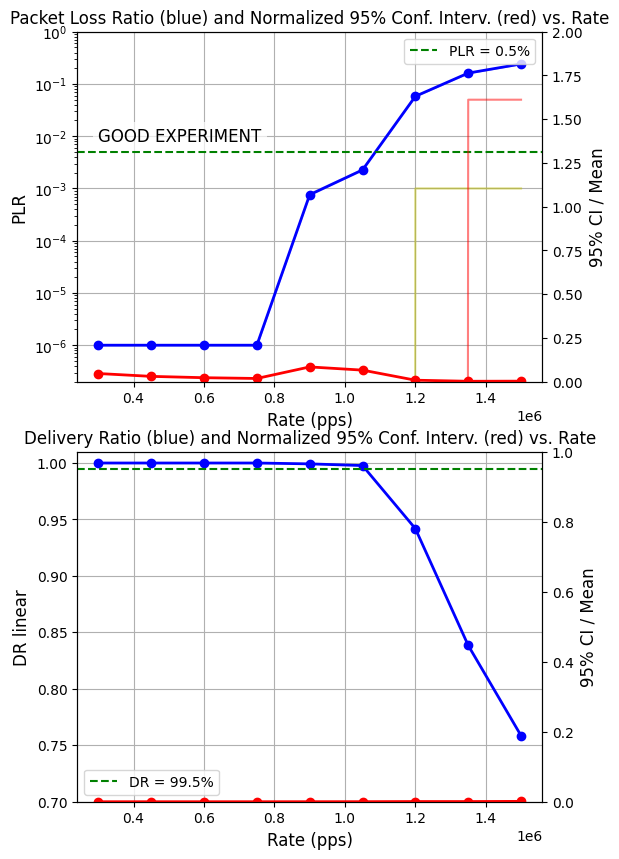}
    \caption{routing, CloudLab, bare metal server, Testbed 3, 50 runs, T = 10s, kernel 5.18, pin-2cpu, LARGE nic buffer, date 0000-00-00, version 00}
    \label{fig:clab_tb3_raw_50_exp_t_10_k5.18_intel_bare-metal_nic_buf_4096_irq_pf0-4_pf1-6}
\end{figure}

\begin{figure}[ht!]
    \centering
    \includegraphics[width=\linewidth]{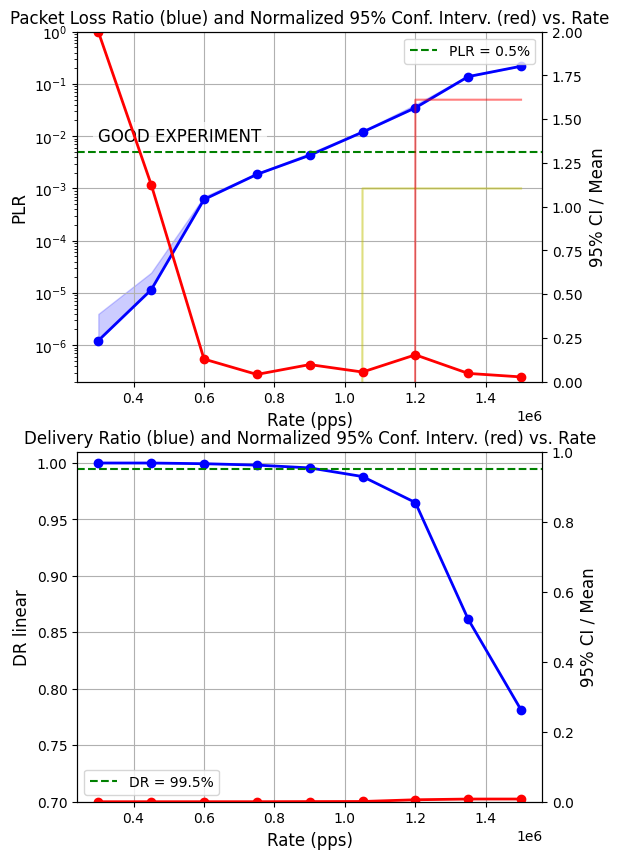}
    \caption{routing, CloudLab, virtual machine, Testbed 0, 50 runs, T = 10s, kernel 5.15, unpinned, SMALL nic buffer, date 0000-00-00, version 00}
    \label{fig:clab_raw_50_k5.15_intel_vm_as-is}
\end{figure}

\begin{figure}[ht!]
    \centering
    \includegraphics[width=\linewidth]{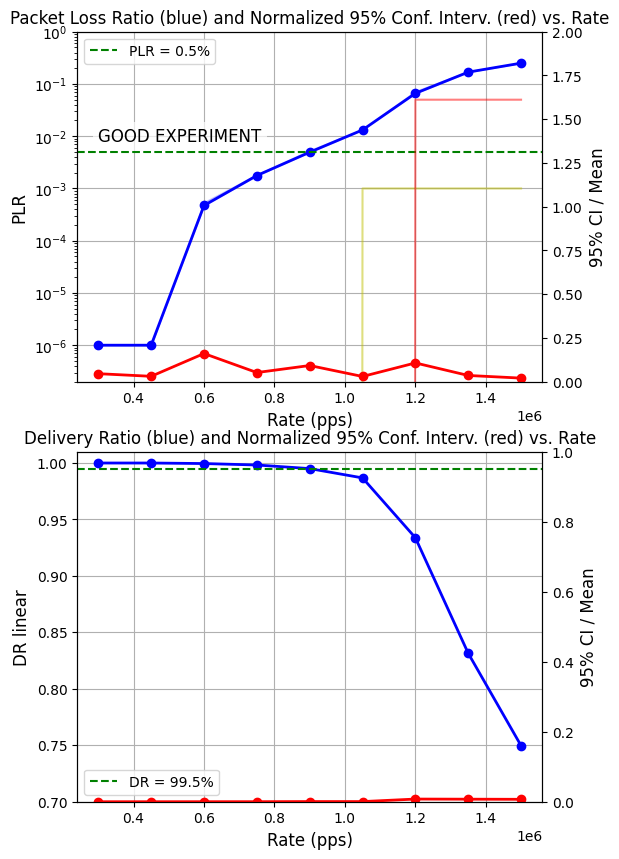}
    \caption{routing, CloudLab, virtual machine, Testbed 0, 50 runs, T = 10s, kernel 5.15, unpinned, LARGE nic buffer, date 0000-00-00, version 00}
    \label{fig:clab_raw_50_k5.15_intel_vm_nic_buf_4096_2nd_exec}
\end{figure}

\begin{figure}[ht!]
    \centering
    \includegraphics[width=\linewidth]{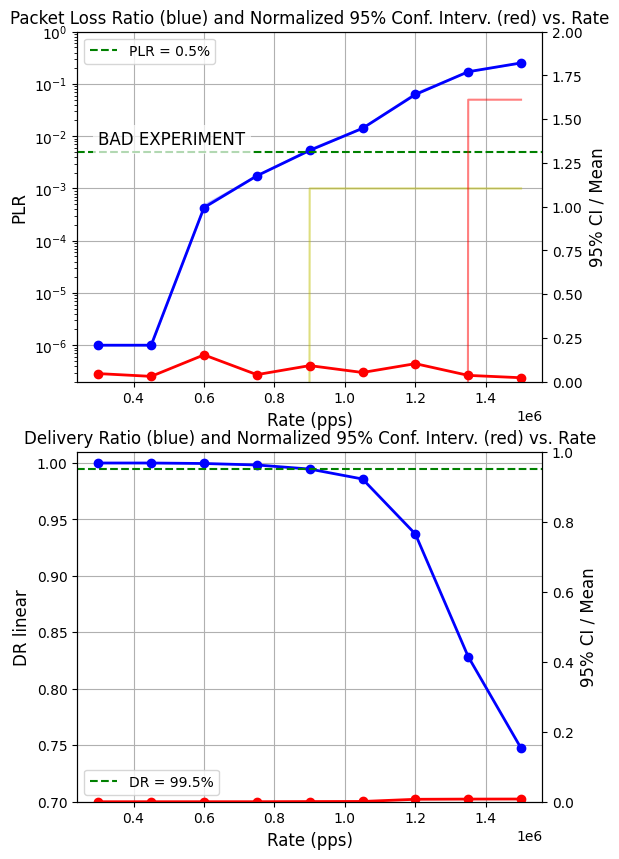}
    \caption{routing, CloudLab, virtual machine, Testbed 0, 50 runs, T = 10s, kernel 5.15, unpinned, LARGE nic buffer, date 0000-00-00, version 00}
    \label{fig:clab_raw_50_k5.15_intel_vm_nic_buf_4096}
\end{figure}

\begin{figure}[ht!]
    \centering
    \includegraphics[width=\linewidth]{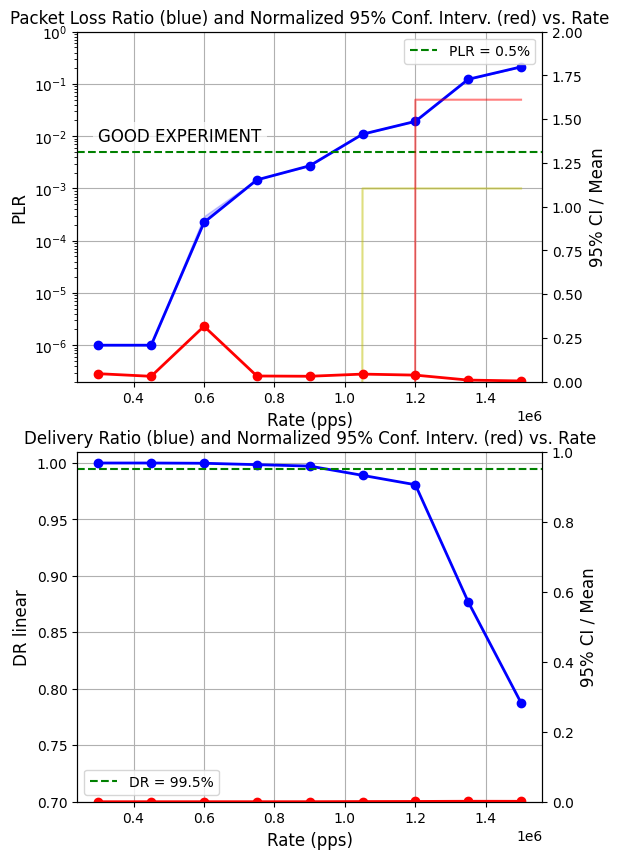}
    \caption{routing, CloudLab, virtual machine, Testbed 0, 50 runs, T = 10s, kernel 5.15, pin-1cpu, LARGE nic buffer, date 0000-00-00, version 00}
    \label{fig:clab_raw_50_k5.15_intel_vm_nic_buf_4096_irq_pf0-1_pf1-1}
\end{figure}

\begin{figure}[ht!]
    \centering
    \includegraphics[width=\linewidth]{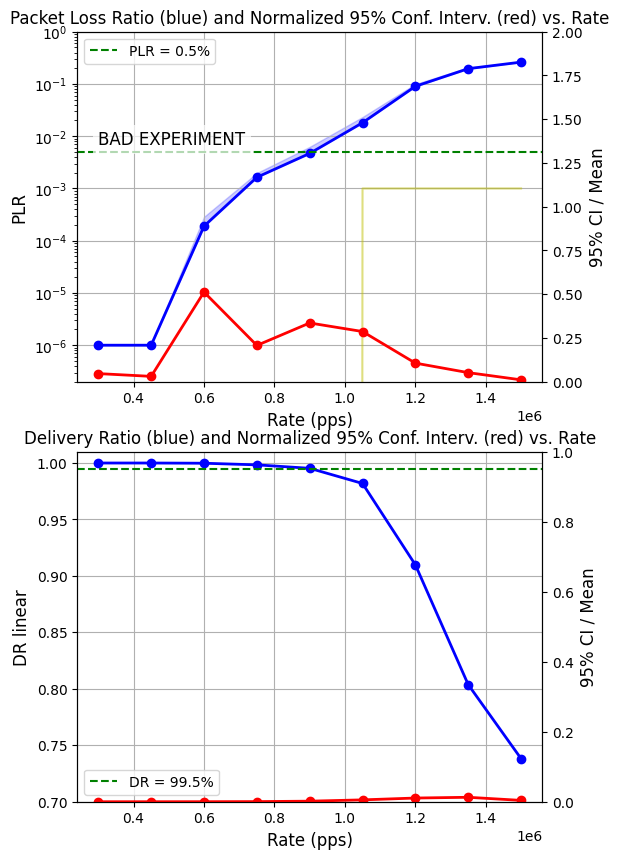}
    \caption{routing, CloudLab, virtual machine, Testbed 0, 50 runs, T = 10s, kernel 5.15, pin-2cpu, LARGE nic buffer, date 0000-00-00, version 00}
    \label{fig:clab_raw_50_k5.15_intel_vm_nic_buf_4096_irq_pf0-1_pf1-2}
\end{figure}

\begin{figure}[ht!]
    \centering
    \includegraphics[width=\linewidth]{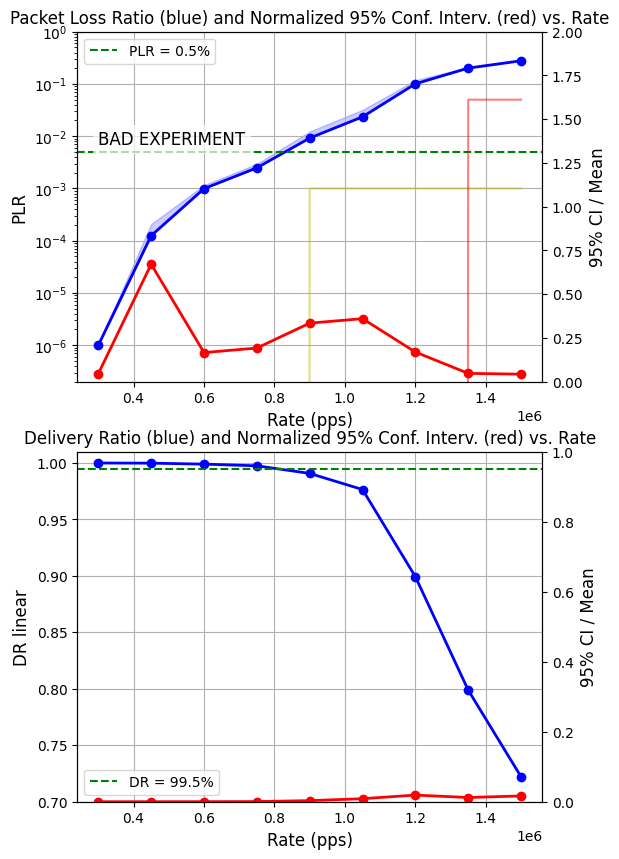}
    \caption{routing, CloudLab, virtual machine, Testbed 0, 50 runs, T = 10s, kernel 6.5, unpinned, SMALL nic buffer, date 0000-00-00, version 00}
    \label{fig:clab_raw_50_k6.5_intel_vm_as-is}
\end{figure}

\begin{figure}[ht!]
    \centering
    \includegraphics[width=\linewidth]{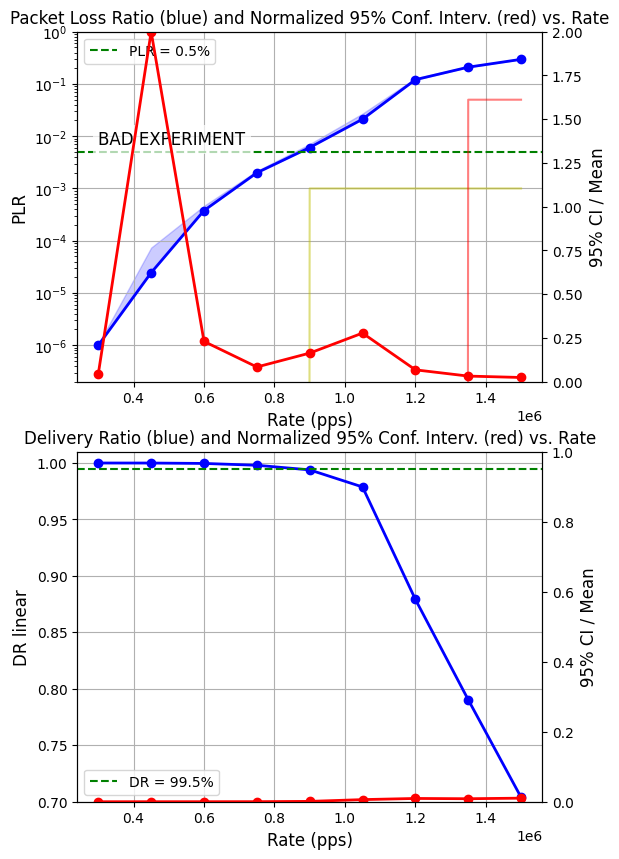}
    \caption{routing, CloudLab, virtual machine, Testbed 0, 50 runs, T = 10s, kernel 6.5, unpinned, LARGE nic buffer, date 0000-00-00, version 00}
    \label{fig:clab_raw_50_k6.5_intel_vm_nic_buf_4096}
\end{figure}

\begin{figure}[ht!]
    \centering
    \includegraphics[width=\linewidth]{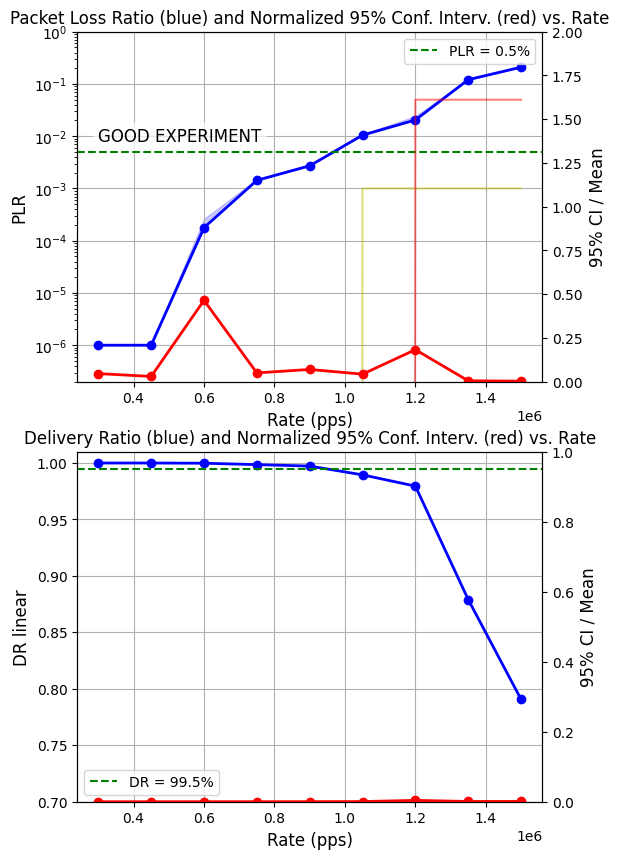}
    \caption{routing, CloudLab, virtual machine, Testbed 0, 50 runs, T = 10s, kernel 6.5, pin-1cpu, LARGE nic buffer, date 0000-00-00, version 00}
    \label{fig:clab_raw_50_k6.5_intel_vm_nic_buf_4096_irq_pf0-1_pf1-1}
\end{figure}

\begin{figure}[ht!]
    \centering
    \includegraphics[width=\linewidth]{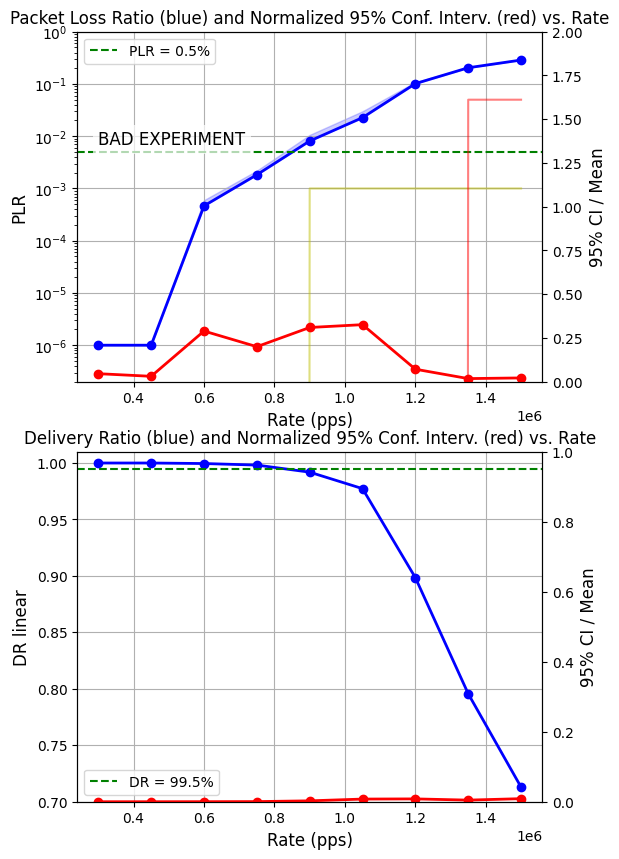}
    \caption{routing, CloudLab, virtual machine, Testbed 0, 50 runs, T = 10s, kernel 6.5, pin-2cpu, LARGE nic buffer, date 0000-00-00, version 00}
    \label{fig:clab_raw_50_k6.5_intel_vm_nic_buf_4096_irq_pf0-1_pf1-2}
\end{figure}

\begin{figure}[ht!]
    \centering
    \includegraphics[width=\linewidth]{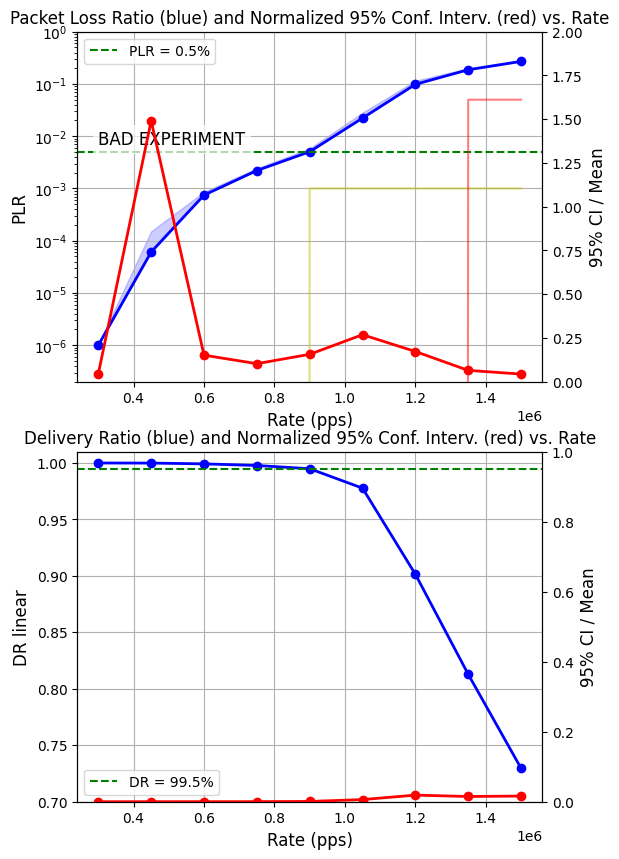}
    \caption{routing, CloudLab, virtual machine, Testbed 0, 50 runs, T = 10s, kernel 6.8, unpinned, SMALL nic buffer, date 0000-00-00, version 00}
    \label{fig:clab_raw_50_k6.8_intel_vm_as-is}
\end{figure}

\begin{figure}[ht!]
    \centering
    \includegraphics[width=\linewidth]{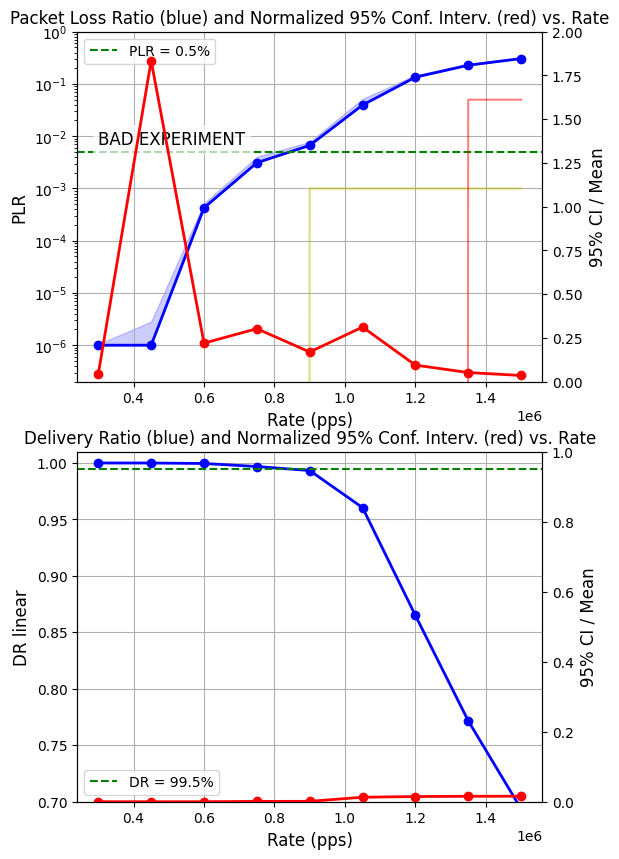}
    \caption{routing, CloudLab, virtual machine, Testbed 0, 50 runs, T = 10s, kernel 6.8, unpinned, LARGE nic buffer, date 0000-00-00, version 00}
    \label{fig:clab_raw_50_k6.8_intel_vm_nic_buf_4096}
\end{figure}

\begin{figure}[ht!]
    \centering
    \includegraphics[width=\linewidth]{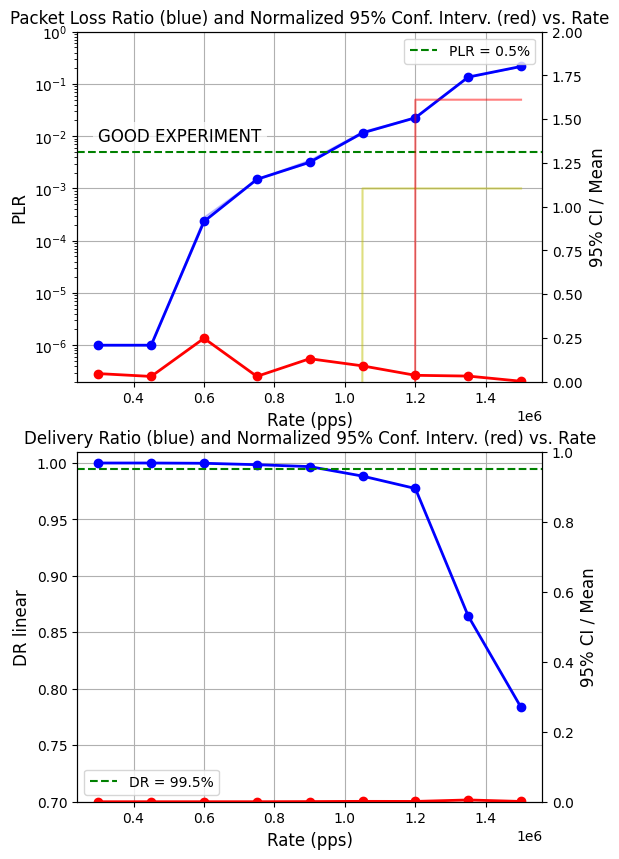}
    \caption{routing, CloudLab, virtual machine, Testbed 0, 50 runs, T = 10s, kernel 6.8, pin-1cpu, LARGE nic buffer, date 0000-00-00, version 00}
    \label{fig:clab_raw_50_k6.8_intel_vm_nic_buf_4096_irq_pf0-1_pf1-1}
\end{figure}

\begin{figure}[ht!]
    \centering
    \includegraphics[width=\linewidth]{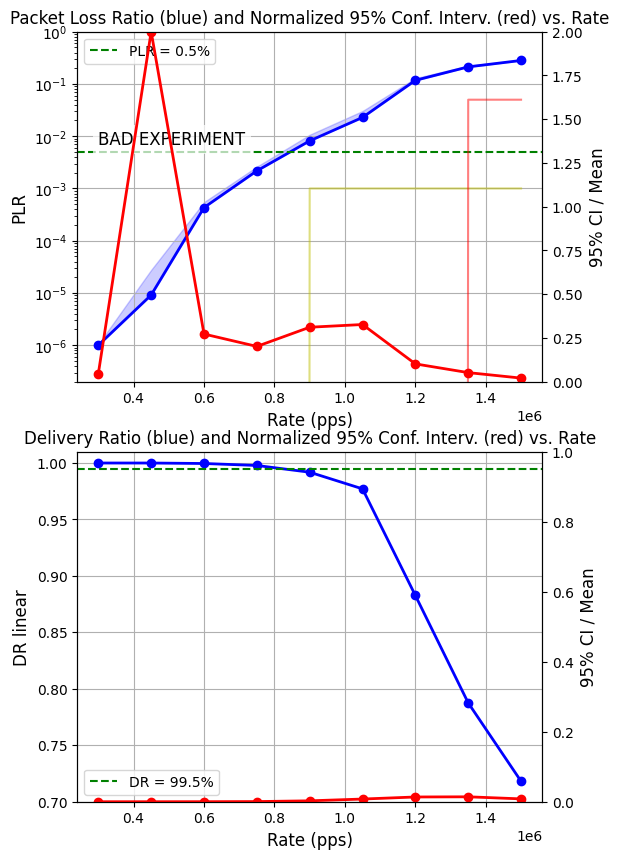}
    \caption{routing, CloudLab, virtual machine, Testbed 0, 50 runs, T = 10s, kernel 6.8, pin-2cpu, LARGE nic buffer, date 0000-00-00, version 00}
    \label{fig:clab_raw_50_k6.8_intel_vm_nic_buf_4096_irq_pf0-1_pf1-2}
\end{figure}

\begin{figure}[ht!]
    \centering
    \includegraphics[width=\linewidth]{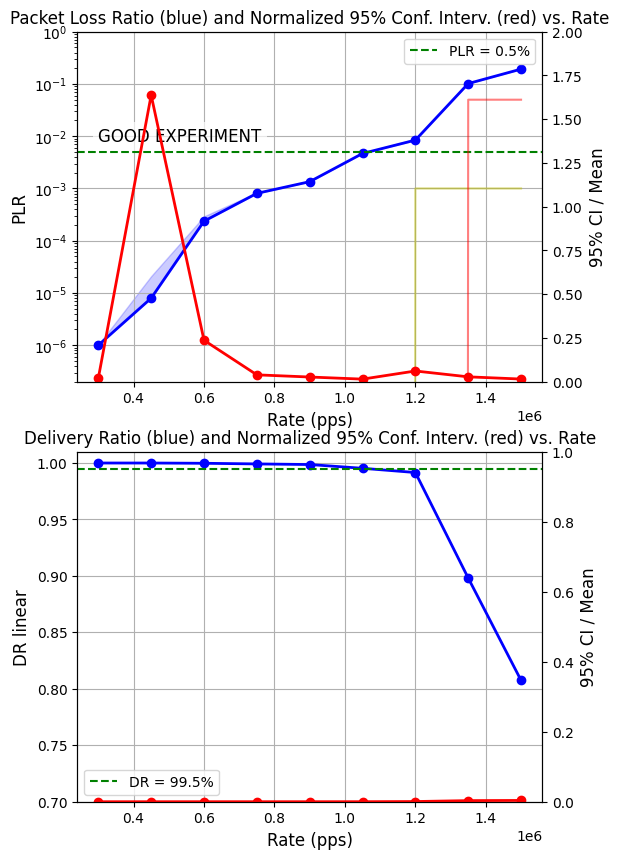}
    \caption{routing, CloudLab, virtual machine, Testbed 0, 50 runs, T = 20s, kernel 5.15, unpinned, SMALL nic buffer, date 0000-00-00, version 00}
    \label{fig:clab_raw_50_exp_t_20_k5.15_intel_vm_as-is}
\end{figure}

\begin{figure}[ht!]
    \centering
    \includegraphics[width=\linewidth]{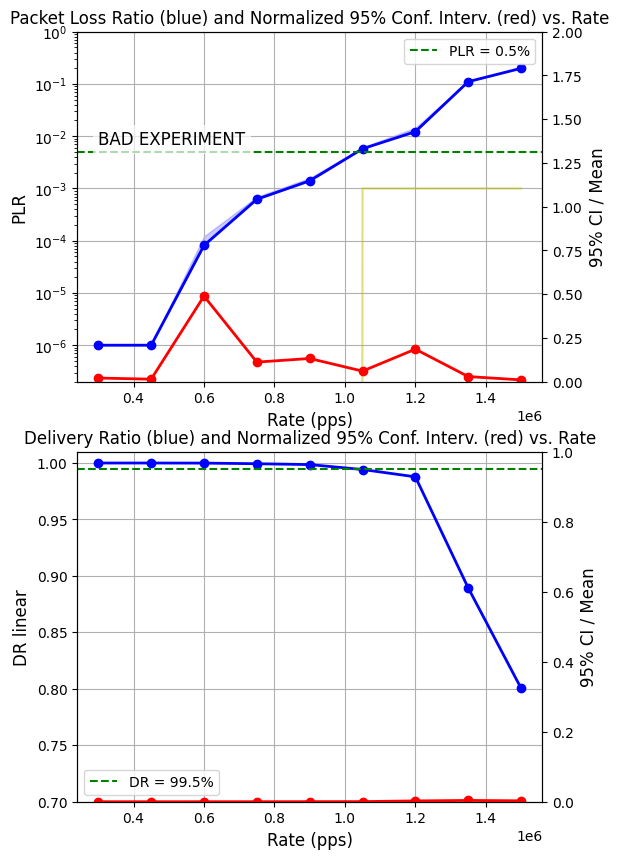}
    \caption{routing, CloudLab, virtual machine, Testbed 0, 50 runs, T = 20s, kernel 5.15, unpinned, LARGE nic buffer, date 0000-00-00, version 00}
    \label{fig:clab_raw_50_exp_t_20_k5.15_intel_vm_nic_buf_4096}
\end{figure}

\begin{figure}[ht!]
    \centering
    \includegraphics[width=\linewidth]{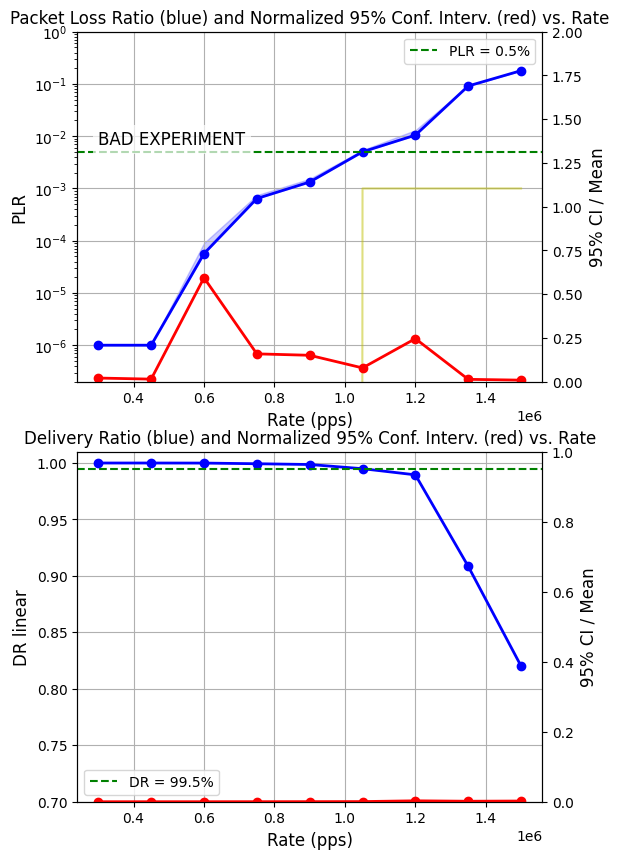}
    \caption{routing, CloudLab, virtual machine, Testbed 0, 50 runs, T = 20s, kernel 5.15, pin-1cpu, LARGE nic buffer, date 0000-00-00, version 00}
    \label{fig:clab_raw_50_exp_t_20_k5.15_intel_vm_nic_buf_4096_irq_pf0-1_pf1-1}
\end{figure}

\begin{figure}[ht!]
    \centering
    \includegraphics[width=\linewidth]{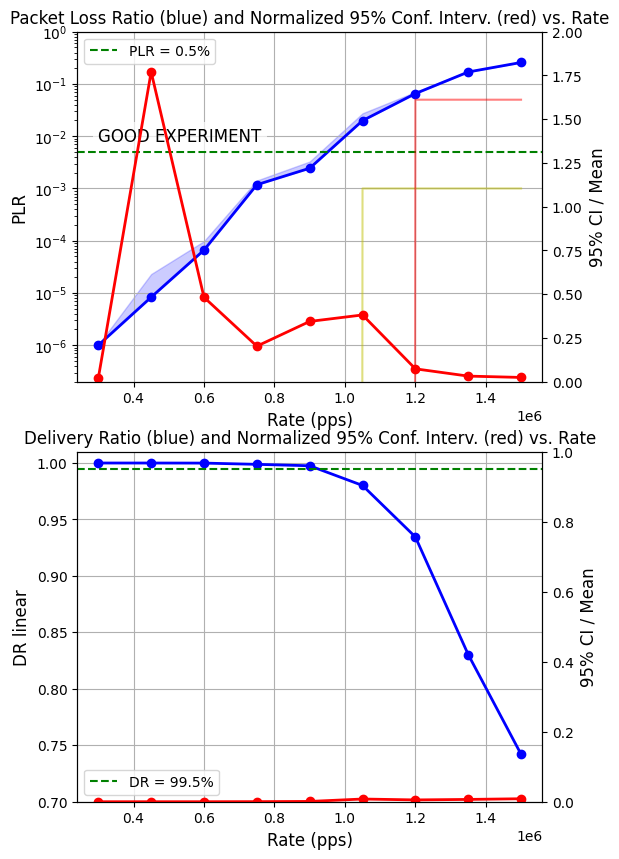}
    \caption{routing, CloudLab, virtual machine, Testbed 0, 50 runs, T = 20s, kernel 5.15, pin-2cpu, LARGE nic buffer, date 0000-00-00, version 00}
    \label{fig:clab_raw_50_exp_t_20_k5.15_intel_vm_nic_buf_4096_irq_pf0-1_pf1-2}
\end{figure}

\begin{figure}[ht!]
    \centering
    \includegraphics[width=\linewidth]{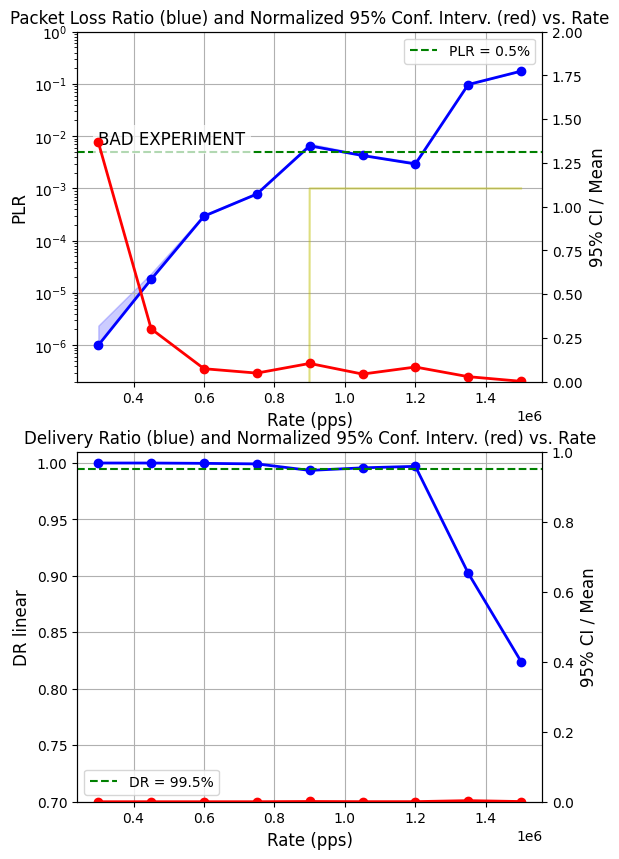}
    \caption{routing, mazinga loc srv, container, Testbed 0, 50 runs, T = 10s, kernel 6.8, unpinned, SMALL nic buffer, date 0000-00-00, version 00}
    \label{fig:mazinga_raw_50_exp_t_10_k6.8_intel_ns_as-is}
\end{figure}

\begin{figure}[ht!]
    \centering
    \includegraphics[width=\linewidth]{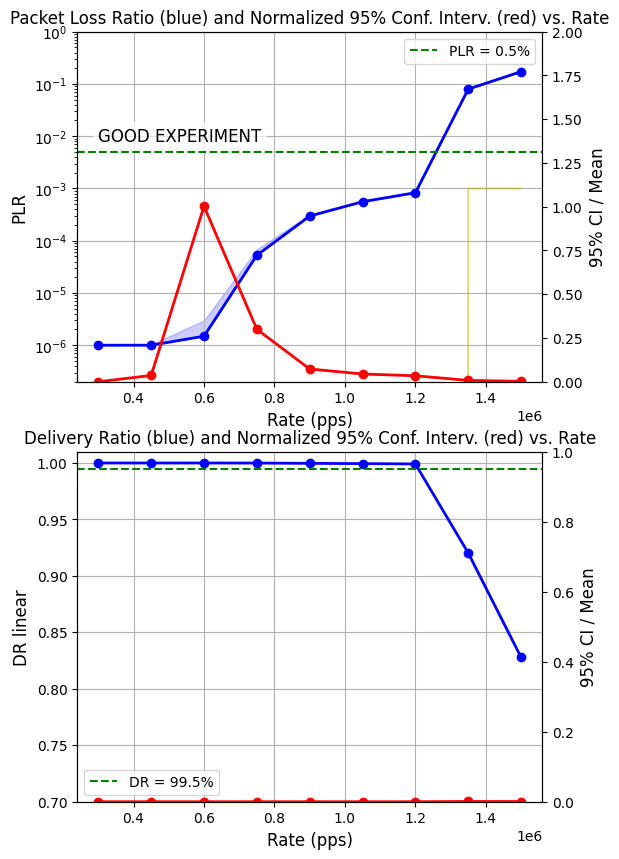}
    \caption{routing, mazinga loc srv, container, Testbed 0, 50 runs, T = 10s, kernel 6.8, unpinned, LARGE nic buffer, date 0000-00-00, version 00}
    \label{fig:mazinga_raw_50_exp_t_10_k6.8_intel_ns_nic_buf_4096}
\end{figure}

\begin{figure}[ht!]
    \centering
    \includegraphics[width=\linewidth]{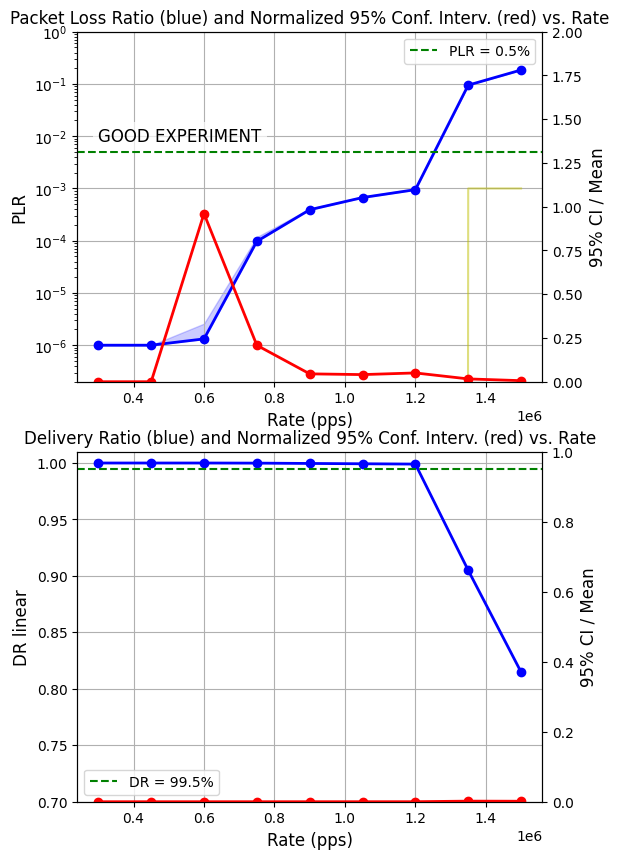}
    \caption{routing, mazinga loc srv, container, Testbed 0, 50 runs, T = 10s, kernel 6.8, pin-1cpu, LARGE nic buffer, date 0000-00-00, version 00}
    \label{fig:mazinga_raw_50_exp_t_10_k6.8_intel_ns_nic_buf_4096_irq_vf0-33_vf1-33}
\end{figure}

\begin{figure}[ht!]
    \centering
    \includegraphics[width=\linewidth]{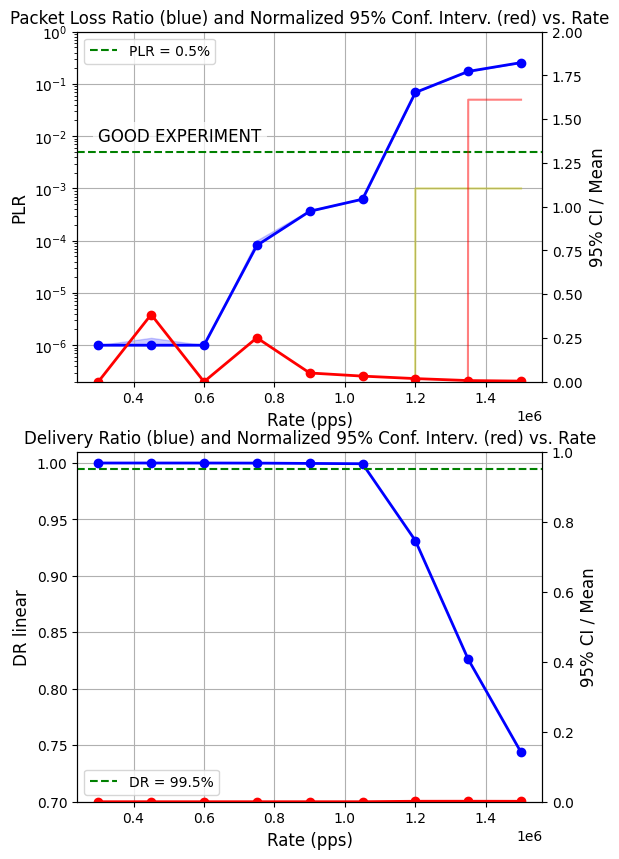}
    \caption{routing, mazinga loc srv, container, Testbed 0, 50 runs, T = 10s, kernel 6.8, pin-2cpu, LARGE nic buffer, date 0000-00-00, version 00}
    \label{fig:mazinga_raw_50_exp_t_10_k6.8_intel_ns_nic_buf_4096_irq_vf0-33_vf1-35}
\end{figure}

\begin{figure}[ht!]
    \centering
    \includegraphics[width=\linewidth]{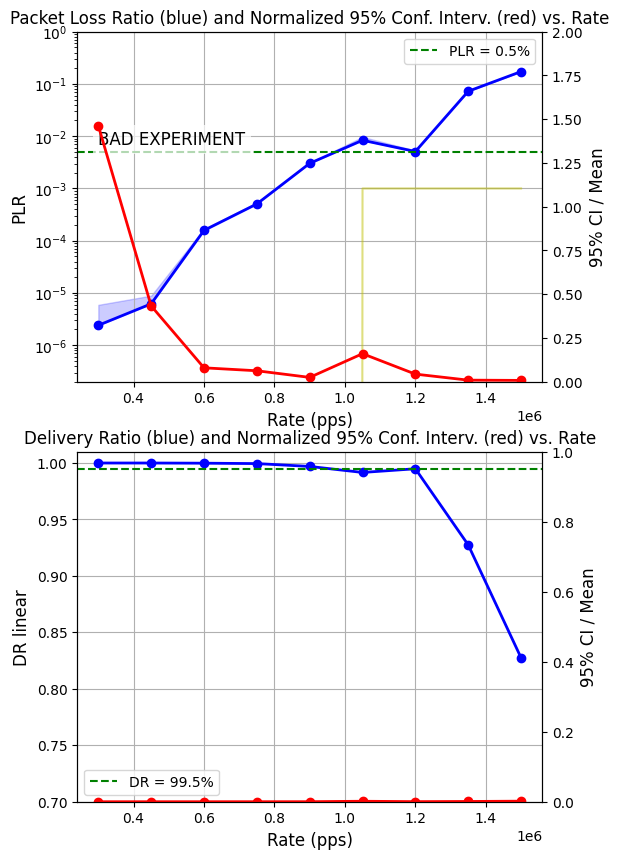}
    \caption{routing, mazinga loc srv, container, Testbed 0, 50 runs, T = 20s, kernel 6.8, unpinned, SMALL nic buffer, date 0000-00-00, version 00}
    \label{fig:mazinga_raw_50_exp_t_20_k6.8_intel_ns_as-is}
\end{figure}

\begin{figure}[ht!]
    \centering
    \includegraphics[width=\linewidth]{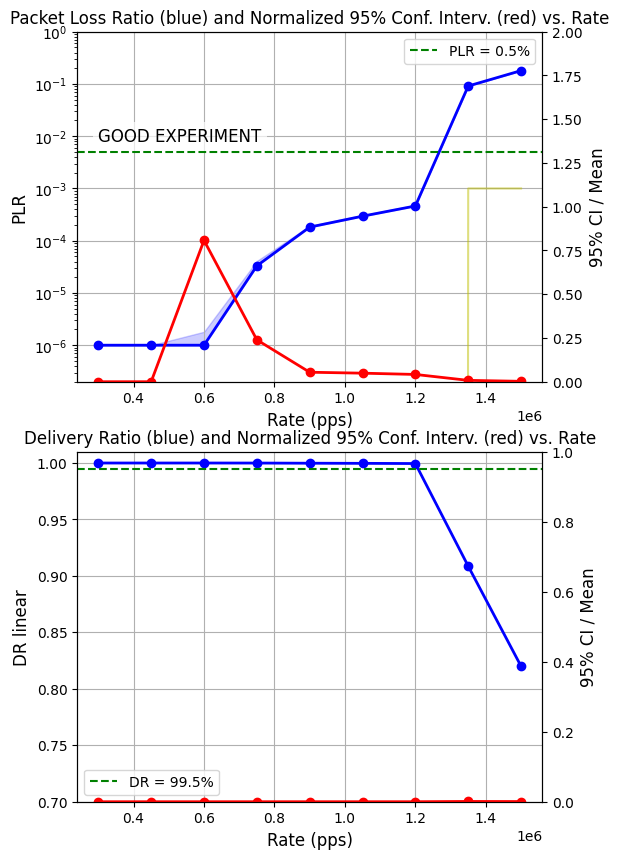}
    \caption{routing, mazinga loc srv, container, Testbed 0, 50 runs, T = 20s, kernel 6.8, unpinned, LARGE nic buffer, date 0000-00-00, version 00}
    \label{fig:mazinga_raw_50_exp_t_20_k6.8_intel_ns_nic_buf_4096}
\end{figure}

\begin{figure}[ht!]
    \centering
    \includegraphics[width=\linewidth]{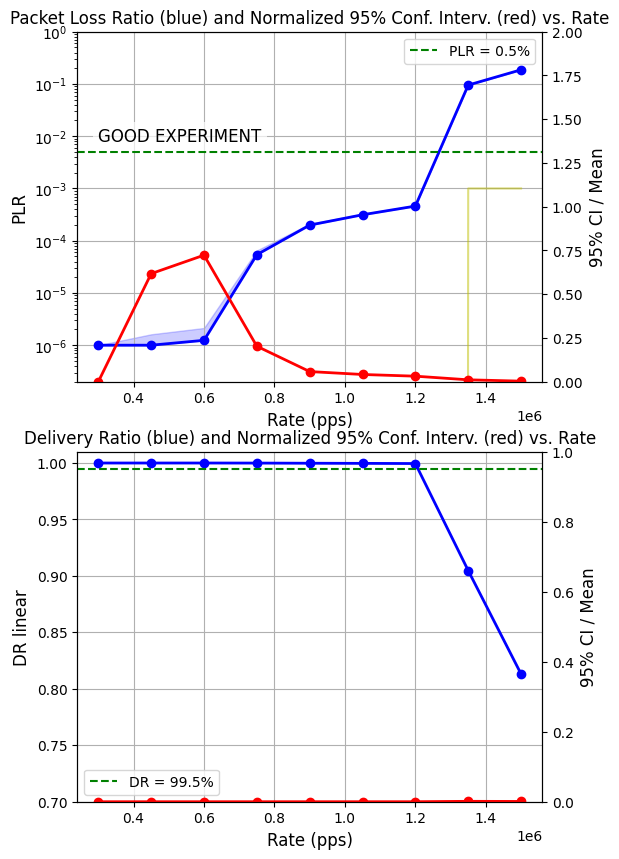}
    \caption{routing, mazinga loc srv, container, Testbed 0, 50 runs, T = 20s, kernel 6.8, pin-1cpu, LARGE nic buffer, date 0000-00-00, version 00}
    \label{fig:mazinga_raw_50_exp_t_20_k6.8_intel_ns_nic_buf_4096_irq_vf0-33_vf1-33}
\end{figure}

\begin{figure}[ht!]
    \centering
    \includegraphics[width=\linewidth]{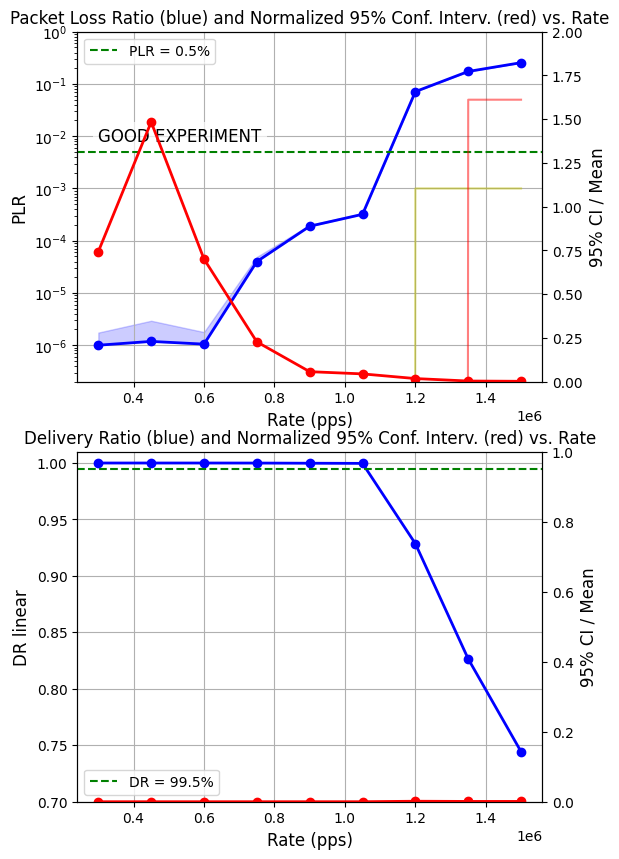}
    \caption{routing, mazinga loc srv, container, Testbed 0, 50 runs, T = 20s, kernel 6.8, pin-2cpu, LARGE nic buffer, date 0000-00-00, version 00}
    \label{fig:mazinga_raw_50_exp_t_20_k6.8_intel_ns_nic_buf_4096_irq_vf0-33_vf1-35}
\end{figure}

}

\end{document}

\section*{Acknowledgment}
    This work has received funding from the Cisco University Research Program Fund.

\ifCLASSOPTIONcaptionsoff
  \newpage
\fi



\bibliographystyle{IEEEtran}
\bibliography{references}

\end{document}